%% file: ms.tex
\documentclass[apj]{emulateapj}
\usepackage{apjfonts}
\usepackage{graphicx}

\newcommand{\teff}{T_{\rm eff}}
\renewcommand{\bv}{B\, -\, V}
\newcommand{\vi}{V\, -\, I_C}
\newcommand{\vk}{V\, -\, K_s}
\newcommand{\jk}{J\, -\, K_s}
\newcommand{\hk}{H\, -\, K_s}

\newcommand{\gr}{g\, -\, r}
\newcommand{\gi}{g\, -\, i}
\newcommand{\gz}{g\, -\, z}
\newcommand{\ebv}{E(B\, -\, V)}
\newcommand{\evi}{E(V\, -\, I_C)}
\newcommand{\evk}{E(V\, -\, K_s)}
\newcommand{\ejk}{E(J\, -\, K_s)}
\newcommand{\ehk}{E(H\, -\, K_s)}
\newcommand{\dmn}{(m\, -\, M)_0}

\slugcomment{Accepted for Publication in the Astrophysical Journal}
\shorttitle{Distances from Main-Sequence Fitting. V}
\shortauthors{An et~al.}

\begin{document}

\title{The Distances to Open Clusters from Main-Sequence Fitting. V.\\
Extension of Color Calibration and Test using Cool and Metal-Rich Stars in NGC~6791}

\author{Deokkeun An\altaffilmark{1},
Donald M.\ Terndrup\altaffilmark{2},
Marc H.\ Pinsonneault\altaffilmark{2},
Jae-Woo Lee\altaffilmark{3}
}

\altaffiltext{1}{Department of Science Education, Ewha Womans University, 52 Ewhayeodae-gil, Seodaemun-gu, Seoul 03760, Korea; deokkeun@ewha.ac.kr}
\altaffiltext{2}{Department of Astronomy, The Ohio State University, 140 West 18th Avenue, Columbus, OH 43210, USA}
\altaffiltext{3}{Department of Physics and Astronomy, Sejong University, 209 Neungdong-ro, Gwangjin-gu, Seoul 05006, Korea}

\begin{abstract}

We extend our effort to calibrate stellar isochrones in the Johnson-Cousins ($BVI_C$) and the 2MASS ($JHK_s$) filter systems based on observations of well-studied open clusters. Using cool main-sequence (MS) stars in Praesepe, we define empirical corrections to the \citeauthor{lejeune:97} color-effective temperature ($\teff$) relations down to $\teff \sim 3600$~K, complementing our previous work based on the Hyades and the Pleiades. We apply empirically corrected isochrones to existing optical and near-infrared photometry of cool ($\teff \la 5500$~K) and metal-rich ([Fe/H]$=+0.37$) MS stars in NGC~6791. The current methodology relies on an assumption that color-$\teff$ corrections are independent of metallicity, but we find that estimates of color-excess and distance from color-magnitude diagrams with different color indices converge on each other at the precisely known metallicity of the cluster. Along with a satisfactory agreement with eclipsing binary data in the cluster, we view the improved internal consistency as a validation of our calibrated isochrones at super-solar metallicities. For very cool stars ($\teff \la 4800$~K), however, we find that $\bv$ colors of our models are systematically redder than the cluster photometry by $\sim0.02$~mag. We use color-$\teff$ transformations from the infrared flux method (IRFM) and alternative photometry to examine a potential color-scale error in the input cluster photometry. After excluding $\bv$ photometry of these cool MS stars, we derive $\ebv=0.105\pm0.014$, [M/H]$=+0.42\pm0.07$, $\dmn = 13.04\pm0.08$, and the age of $9.5\pm0.3$~Gyr for NGC~6791.

\end{abstract}


\section{Introduction}\label{sec:intro}

Although the theory of stellar structure and evolution is considered one of the most successful developments in astrophysics, there still remains a significant mismatch between theoretical stellar models and the observed main-sequence (MS) of the best studied nearby open clusters. Because MS-fitting on color-magnitude diagrams (CMDs) serves as the principal method of constructing an accurate distance scale in the local universe, identifying and removing underlying systematic errors in theoretical models not only helps us understand properties of individual stars, but also has a profound impact on other fields in astronomy. Since the MS fitting can provide information on both distance and foreground reddening, it is an important and useful way of achieving accurate luminosity calibration, which is demanded in many astrophysical problems.

In this series of papers, we have provided an overall assessment of theoretical stellar models and improved the accuracy of models in the Johnson-Cousins and the 2MASS \citep{skrutskie:06} filter systems. We constructed stellar models using the Yale Rotating Evolutionary Code \citep[YREC;][]{sills:00} suite of programs, which provides consistent results with helioseismic observations and solar neutrino fluxes from the Sun \citep{basu:00,bahcall:01,bahcall:04}.  In \citet[][hereafter Paper~I]{pinsonneault:03}, we verified that YREC models are in agreement with the masses and luminosities of the well-studied Hyades eclipsing binary system, V818~Tau \citep[= vB~22;][]{torres:02}. We further demonstrated that our models provide a good match to spectroscopically determined temperatures of individual Hyades members \citep{paulson:03} with good trigonometric parallaxes from {\it Hipparcos} \citep{debruijne:01}.

While YREC models are in satisfactory agreement with the best available data on a theoretical plane (i.e., mass, luminosity, temperature), the match to data becomes poor when models are compared on an observational plane with broadband colors.  In \citet[][hereafter Paper~II]{pinsonneault:04}, we employed the color-effective temperature ($\teff$) relations of \citet{alonso:95,alonso:96} and \citet{lejeune:97,lejeune:98}, to transform a theoretically predicted $\teff$ into observed broadband colors.  The former color-$\teff$ relation is based on the infrared flux method (IRFM), while the latter is from theoretical spectra with empirical corrections. We found that all of these widely used relations fail to reproduce the observed MS of the Hyades, with differences in colors as large as $0.1$~mag in the Johnson-Cousins and 2MASS filter systems.  Since YREC correctly predicts the mass-luminosity-$\teff$ relation for these stars, a natural explanation for the observed offset is a large systematic error in the adopted color-$\teff$ relations, at least in the parameter space covered by the Hyades members.

To ease the tension between models and observations, we introduced empirical corrections for the color-$\teff$ relations of \citet{lejeune:97,lejeune:98} to match photometry of the Hyades' MS (Paper~II). This is a much simpler, but more practical way of overcoming the difficulty than directly examining stellar atmosphere models that have large theoretical complexities and uncertainties.  Our correction scheme has important limitations, but we demonstrated that models with the Hyades-based color corrections successfully reproduce observed MSs of other nearby, well-studied open clusters (the Pleiades, Praesepe, M67, and NGC~2516) and provide accurate estimates on cluster's distances, reddenings, and metal abundances from photometry \citep[][hereafter Paper~III]{an:07b}.

We extended the Hyades-based color-$\teff$ corrections to hotter and brighter stars using the Pleiades \citep[][hereafter Paper~IV]{an:07a}.  As described in Paper~IV we applied the calibrated models to estimate distances and reddenings for young Galactic open clusters with Cepheid variables, and performed the luminosity calibration of Leavitt's Cepheid period-luminosity (PL) relations.  The PL relations constructed in this way provided a distance to the active galaxy NGC~4258 consistent with its maser-based distance estimate \citep[see also][for a recent revision]{humphreys:13}, essentially closing a loop of the distance scale in the local universe.

\input{tab1.tex}

Table~\ref{tab:library} summarizes a library of stellar isochrones with the previous and current empirical corrections in this series and the accompanying series of papers in the Sloan Digital Sky Survey \citep[SDSS;][]{dr10} $ugriz$ system. The Hyades-based corrections in Paper~II are valid only at $M_V \la 8$ or $\teff \ga 4000$~K because of the magnitude limit of {\it Hipparcos} parallaxes for faint Hyades stars \citep{perryman:98}. The Pleiades could not be used to extend the calibration because its lower mass stars have not yet arrived on the MS, and K dwarfs in the cluster seem anomalously blue in broadband colors \citep[][see also Paper~III]{stauffer:03}.  In \S~\ref{sec:corr} we describe an extension of empirical color calibration to cooler stars ($\teff \ga 3600$~K) using low-mass stars in Praesepe.  This complements our earlier color corrections to YREC, covering a wider range of effective temperature of stars ($3600$~K $\la \teff \la 12000$~K).

\input{tab2.tex}

In \S~\ref{sec:phot} through \S~\ref{sec:msfitting}, we describe the application of the calibrated set of isochrones to NGC~6791, a benchmark metal-rich system in the Galaxy, taking advantage of high-quality optical and infrared (IR) photometry \citep{stetson:03,carney:05} and precise measurements of eclipsing binary systems \citep{brogaard:11,brogaard:12}.  There are a considerable number of studies on astrophysical parameters of the cluster in the literature.  Early photometric studies suggested that the cluster's metallicity is near solar \citep[e.g.,][among others]{twarog:85,geisler:91}, while more recent studies based on low-resolution spectra indicated [Fe/H]$\sim+0.2$ \citep[e.g.,][among others]{garnavich:94,friel:02}. However, the most recent efforts using high-resolution spectroscopy consistently found [Fe/H]$\sim+0.4$.  Table~\ref{tab:feh} summarizes high-resolution spectroscopic metallicity estimates and their errors in the literature. Most of the studies found an $\sim0.1$~dex error in the mean [Fe/H] of the cluster, with additional systematic errors of $\sim0.1$~dex as shown in the last column \citep{gratton:06,origlia:06,brogaard:11,boesgaard:15}.  Throughout the paper we refer to an unweighted mean and a standard deviation ([Fe/H]$=+0.37\pm0.07$) of these measurements as a spectroscopic metallicity of the cluster.

The $\alpha$-element abundances of stars in NGC~6791 are essentially solar \citep{carraro:06,origlia:06,carretta:07,brogaard:11,boesgaard:15}.  Although there is some evidence that [C/Fe] is below the solar value \citep{gratton:06,origlia:06}, a sub-solar [O/Fe] among cluster members is still in dispute \citep[e.g.,][]{gratton:06,origlia:06,geisler:12,boesgaard:15,cunha:15}. In any case, the effect of CNO abundances is likely to be small in most parts of MS compared to a change in the bulk metallicity \citep[e.g.,][]{brogaard:12}. We assumed a scaled solar abundance of the cluster, and used [M/H] and [Fe/H] interchangeably to indicate the bulk metallicity of its stars.

While there has been a convergence in the cluster's metallicity, the foreground reddening is not well constrained. A group of studies found that the mean foreground reddening of NGC~6791 is around $\ebv=0.10$ \citep[e.g.,][among others]{stetson:03,carraro:06}. On the other hand, there is a broad distribution of $\ebv$ estimates centered at $\ebv\sim0.15$ \citep[e.g.,][among others]{kaluzny:95,twarog:07}.  Meanwhile, dust emission maps \citep{schlegel:98} set the integrated line-of-sight extinction toward NGC~6791 ($|b|\approx11\arcdeg$), which is $\ebv\approx0.133$ after a $14\%$ downward revision in the reddening scale \citep{schlafly:10}. Recent estimates of the true distance modulus of the cluster range from $\dmn\sim12.9$ to $\dmn\sim13.1$ \citep[e.g.,][among others]{salaris:04,carney:05,twarog:07,brogaard:11}. A dispersion in these distance estimates is partly due to uncertainty in reddening, since $\Delta \ebv = 0.05$ is translated into $\Delta \dmn \approx 0.12$ in the MS fitting. When it comes to a determination of absolute magnitudes ($M_V$) of stars in the cluster, errors in $M_V$ from distance [$\sigma (M_V) = \sigma_{\dmn} \approx 0.12$] and extinction errors [$\sigma (M_V) = R_V \sigma_{\ebv} \approx 0.16$], both of which are induced by the error in reddening [$\sigma_{\ebv} = 0.05$], are added together (positively correlated) to give $\sigma(M_V)\approx0.28$~mag, where $R_V = 3.26$ is the ratio of total to selective extinction. Certainly, more work is needed to shrink the size of the error in the cluster's $\ebv$.

The primary goal of this paper is to extend empirical color-$\teff$ corrections using cool MS stars in Praesepe and to validate the calibration at super-solar metallicity using multicolor photometry of NGC~6791. In the following section, we describe the extension of color-$\teff$ corrections. In \S~\ref{sec:phot} we describe how we collected and combined optical and IR photometry of NGC~6791. Before we apply color-calibrated models to the photometric data, we compare, in \S~\ref{sec:ecbinary}, theoretical stellar models against precise measurements of eclipsing binary systems in NGC~6791, and show that our interior models are in satisfactory agreement with the data with reasonable assumptions on $\teff$ scale and helium abundances. In \S~\ref{sec:msfitting} we conduct tests on the calibrated models using photometry of NGC~6791 with the cluster's well-known spectroscopic metallicity. We also report our best-fitting photometric estimates of the cluster's reddening, distance, metallicity, and age, along with a thorough evaluation of their errors. These parameters will be used in a subsequent paper (D.\ An et~al.\ 2015, in preparation) to derive accurate luminosities and masses of red-clump stars, providing useful information on the amount of mass loss along the red giant branch in NGC~6791.

\section{Extension of Empirical Color Calibration}\label{sec:corr}

The general scheme of our calibration procedure is described in Paper~II, where we forced a match between computed and observed cluster sequences on CMDs by adjusting color-$\teff$ relations in \citet{lejeune:97,lejeune:98}. Empirically corrected isochrones constructed in this way were designed to match a MS of a calibrating cluster system.  In Paper~II we used the Hyades in the calibration, but the Hyades-based corrections were defined at $M_V \la 8$ or $\teff \ga 4000$~K because of the lack of faint Hyades members with good {\it Hipparcos} parallaxes \citep{perryman:98}. Our new empirical color correction procedure described below extends the calibration further down below this limit by employing CMDs of Praesepe. Praesepe has a similar age ($550$~Myr)\footnote{The age of the Hyades is about $625$~Myr based on models with convective overshooting \citep{perryman:98}, but a difference in age has little impact for low-mass MS stars.} and metallicity ([Fe/H]$=+0.14$, see below) as the Hyades, but exhibits a negligible depth effect at its intermediate distance from the Sun [$\dmn=6.33\pm0.04$; Paper III]. At the hot end ($\teff \ga 8500$~K), the Hyades-based calibration was limited due to the relatively old age of the cluster, so we used young Pleiades stars to extend color corrections to hotter stars (Paper~IV). Our revised color calibrations presented in this paper are based on all of these benchmark open clusters, and supersede the previous color corrections.  Luminosities of the models in these studies are ultimately tied to the {\it Hipparcos} parallaxes to the Hyades $\dmn=3.33\pm0.01$ \citep{perryman:98}.

\subsection{Base Models}\label{sec:model}

The isochrone method assumes that the luminosity-$\teff$ relationship in the models is correct, and maps temperature onto color via the inferred distance, magnitudes, and bolometric corrections. In this paper, we used an updated version of theoretical isochrones generated using the YREC stellar evolutionary code (F.\ Delahaye \& M.\ H.\ Pinsonneault, in preparation). These theoretical models depend on the choice of input physics, in particular the opacity tables and chemical mixture. In this work we have switched to using the atomic opacity tables provided by the Opacity Project \citep[OP;][]{badnell:05} from the OPAL tables \citep{iglesias:96}, which slightly alters the underlying isochrones in the theoretical plane (see the second column in Table~\ref{tab:library}). The current models supercede the prior ones. Detailed information on the OP-based YREC models can be found in \citet{delahaye:05,delahaye:06}; see also \citet{vansaders:12} for an updated discussion of input physics. Even though the opacity approaches are different, they have similar Rosseland means for solar conditions and thus yield similar solar calibrations. In the temperature range where these models were compared to MS stars in NGC~6791 ($3600$~K $\la \teff \la 5600$~K), the biggest difference ($\Delta\teff\sim100$~K) is found at $\teff\sim4600$~K in the sense that the OP-based models predict hotter temperatures. Below we repeated our earlier color calibrations based on the Hyades (Paper~II) and the Pleiades (Paper~IV) using OP-based models for an internal consistency.

The remaining parameters in the models, including the initial chemical mixture \citep{grevesse:98}, mixing length ($\alpha = 1.72$), and helium enrichment parameter ($\Delta Y/\Delta Z = 1.2$), remained the same as in our previous papers of this series. The helium enrichment parameter was derived from the primordial helium abundance \citep[$Y_p = 0.245$;][and references therein]{bono:02,thuan:02} and solar models that lack microscopic diffusion. As shown in Paper~I, models without diffusion correctly predicted luminosities of the vB~$22$ eclipsing binary system in the Hyades.  However, the microscopic diffusion of heavy elements and helium could be important at the age of NGC~6791. Nevertheless, we neglected microscopic diffusion in our analysis, because there is as yet no strong observational evidence of such effects in the cluster \citep[see discussions in][]{brogaard:12}. In any case, models with microscopic diffusion are only few tens of Kelvin hotter in most parts of the cluster's MS \citep[e.g.,][]{chaboyer:01}, while a change of heavy metals in the surface of stars is likely to be small ($\Delta {\rm [Fe/H]} \la 0.05$) along the MS \citep[e.g.,][see their Figure~7]{brogaard:12}.

The YREC evolutionary tracks were interpolated to generate theoretical isochrones at stellar ages from $1$~Gyr to $16$~Gyr.  Models were constructed over $-0.3 \leq {\rm [Fe/H]} \leq +0.5$ in $0.1$~dex increments and at ${\rm [Fe/H]} = +0.75$. Luminosities and $\teff$ were initially converted to $V$, $\bv$, $\vi$, $\vk$, $\jk$, and $\hk$ in the Johnson-Cousins and 2MASS systems using the \citet{lejeune:97,lejeune:98} color-$\teff$ table, where ``merged'' color-$\teff$ relations of their original tables (see Paper~IV) were used in order to produce smooth base isochrones.

\begin{figure*}
\epsscale{0.8}
\plotone{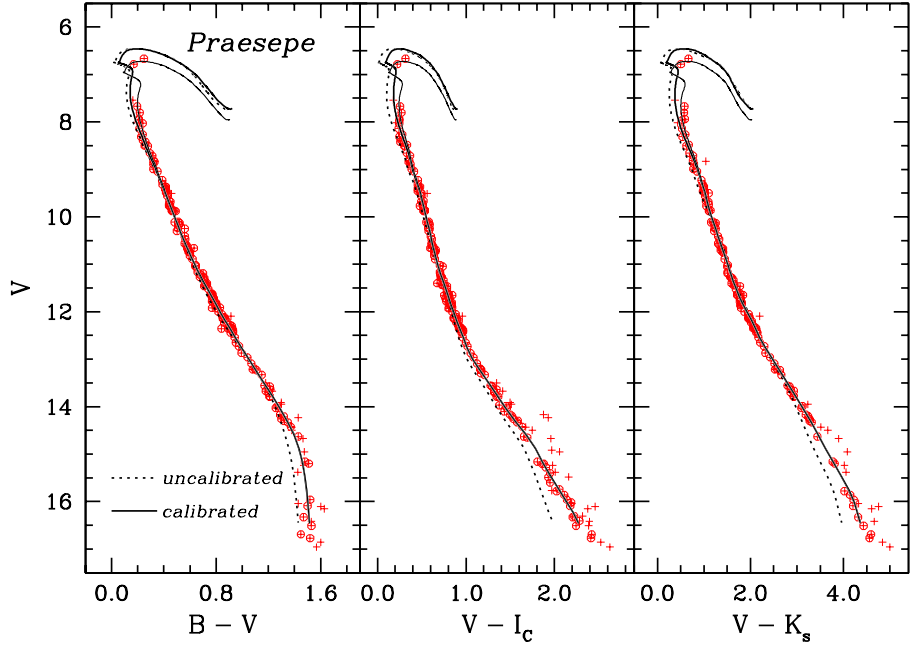}
\caption{CMDs of Praesepe. Plus signs are cluster members after excluding previously known binaries. Open circles are those remained from a statistical filtering routine, which was designed to identify and remove unresolved binaries and/or blended sources independently of models.  Dotted lines are $550$~Myr old models at [Fe/H]$=+0.14$ using \citet{lejeune:97,lejeune:98} color-$\teff$ relations, and thick solid lines represent models with empirical corrections derived in this paper.  For the purpose of comparison, calibrated isochrones at $650$~Myr are additionally shown as a thin solid line. Isochrones are placed at $\dmn=6.33$ assuming $\ebv=0.006$.  \label{fig:prcmd}} \end{figure*}

\begin{figure*}
\epsscale{0.7}
\plotone{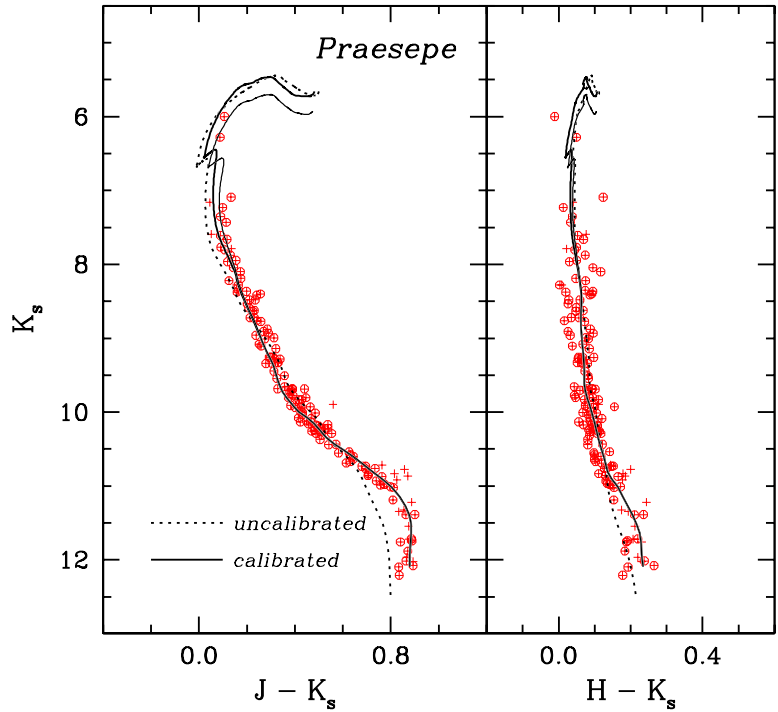}
\caption{Same as in Figure~\ref{fig:prcmd}, but CMDs with $\jk$ and $\hk$ indices.
\label{fig:prcmdjhk}}
\end{figure*}

In Figures~\ref{fig:prcmd} and \ref{fig:prcmdjhk} isochrones are compared to CMDs of Praesepe with $\bv$, $\vi$, $\vk$, $\jk$, and $\hk$ as the color index (hereafter $BV$, $VI_C$, $VK_s$, $JK_s$, and $HK_s$ CMDs, respectively), where we used the same cluster photometry as in Paper~III. Here, we present CMDs of Praesepe to highlight the extension of color-$\teff$ calibration in the lower MS. Interested readers are referred to papers in this series for CMDs of the Hyades and the Pleiades used in the following analysis.

The plus symbols in Figures~\ref{fig:prcmd} and \ref{fig:prcmdjhk} represent cluster members after excluding previously known binaries. Because of extensive membership and binarity data in the literature (see references in Paper~III), most parts of the cluster's CMDs are relatively free of cluster binaries and foreground/background stars except at the bottom of MS. We employed a photometric filtering routine that utilized photometric data in various color indices to identify and reject likely cluster binaries and/or background stars independently of theoretical isochrones (see Paper~III).  Open circles are those remained after the photometric filtering, and were used in the following calibration process.

The dotted lines in Figures~\ref{fig:prcmd} and \ref{fig:prcmdjhk} are YREC isochrones with \citet{lejeune:97,lejeune:98} colors.  The solid lines are models with our empirical color corrections that are described below in detail.  We assumed the same cluster age with no core overshoot ($550$~Myr) as the Hyades \citep[e.g.,][]{mermilliod:81}, and adopted the best estimates on the cluster metallicity (${\rm [Fe/H]} = +0.14\pm0.02$) and foreground reddening [$\ebv = 0.006\pm0.002$] in Paper~III.  More specifically, the cluster metallicity is the weighted mean of our spectroscopic (${\rm [Fe/H]}=+0.11\pm0.03$) and photometrically derived metallicity (${\rm [M/H]}=+0.20\pm0.04$). The former is based on \ion{Fe}{1} and \ion{Fe}{2} line measurements of four MS stars in the cluster, using the Magellan Inamori Kyocera Echelle (MIKE) spectrograph \citep{bernstein:03} on the Magellan 6.5-m Clay telescope.  The photometric metallicity was determined using isochrones with the Hyades-based color calibration. Our adopted metallicity of Praesepe is consistent with more recent determinations in \citet[][${\rm [Fe/H]}= +0.16\pm0.05$]{carrera:11} and \citet[][${\rm [Fe/H]}= +0.12\pm0.04$]{boesgaard:13}, but is lower than the value in \citet[][${\rm [Fe/H]}=+0.27\pm0.10$]{pace:08}.  Our adopted cluster reddening is the average of $\ebv$ estimates in the literature and our own photometric estimates using calibrated isochrones in Paper~III. We also took the cluster's distance [$\dmn = 6.33\pm0.04$] from Paper~III, as derived using the Hyades-based calibrated models with the above mean metallicity and reddening. In Figures~\ref{fig:prcmd} and \ref{fig:prcmdjhk}, $650$~Myr old models with color corrections are shown in thin solid lines to illustrate the age dependence of models.

Throughout the paper, we adopted extinction laws in Paper~IV (equations $2$--$7$). Briefly, these equations have color-excess ratios for zero-color stars from \citet{cardelli:89}, i.e., $\evi/\ebv=1.30$, $\evk/\ebv=2.88$, $\ejk/\ebv=0.56$, and $\ehk/\ebv=0.20$ with $R_V = 3.26$ for the ratio of total-to-selective extinction. They include color terms in the color-excess ratios and $R_V$ from \citet{bessell:98} to take into account effective wavelength shifts in broadband filters for cooler stars. The cluster's $\ebv$ in the following analysis refers to a value for zero-color stars.

\subsection{Empirical Color Calibration}\label{sec:empirical}

\begin{figure*}
\epsscale{0.85}
\plotone{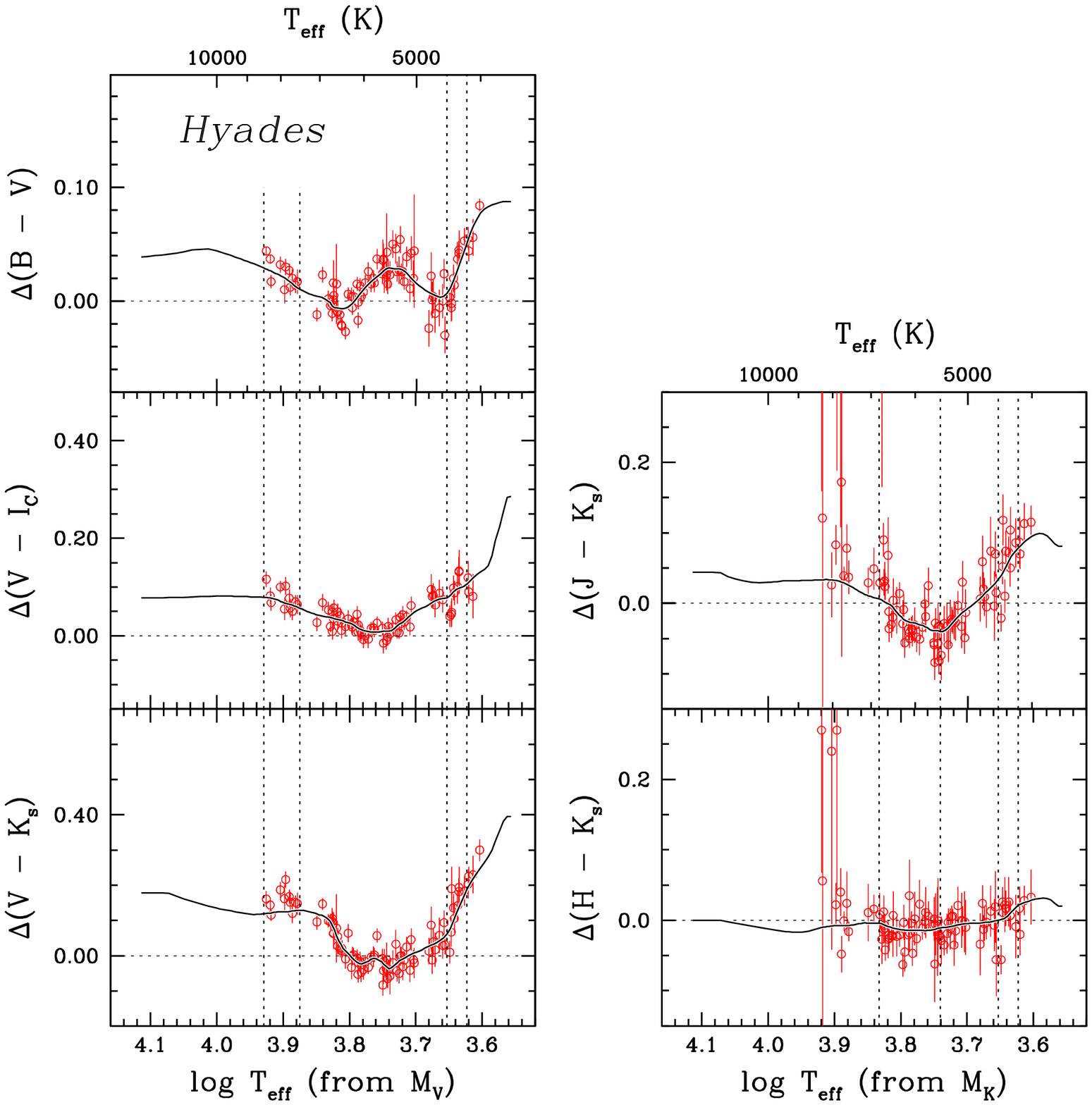}
\caption{Color differences between models using \citet{lejeune:97,lejeune:98} relations and single cluster members in the Hyades with accurate parallaxes. A $550$~Myr old model with [Fe/H]$=+0.13$ was used at $\dmn=3.33$ without foreground dust extinction correction. Color differences were computed at constant $M_V$ for $\bv$, $\vi$, and $\vk$, and those at constant $M_K$ for $\jk$ and $\hk$, in the sense of observed data minus model values.  Solid lines are empirical color-$\teff$ corrections, or moving-averaged points of the systematic color difference, constructed from the three calibrating cluster systems: the Hyades ($4500$~K $\leq \teff \leq 7500$~K, or $\teff \leq 5500$~K in $\jk$ and $\hk$), the Pleiades ($\teff \geq 8500$~K or $6800$~K in $\jk$ and $\hk$; see Figure~\ref{fig:corr_pl}), and Praesepe ($\teff \leq 4200$~K; see Figure~\ref{fig:corr_pr}). The vertical dashed lines indicate these $\teff$ divisions.\label{fig:corr}} \end{figure*}

\begin{figure*}
\epsscale{0.85}
\plotone{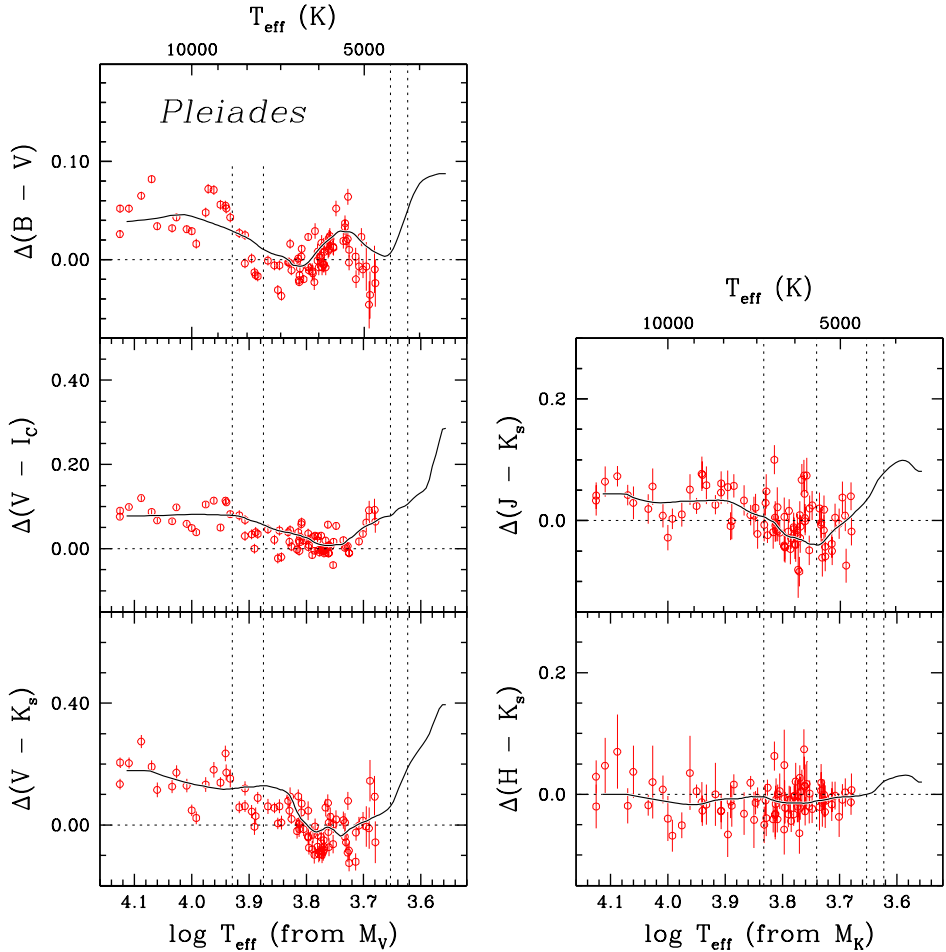}
\caption{Same as in Figure~\ref{fig:corr}, but showing color differences between model and observed colors of single cluster members in the Pleiades.  Stars below $\teff \sim 5000$~K are not displayed because of anomalous colors related to the young age of the cluster.  A $100$~Myr old model at [Fe/H]$=+0.04$ was used with $\dmn=5.63$ and $\ebv=0.032$. \label{fig:corr_pl}} \end{figure*}

\begin{figure*}
\epsscale{0.85}
\plotone{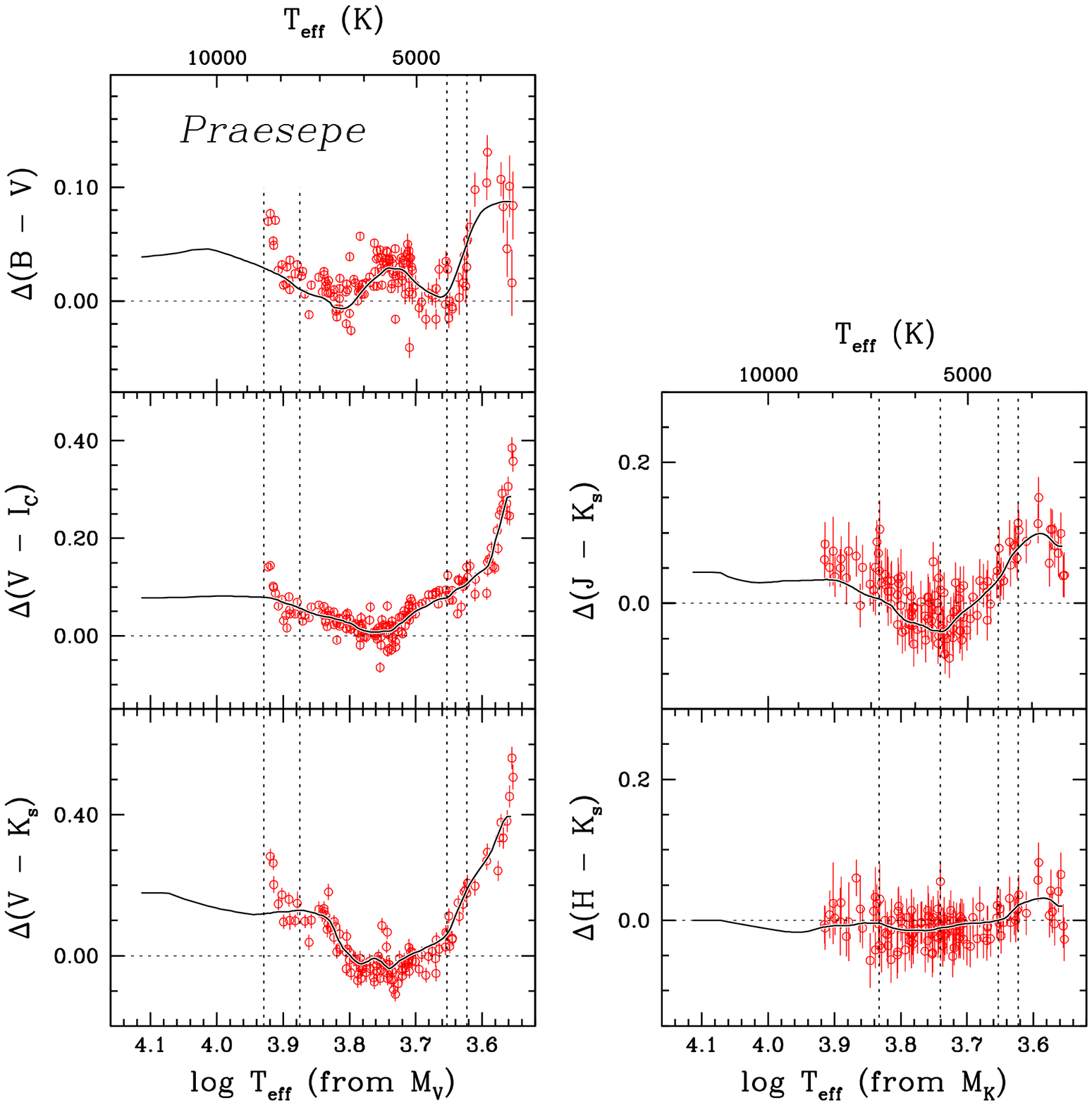}
\caption{Same as in Figure~\ref{fig:corr}, but showing color differences between model and observed colors of single cluster members in Praesepe.  A $550$~Myr old model at [Fe/H]$=+0.14$ was used with $\dmn=6.33$ and $\ebv=0.006$.  \label{fig:corr_pr}} \end{figure*}

Figures~\ref{fig:corr}--\ref{fig:corr_pr} show differences between theoretical and observed colors of individual stars in the Hyades, the Pleiades, and Praesepe, respectively. Color differences (in the sense of observed colors minus model values) were computed using models without color calibration (e.g., dotted lines in Figures~\ref{fig:prcmd} and \ref{fig:prcmdjhk}) at constant $M_V$ in $\bv$, $\vi$, and $\vk$ colors, and at constant $M_K$ in $\jk$ and $\hk$.  The temperatures on the abscissa are $\teff$ inferred from $M_V$ or $M_K$ in the models. We adopted the same cluster parameters (age, metallicity, reddening, and distance) of the Hyades and the Pleiades as in the previous papers of this series: $550$~Myr, ${\rm [Fe/H]}=+0.13\pm0.01$, $\ebv = 0.000\pm0.002$, $\dmn = 3.33\pm0.01$ for the Hyades, and $100$~Myr, ${\rm [Fe/H]}=+0.04\pm0.02$, $\ebv = 0.032\pm0.003$, $\dmn = 5.63\pm0.02$ for the Pleiades. The Pleiades' distance is the average geometric distance to the cluster (see Paper~III), and is consistent with the most recent determination of trigonometric parallax using very long baseline radio interferometry \citep{melis:14}. Single cluster members of the Hyades after excluding known binaries are shown in Figure~\ref{fig:corr}.  Photometrically selected single stars are shown for the Pleiades in Figure~\ref{fig:corr_pl} after rejecting known cluster binaries and non-members (see references in Paper~III).  Cool MS stars in the Pleiades with $\teff \la 5000$~K are not displayed, because the young age of the cluster makes broadband colors of its low-mass stars deviate from those observed in older systems \citep[][see also Paper~III]{stauffer:03}.

As shown in Figures~\ref{fig:corr}--\ref{fig:corr_pr}, color deviations of the models from the three cluster systems look similar, suggesting a systematic origin of these offsets.  We defined empirical corrections for the \citet{lejeune:97,lejeune:98} color-$\teff$ relations as these systematic color residuals, and are shown as solid lines. The detailed steps in constructing these curves are described below.

The vertical dotted lines in Figures~\ref{fig:corr}--\ref{fig:corr_pr} show $\teff$ ranges, in which each of the calibrating cluster systems was employed in the empirical color-$\teff$ corrections.  At $4500$~K $\leq \teff \leq 7500$~K, we employed the Hyades, and computed moving averages of color differences using $4$--$8$ data points in each moving window, with a $50\%$ overlap between adjacent windows. The upper $\teff$ limit was set to $\teff=5500$~K in $JK_s$ and $HK_s$ CMDs, because of large photometric errors of saturated stars in 2MASS. We used $7$-point moving averages of the Pleiades stars with $50\%$ overlaps to define color corrections at $\teff \geq 8500$~K in $BV$, $VI_C$, and $VK_s$ CMDs, and at $\teff \geq 6800$~K in the $JK_s$ and $HK_s$ CMDs. At $7500$~K $\leq \teff \leq 8500$~K, the Pleiades stars show smaller color deviations than those in the Hyades, which may be due to a younger age and/or a lower metallicity of the Pleiades than the Hyades, but other error sources, such as a photometric zero-point error, may also be responsible for the discrepancy. We used a linear ramp to bridge over the gap ($7500$~K $< \teff < 8500$~K or $5500$~K $< \teff < 6800$~K), where the ramped curves simply represent intermediate color deviations from the two clusters.

The systematic trend observed in Praesepe (Figure~\ref{fig:corr_pr}) is in good agreement with that of the Hyades (Figure~\ref{fig:corr}) over $4000$~K $\la \teff \la 8000$~K. Given the similar age and metallicity of the two systems, the observed similarities strongly suggest that the color residuals are systematic in nature. We employed Praesepe to define color corrections at $\teff \le 4200$~K with $4$-point moving averages and $50\%$ overlaps between adjacent moving windows.  A linear ramp was used over $4200$~K $\leq \teff \leq 4500$~K to combine color corrections from Praesepe and the Hyades.

\input{tab3.tex}

We smoothed the moving-averaged points by taking a linear interpolation of five neighboring points. We used three moving-averaged points in the smoothing in the upper and the lower ends of $\teff$. Our final set of empirical corrections are shown as solid lines in Figures~\ref{fig:corr}--\ref{fig:corr_pr}, and are tabulated in Table~\ref{tab:corr}. We assumed that our empirical corrections are independent of metallicity and age (see below), and applied to all isochrones generated using YREC and the \citet{lejeune:97,lejeune:98} color transformations.\footnote{Available at http://home.ewha.ac.kr/\~{}deokkeun/astro/isochrone.html.}

In Paper~III we already demonstrated that isochrones with the Hyades-based corrections provide excellent matches to the MS of nearby open clusters, including the Pleiades (see Figure~\ref{fig:corr_pl}), Praesepe (see Figure~\ref{fig:corr_pr}), M67 ([Fe/H]=$+0.00\pm0.01$, $4$~Gyr), and NGC~2516 ([Fe/H]$=-0.04\pm0.05$, $140$~Myr). This implies that our empirical color corrections are primarily a function of $\teff$ in the metallicity range covered by these systems ($-0.04 \leq {\rm [Fe/H]} \leq +0.14$). However, it is necessary to check and validate models beyond the above metallicity range. The main goal of this work is to perform a stringent test of empirical corrections at the high metallicity end using NGC~6791.

In Figures~\ref{fig:prcmd} and \ref{fig:prcmdjhk}, isochrones incorporating these corrections are shown as solid lines on the Praesepe CMDs. Praesepe was solely used to define color offsets at $\teff < 4200$~K, which correspond to $\bv\approx1.3$, $\vi\approx1.6$, $\vk\approx3.3$, and $\jk\approx0.8$.  Colors are based only on the Hyades stars at $0.3\la\bv\la1.2$, $0.4\la\vi\la1.3$, $0.8\la\vk\la2.8$, and $0.4\la\jk\la0.7$.  Model colors near the cluster's MS turn-off hinge on calibrations from both the Hyades and the Pleiades. As noted above, the two systems predict slightly different color corrections in this region, which partly explains why our $550$~Myr old model shows a color offset near the cluster's turn-off.  However, this offset should not affect our results in the following analysis, because old systems like NGC~6791 have MS stars cooler than $\teff\sim7000$~K.

\subsection{Comparison with IRFM Color-$\teff$ Relations}

The IRFM has been widely used to obtain temperatures of stars based on photometric observations, with little dependence on theoretical stellar atmosphere models \citep[see references in][]{casagrande:10}. Considering that $\teff$ is a defined quantity from $L=4\pi R^2 \sigma \teff^4$, the IRFM temperatures are arguably closer to a fundamental scale than spectroscopic temperatures, either based on an excitation equilibrium of \ion{Fe}{1} lines or Balmer line profiles. On the other hand, a fundamental scale can be obtained independently using observations of well-studied cluster systems with accurate distance, reddening, and metallicity, as described in this series of papers.

The two independent approaches yield the same colors of the Sun within errors. At the age of the Sun ($4.57$~Gyr), colors from our calibrated solar metallicity isochrone are $\bv=0.651$, $\vi=0.704$, $\vk=1.563$, and $\jk=0.336$ at $T_{{\rm eff},\odot}=5777$~K. The IRFM colors of the Sun in \citet{casagrande:10} are $\bv=0.641\pm0.024\pm0.004$, $\vi=0.690\pm0.016\pm0.004$, $\vk=1.544\pm0.018\pm0.010$, and $\jk=0.362\pm0.029\pm0.003$, where errors in each quantity represent a random and a systematic error, respectively.

\begin{figure*}
\epsscale{0.9}
\centering
\plottwo{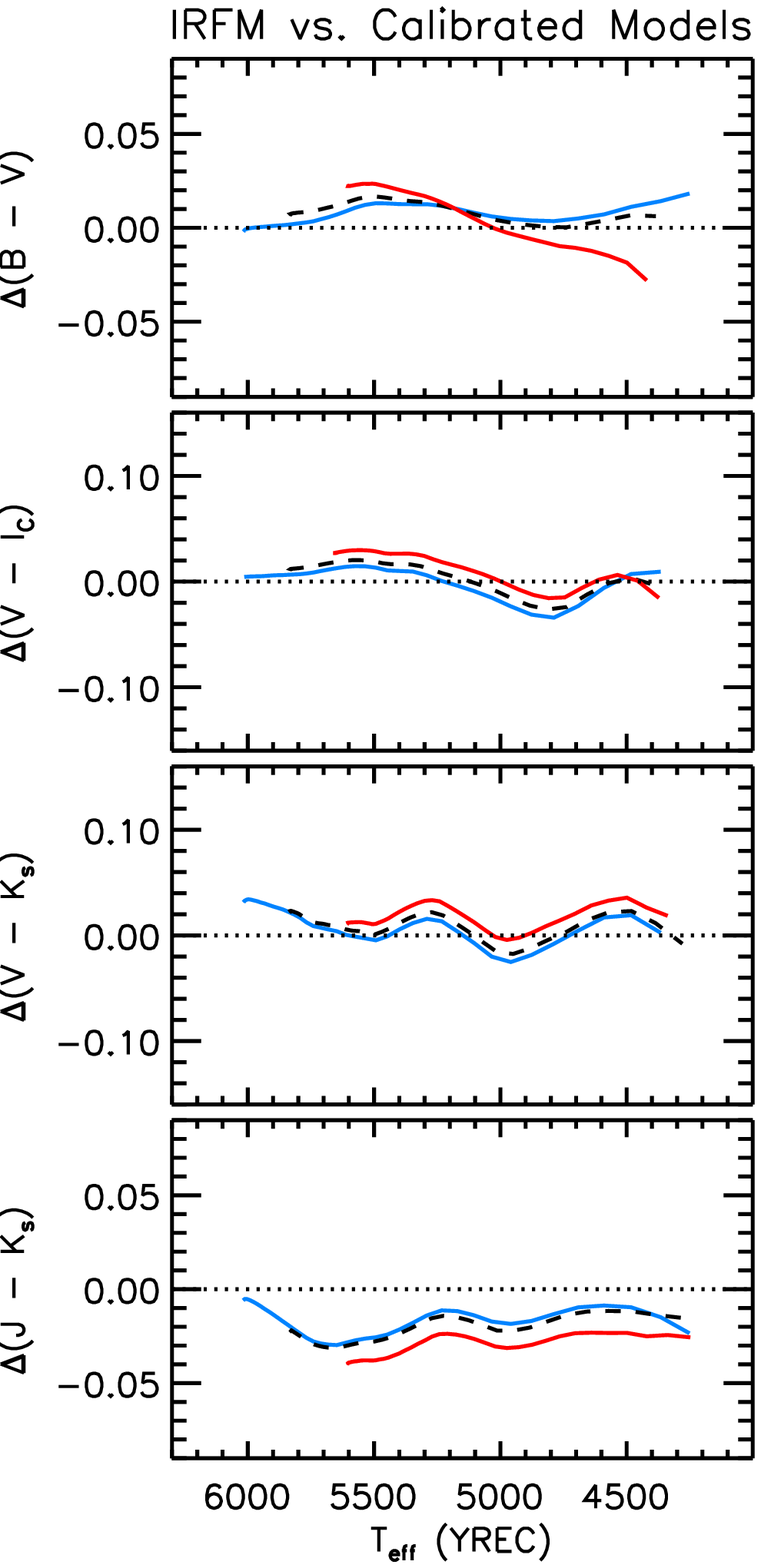}{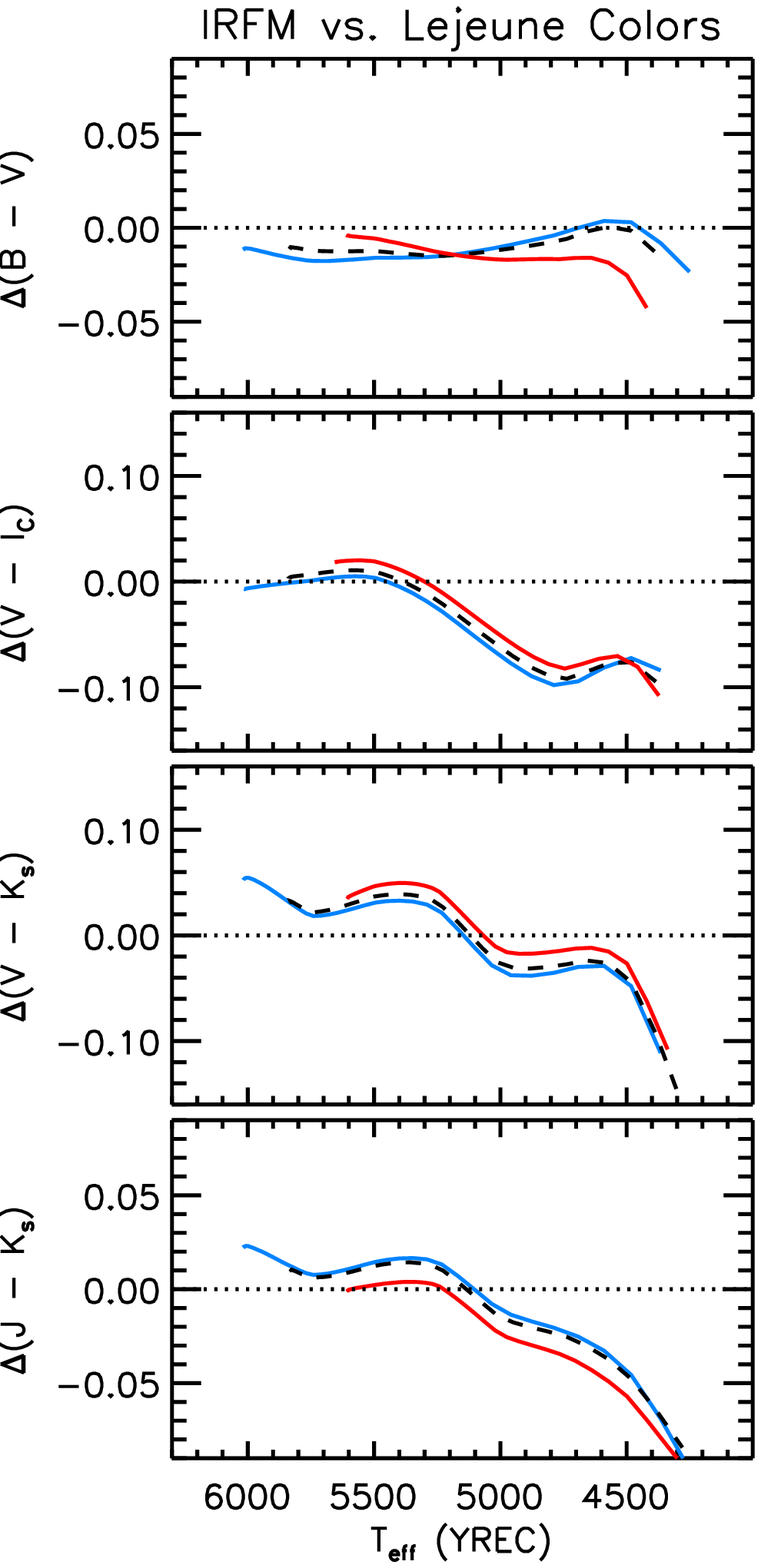}
\caption{Comparisons between IRFM \citep{casagrande:10} colors and those from 9~Gyr old models in $\bv$, $\vi$, $\vk$, and $\jk$, in the sense of isochrone colors minus IRFM values.  Color differences from calibrated models and those using \citet{lejeune:97,lejeune:98} colors are displayed in the left and right panels, respectively. Comparisons are shown at three different metallicities: [Fe/H]$=-0.3$ (blue solid line), [Fe/H]$=0.0$ (black dashed line), and [Fe/H]$=+0.4$ (red solid line; [Fe/H]$=+0.3$ in $\Delta \vi$). \label{fig:irfmcomp}} \end{figure*}

In the left panels of Figure~\ref{fig:irfmcomp} our calibrated color-$\teff$ relations are compared to the IRFM relations in \citet{casagrande:10} over a wide range of $\teff$ in $\bv$, $\vi$, $\vk$, and $\jk$, respectively. In these comparisons, IRFM colors were estimated directly from $\teff$ in the YREC isochrones. We used $9$~Gyr old models, but the comparisons in Figures~\ref{fig:irfmcomp} largely represent differences between the two color-$\teff$ relations, independently of the adopted age of models.  Color differences are shown at [Fe/H]$=-0.3$ (blue solid line), [Fe/H]$=0.0$ (black dashed line), and [Fe/H]$=+0.4$ (red solid line).  Since the IRFM color-$\teff$ relation in $\vi$ is defined up to [Fe/H]$=+0.3$, the $\vi$ comparison is shown at [Fe/H]$=+0.3$ by the red solid line.

As seen in these comparisons, IRFM relations show better agreement with our empirically corrected color-$\teff$ relations than the original \citet{lejeune:97,lejeune:98} relations. In particular, color differences are significantly reduced to $\sim0.02$--$0.04$~mag for cool MS stars ($\teff < 5000$~K). Both IRFM and the cluster-based approaches are limited in the super metal-rich regime by a small number of calibration stars, but the agreement in Figure~\ref{fig:irfmcomp} shows that the two independent color-$\teff$ relations are now converging on a consistent temperature scale.

\begin{figure}
\centering
\includegraphics[scale=0.68]{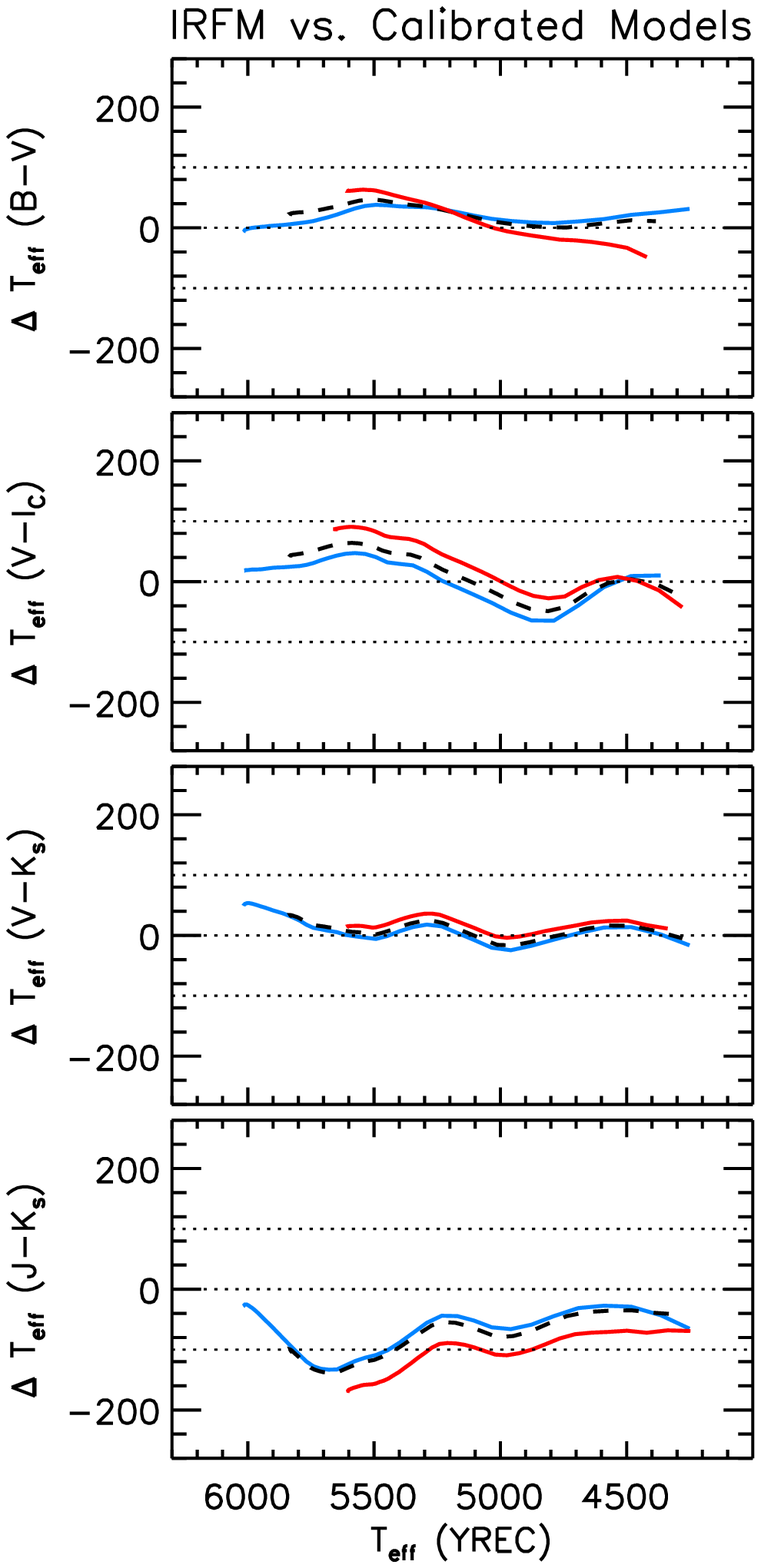}
\caption{Same as in Figure~\ref{fig:irfmcomp}, but showing $\teff$ differences between calibrated isochrones and IRFM relations, in the sense of the former minus latter. The dotted lines indicate $\Delta \teff = 0, \pm100$~K.\label{fig:irfmcomp2}} \end{figure}

The $\teff$ differences between the IRFM and the calibrated stellar isochrones are shown in Figure~\ref{fig:irfmcomp2} in each color index. The temperatures from these two approaches are nearly on the same scale, although isochrones in $\jk$ show systematically higher temperatures at all colors.  Nevertheless, there still remain small-scale structures in the color difference between IRFM and the calibrated isochrones, which may indicate errors in either of these two approaches. Below we directly compare isochrones with an observed MS of NGC~6791 to test the accuracy of color-$\teff$ relations (\S~\ref{sec:msfitting}).

In Figures~\ref{fig:irfmcomp} and \ref{fig:irfmcomp2} the sets of lines with different metallicities are not identical, indicating that our model and IRFM relations have different degrees of sensitivity to metallicity. Since our empirical corrections do not have an explicit dependence on metallicity, the differences show that the metallicity dependence of the \citet{lejeune:97,lejeune:98} relations differs from that of IRFM colors.  The differences are relatively small, but they are of the order of systematic color or $\teff$ differences for a given metallicity. Photometric estimates of cluster parameters (distance, reddening, and metallicity) depend on the adopted sensitivity to metallicity of models, so these two systems should give slightly different photometric solutions. In \S~\ref{sec:msfitting} we discuss the metallicity dependence of models using NGC~$6791$.

Since IRFM relations in \citet{casagrande:10} treat both MS and subgiant stars equally without involving gravity terms, we did not include color or $\teff$ comparisons for subgiant stars in Figures~\ref{fig:irfmcomp} and \ref{fig:irfmcomp2}. The effect of surface gravity on stellar colors is generally milder than that of a temperature or a metallicity change. In \citet{lejeune:97,lejeune:98}, this amounts to $\sim0.01$--$0.02$~mag between MS and subgiant in the above broadband colors. In this paper, we derived empirical corrections based on MS stars, and applied them to subgiant branches in the models with \citet{lejeune:97,lejeune:98} colors. Empirical correction terms that explicitly depend on gravity are expected to be less than $\sim0.01$~mag, if any. Since the following analysis is mainly concerned with MS stars in NGC~6791, systematic errors from surface gravities can be neglected.

\section{Photometry of NGC~6791}\label{sec:phot}

\begin{figure*}
\epsscale{0.95}
\plotone{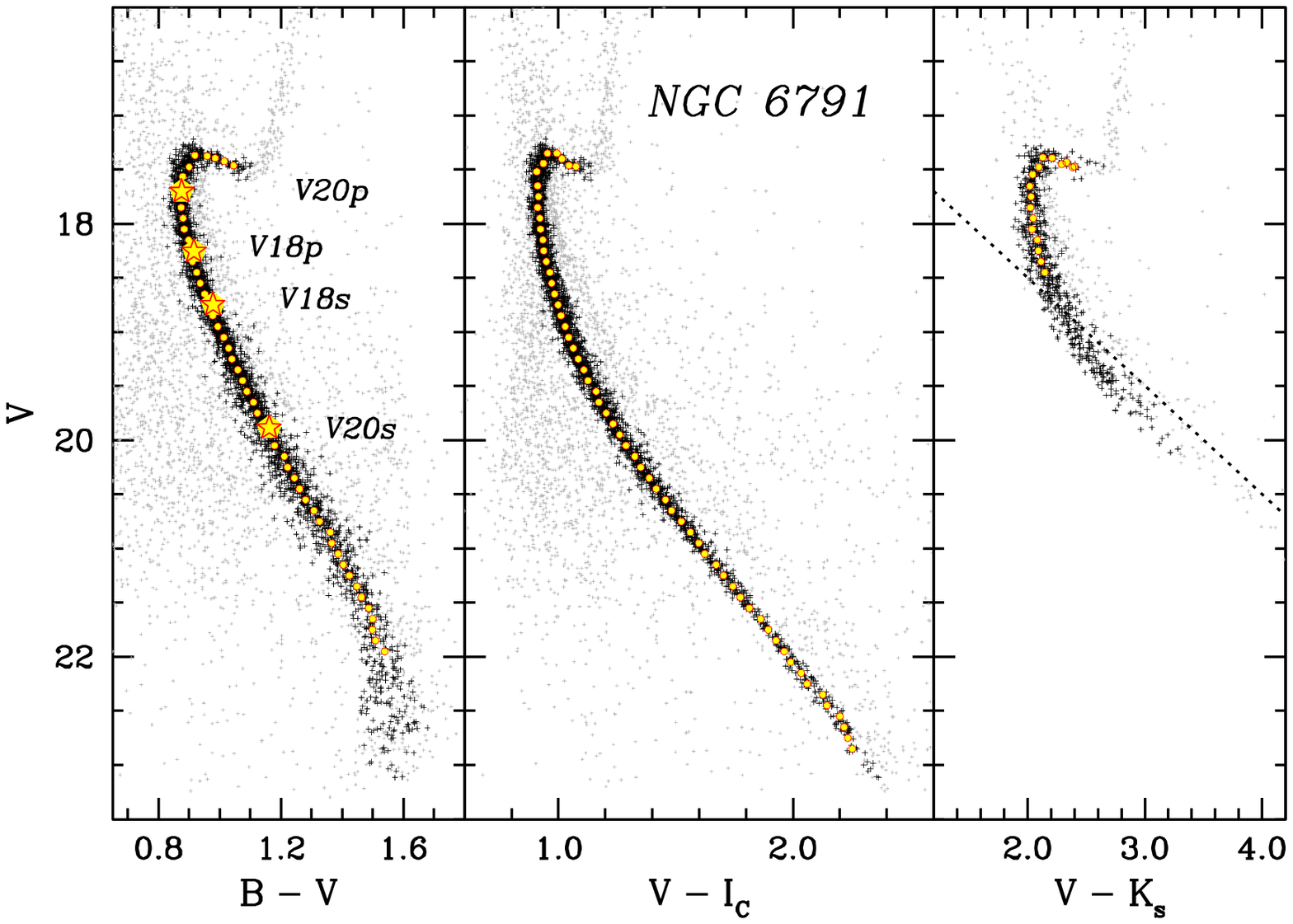}
\caption{CMDs of NGC~6791. The $BVI_C$ data are from \citet{stetson:03}, and $K_s$-band data in the right panel are originally from \citet{carney:05} after transforming their CIT $K$ into $K_s$ magnitudes in the 2MASS system.  Black plus points are stars remaining after the photometric filtering, which trace a single-star sequence of the cluster with little influence from cluster binaries and background stars.  Open circles are cluster sequences derived from these points.  Locations of individual members of V18 and V20 eclipsing binary systems are marked with star symbols. The dotted diagonal line on the right panel represents a completeness limit in $K_s$.  \label{fig:cmd}} \end{figure*}

In our analysis, we utilized optical and near-IR photometry of NGC~$6791$ in the literature.  Figure~\ref{fig:cmd} shows the cluster's CMDs, where $BVI_C$ photometry originally from \citet{stetson:03} is combined with $K$-band magnitudes in \citet{carney:05}. \citet{stetson:05} found color-scale errors in his standard star database, and subsequently reported the update of the cluster photometry in \citet{brogaard:12} with the newer calibration. We took the updated photometry table from Stetson's homogeneous photometry website\footnote{http://www3.cadc-ccda.hia-iha.nrc-cnrc.gc.ca/community/STETSON/} and utilized it throughout the analysis. In this paper, we keep the original reference \citep{stetson:03} to indicate the updated photometric table. Taking the same selection criteria as in \citet{stetson:03} for good photometry, we applied cuts based on $\chi$, {\tt sharp}, {\tt separation} indices, and photometric errors.  We assigned minimum errors of $0.002$~mag in $V$, $\bv$, and $\vi$.  We combined the optical photometry with near IR photometry after transforming $J$- and $K$-band measurements by \citet{carney:05} in the CIT system into $JK_s$ in 2MASS, using a color transformation equation found in \citet{carpenter:01}.\footnote{Updated color transformations can be found at\\ http://www.astro.caltech.edu/~jmc/2mass/v3/transformations/.} The sloping dotted line in the $VK_s$ CMD in Figure~\ref{fig:cmd} represents the completeness limit in \citet{carney:05} at $K=16.5$~mag, which is $\sim2$~mag deeper than the $99.9\%$ completeness limit in 2MASS ($K_s\sim14.3$~mag).  Although the $VK_s$ CMD is limited to the upper part of the cluster's MS, the observed sequence provides useful constraints on the cluster's reddening and metallicity (\S~\ref{sec:prelm}).

Unresolved cluster binaries (or photometric blends) and cluster non-members are sources of bias in the MS fitting. There exists useful information on the cluster membership from proper-motion measurements \citep{platais:11}, but there is no detailed census taken on binaries along the cluster's MS. In order to identify and remove outliers from the cluster MS, we applied a photometric filtering routine (Paper~III) to the $BV$, $VI_C$, and $VK_s$ CMDs.  This step was done independently of the models.  Stars tagged as outliers in one of the CMDs were rejected from all of the CMDs in the above filtering procedure. Black cross points in Figure~\ref{fig:cmd} are those remaining after the photometric filtering, with a final $\chi^2$ threshold value corresponding to $2.5\sigma$ (see Paper~III). As seen in Figure~\ref{fig:cmd}, the photometric filtering routine provides a robust detection and removal of binaries and background stars, even without cluster membership information.  On the other hand, lower mass-ratio binaries could still remain after the filtering process, because they lie near a single-star MS.  We included a systematic error induced by these leftovers in the total error budget of the cluster parameter estimates (see \S~\ref{sec:error}). Because the filtering routine was designed to select single stars on the MS, we rejected outliers by hand near MS turn-off ($V < 17.6$~mag) and on the subgiant branch.

Open circles in Figure~\ref{fig:cmd} represent single-star cluster sequences that were derived using the above remaining stars. At $V \geq 17.5$~mag, we computed the median color of these stars in each magnitude bin having $\Delta V = 0.1$~mag.  We constructed a cluster sequence near MS turn-off by picking a number density peak along the observed sequence.  In the subgiant region, we computed a median $V$ magnitude in each color bin with a bin size of $0.03$, $0.03$, and $0.08$~mag in the $BV$, $VI_C$, and $VK_s$ CMDs, respectively.  These cluster sequences were used in a joint determination of the cluster's distance, metallicity, reddening, and age (\S~\ref{sec:chi}).

\section{Test of Models with Eclipsing Binaries in NGC~6791}\label{sec:ecbinary}

In Paper~I we compared the YREC models with eclipsing binary data for the well-studied system (vB~$22$) in the Hyades, from which we found a reasonably good agreement of theoretical models with observed mass-luminosity and mass-radius relations. This result provided a basis for the empirical corrections to $\teff$-to-color transformations, which improved a match between theoretical and observed colors of stars in nearby clusters. However, NGC~6791 is a more metal-rich system than the Hyades and the other nearby clusters used in our calibration. Since systematic errors in the interior models can be propagated into errors in the stellar colors and magnitudes, a verification of our empirical corrections should be preceded by a test of interior models at the metallicity of NGC~6791.

\begin{figure*}
\epsscale{0.8}
\centering
\plottwo{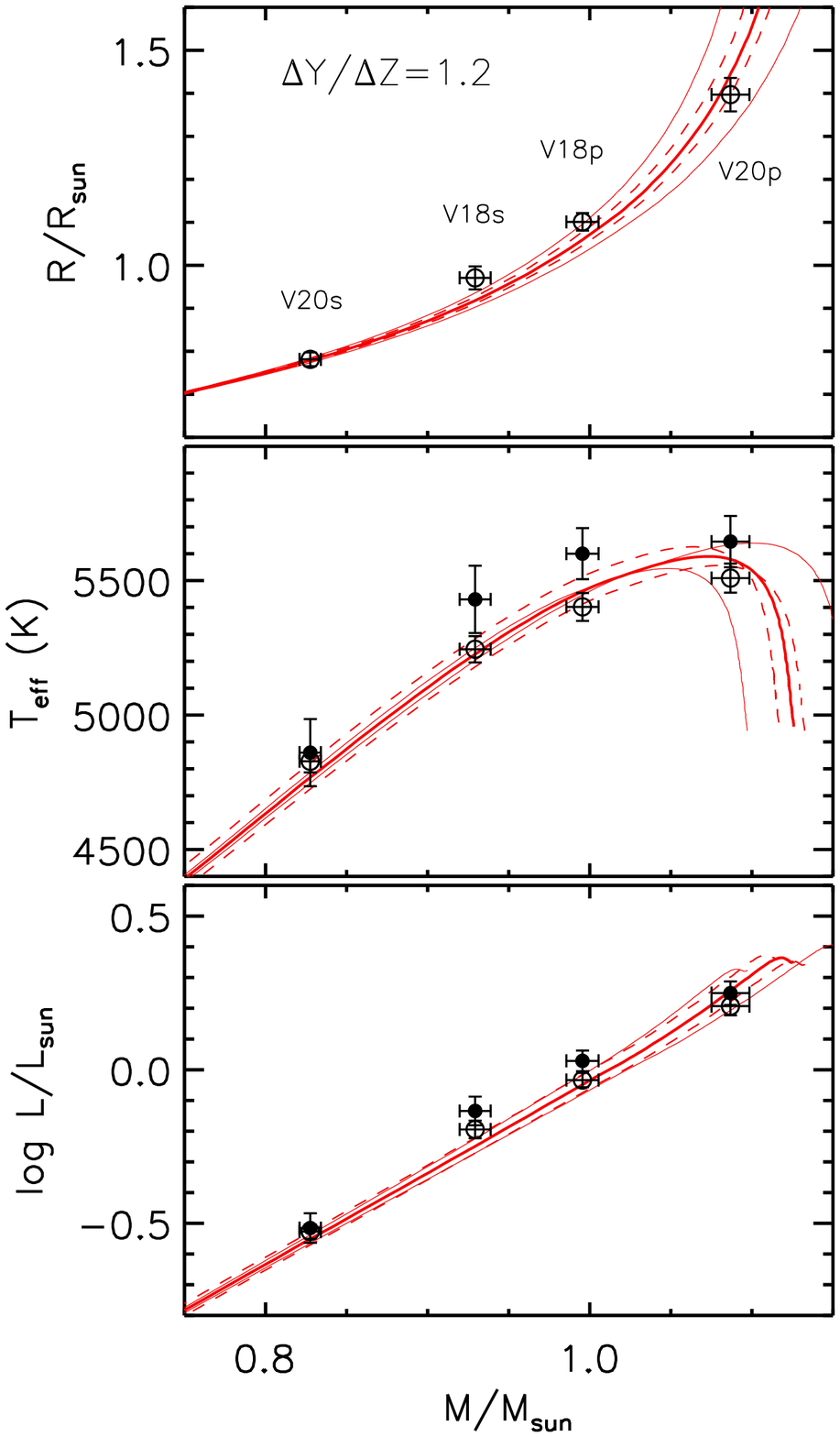}{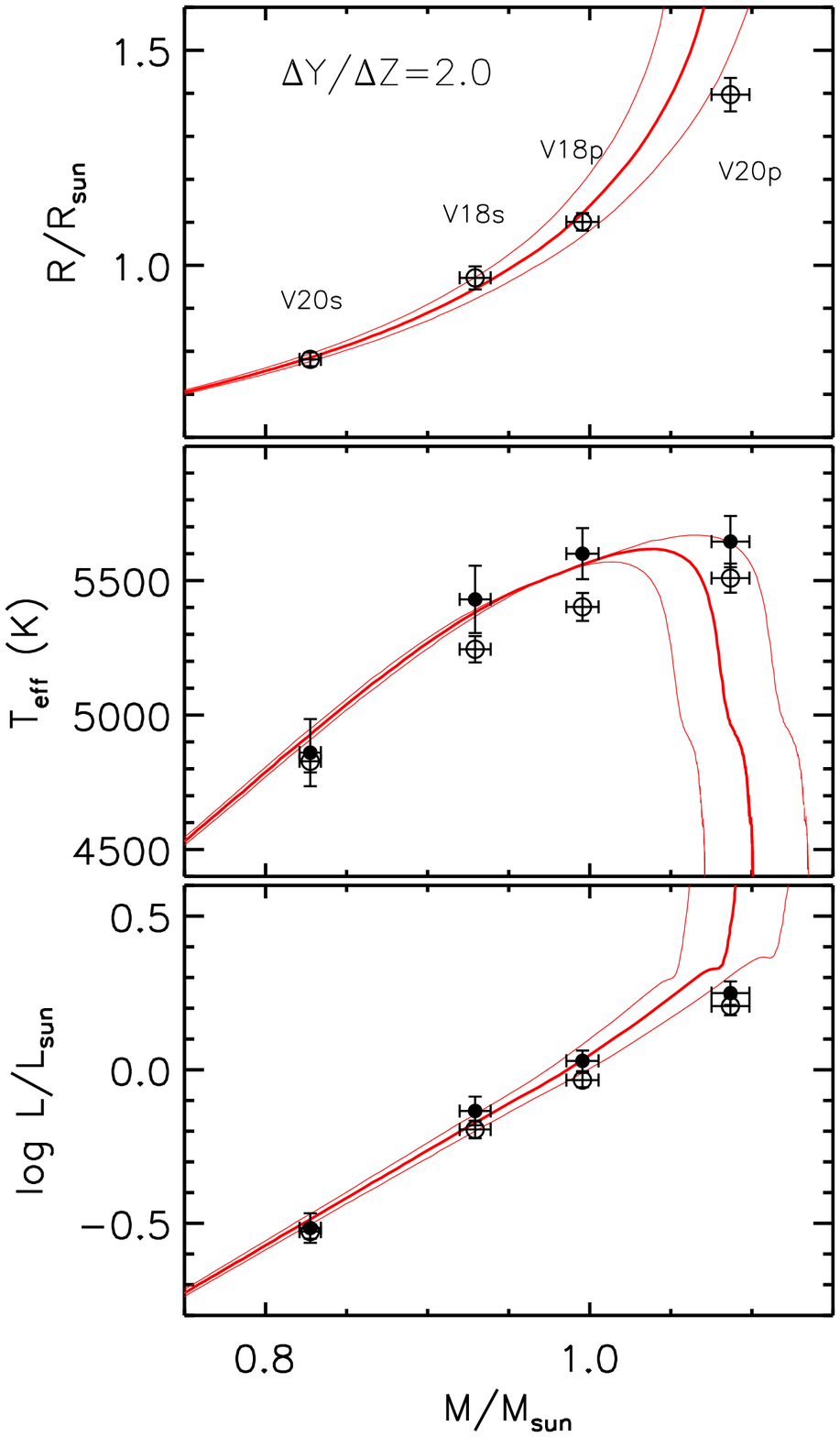}
\caption{Comparisons of theoretical models with observed mass-radius (top), mass-temperature (middle), and mass-luminosity (bottom) relations for V18 and V20 eclipsing binary systems in NGC~6791 \citep{brogaard:11,brogaard:12}. Error bars are shown for $3\sigma$ errors in mass and radius, and $1\sigma$ errors in $\teff$. The error bars in luminosity include a $1\sigma$ error in $\teff$ and a $3\sigma$ error in radius.  {\it Left:} solid lines are $8.5$, $9.5$, and $10.5$~Gyr old isochrones at the spectroscopic metallicity of NGC~6791 ([Fe/H]$=+0.37)$, while dashed lines show $9.5$~Gyr old models with a $\pm1\sigma$ change in metallicity ($\sigma = 0.07$~dex). In the middle and bottom panels, filled circles are measurements based on spectroscopic $\teff$ in \citet{brogaard:12}, while open circles show the case when $\teff$ is estimated using the \citet{casagrande:10} IRFM relation at [M/H]$=+0.37$ and $\ebv=0.11$. {\it Right:} same as in the left panels, but showing helium-enhanced models with $\Delta Y/\Delta Z=2.0$ (see text).}\label{fig:ecbinary} \end{figure*}

Recently, \citet{brogaard:11,brogaard:12} analyzed light curves of V18 and V20, two of the previously known eclipsing binary systems in NGC~6791 with orbital periods of $18.8$~days and $14.5$~days, respectively \citep{rucinski:96}. Their measurements of masses and radii of the individual binary components are shown by open circles in the top left panel of Figure~\ref{fig:ecbinary}. Their reported $1\sigma$ errors are $0.3\%$ in mass and $0.6\%$--$0.9\%$ in radius, but $3\sigma$ measurement uncertainties in masses and radii are displayed as in \citet{brogaard:12}.  They also provided spectroscopic $\teff$ for the individual components, which are shown by filled circles in the middle left panel with $\pm1\sigma$ error bars. In \citet{brogaard:12}, $\teff$ for the secondary of V20 was not directly determined from the spectra, but estimated assuming the same distance as for the primary. In the bottom left panel, filled circles show luminosities computed from their measured radii and spectroscopic $\teff$ measurements.  Errors in luminosity are those propagated from a $3\sigma$ error in radius and a $1\sigma$ error in $\teff$.

The solid lines in the left panels of Figure~\ref{fig:ecbinary} are theoretical YREC models at $8.5$~Gyr, $9.5$~Gyr, and $10.5$~Gyr, respectively, at the mean spectroscopic metallicity of NGC~6791 ([Fe/H]$=+0.37\pm0.07$). The two dashed lines additionally show $9.5$~Gyr models with $\pm1\sigma$ changes in the adopted metallicity.  Models are shown for the MS only, since individual members of V18 and V20 are on the MS (Figure~\ref{fig:cmd}). As shown in the top left panel, the observed mass-radius relation can be matched satisfactorily with $9.5$ or $10.5$~Gyr models, when a $3\sigma$ allowance is made in the comparison with the binary data. If one adopts a model that matches the mass-radius relation of V20 components ($9.5$~Gyr at [Fe/H]$=+0.37$), both V18p and V18s lie systematically above the model line, indicating that our model slightly underestimated  stellar radii. The observed offset resembles earlier findings that standard one-dimensional stellar interior models generally predict $\sim3\%$ smaller radii than actually measured for low-mass stars \citep[][and references therein]{spada:13}, although the binary components in NGC~6791 are more massive than these stars ($\sim0.2\ M_\odot$--$0.8\ M_\odot$).

In the middle left panel of Figure~\ref{fig:ecbinary}, open circles indicate $\teff$ inferred from the IRFM relations in \citet{casagrande:10}.  Since the IRFM relations in $\vi$ are defined only up to [Fe/H]$=+0.3$, we computed IRFM-based $\teff$ at [Fe/H]$=+0.37$ from $\bv$ colors \citep[see][]{brogaard:11} for individual components in the binary systems.  We took errors of $0.02$~mag in $\bv$, and corrected for the foreground extinction using $\ebv=0.11$ (see below). Except for the V20s, IRFM estimates are $\sim130$--$190$~K cooler than the spectroscopic values reported by \citet[][filled circles]{brogaard:11,brogaard:12}, but lower temperatures show better agreement with our models.  As shown in the bottom left panel of Figure~\ref{fig:ecbinary}, larger radii and higher $\teff$ in \citet{brogaard:11,brogaard:12} yield larger luminosities of V18 components than our model predictions.

In fact, the differences between the filled and open circles in Figure~\ref{fig:ecbinary} essentially show a range of solutions available for eclipsing binary systems from independent $\teff$ estimates, since each of the methods is subject to systematic uncertainties in the parameter estimation.  Photometric temperatures from IRFM depend on the foreground extinction correction, but $\teff$ for both V18 components is still $\sim70$~K cooler than spectroscopic estimates, even if we assumed the same reddening [$\ebv=0.16$] as in \citet{brogaard:11}. The IRFM relations are also weakly constrained in this regime, where true empirical constraints would be welcome and are needed for cool, metal-rich stars.  Spectroscopic methods provide an independent check on photometric temperatures, but there could be systematic errors unaccounted for when separating individual components of unresolved binary systems from their spectra.  Given the above limitations in each of these approaches, we conclude that our models are in reasonable agreement with the observed binary data in NGC~6791.

The right panels in Figure~\ref{fig:ecbinary} show the same binary data as in the left panels, but display a set of YREC models at $8.5$, $9.5$, and $10.5$~Gyr assuming $\Delta Y/\Delta Z = 2.0$ \citep[G. Newsham \& D. M. Terndrup 2007, private communication; see also][]{newsham:07}. The corresponding helium mass fraction is $Y=0.32$ at [Fe/H]$=+0.37$ ($Z=0.038$), as opposed to $Y=0.29$ in our base models with $\Delta Y/\Delta Z = 1.2$.  Comparisons with the binary data show that helium-enhanced isochrones better match data with spectroscopic temperatures (filled circles in the middle and bottom panels). However, $\Delta Y/\Delta Z = 2.0$ is significantly steeper than our value constrained from the primordial and the solar helium abundances (see \S~\ref{sec:model}). The amount of helium required is also larger than that of \citet[][$Y=0.30\pm0.01$]{brogaard:12} based on an improved set of Victoria-Regina models \citep{vandenberg:14}. Fortunately, the MS-fitting method, and thus our calibration, is relatively insensitive to the adopted $\Delta Y/\Delta Z$ as discussed in \S~\ref{sec:error}.

Although there still remain unresolved issues about stellar radii, similar quality fits were obtained using other independent models \citep{brogaard:11,brogaard:12,vandenberg:14}. The comparison with the binary data suggests that YREC interior models are in satisfactory agreement with observations. In the next section, we apply color-$\teff$ relations and the proposed empirical color-$\teff$ corrections to YREC models, and proceed to evaluate the accuracy of our color-calibrated YREC models by inspecting the internal consistency of MS-fitting results on CMDs with different color indices. The goal of this exercise is to verify that empirical corrections do not depend, or depend only weakly, on metallicity, and that the corrections derived using the Hyades and Praesepe are valid for NGC~6791.

\section{Test of Models from MS Fitting}\label{sec:msfitting}

In this section, we describe our effort to test the accuracy of the calibrated isochrones using CMDs of NGC~6791 in different colors. We begin our discussion by inspecting the consistency of isochrone fitting in narrow color ranges, where a pair of cluster parameters (reddening versus metallicity and distance modulus versus metallicity) can be constrained with a little dependence on others (\S~\ref{sec:prelm}). We also use IRFM color-$\teff$ relations (\S~\ref{sec:irfm}), independent cluster photometry (\S~\ref{sec:kr95}), and cluster photometry and models in the SDSS filter system (\S~\ref{sec:sdss}) to check the accuracy of our models and the input cluster photometry. We present a global search for the cluster parameters and robust error estimates in Sections~\ref{sec:chi} and \ref{sec:error}, respectively.

\subsection{Preliminary Fitting with Calibrated Isochrones}\label{sec:prelm}

\begin{figure}
\centering
\includegraphics[scale=0.5]{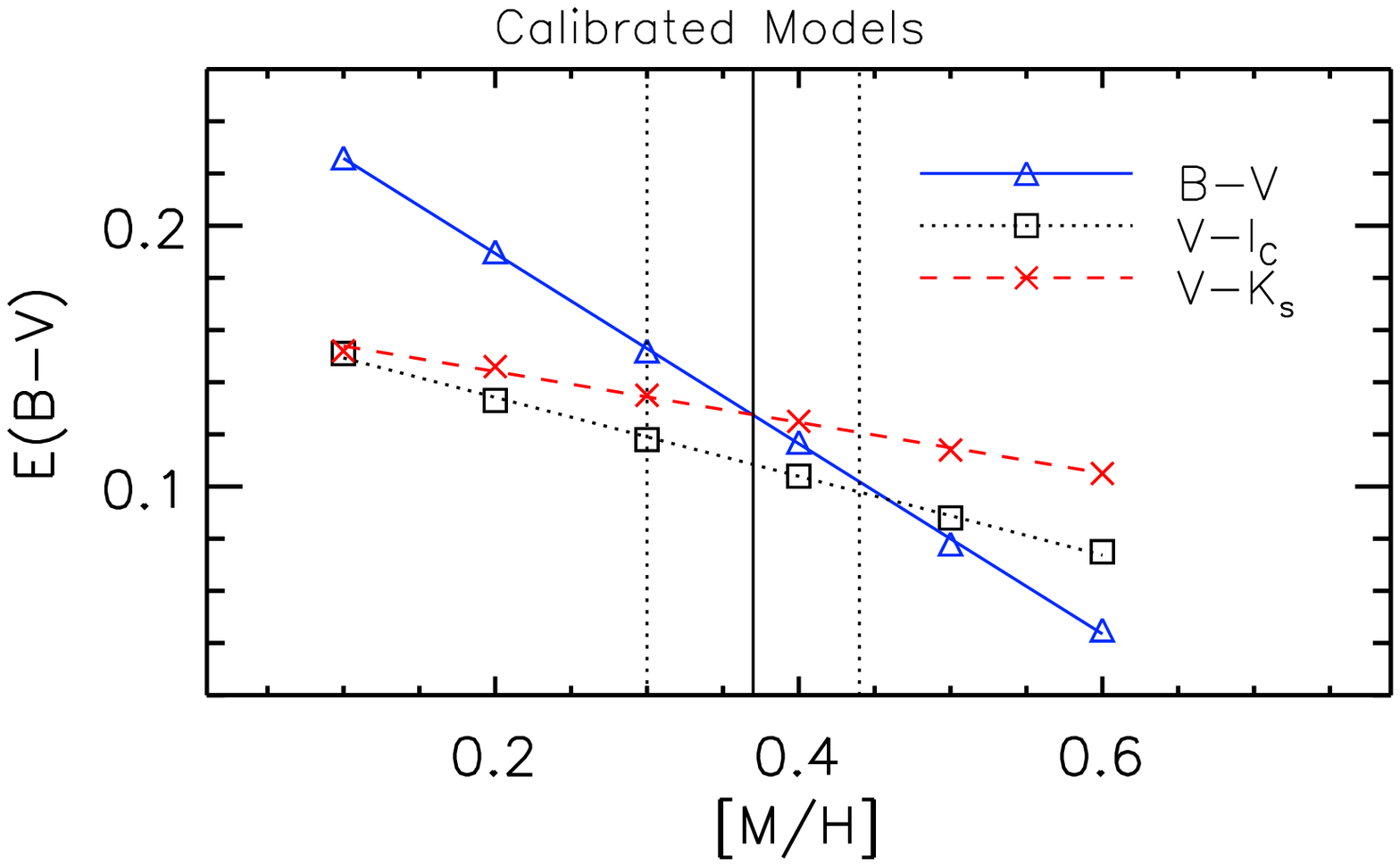}
\includegraphics[scale=0.5]{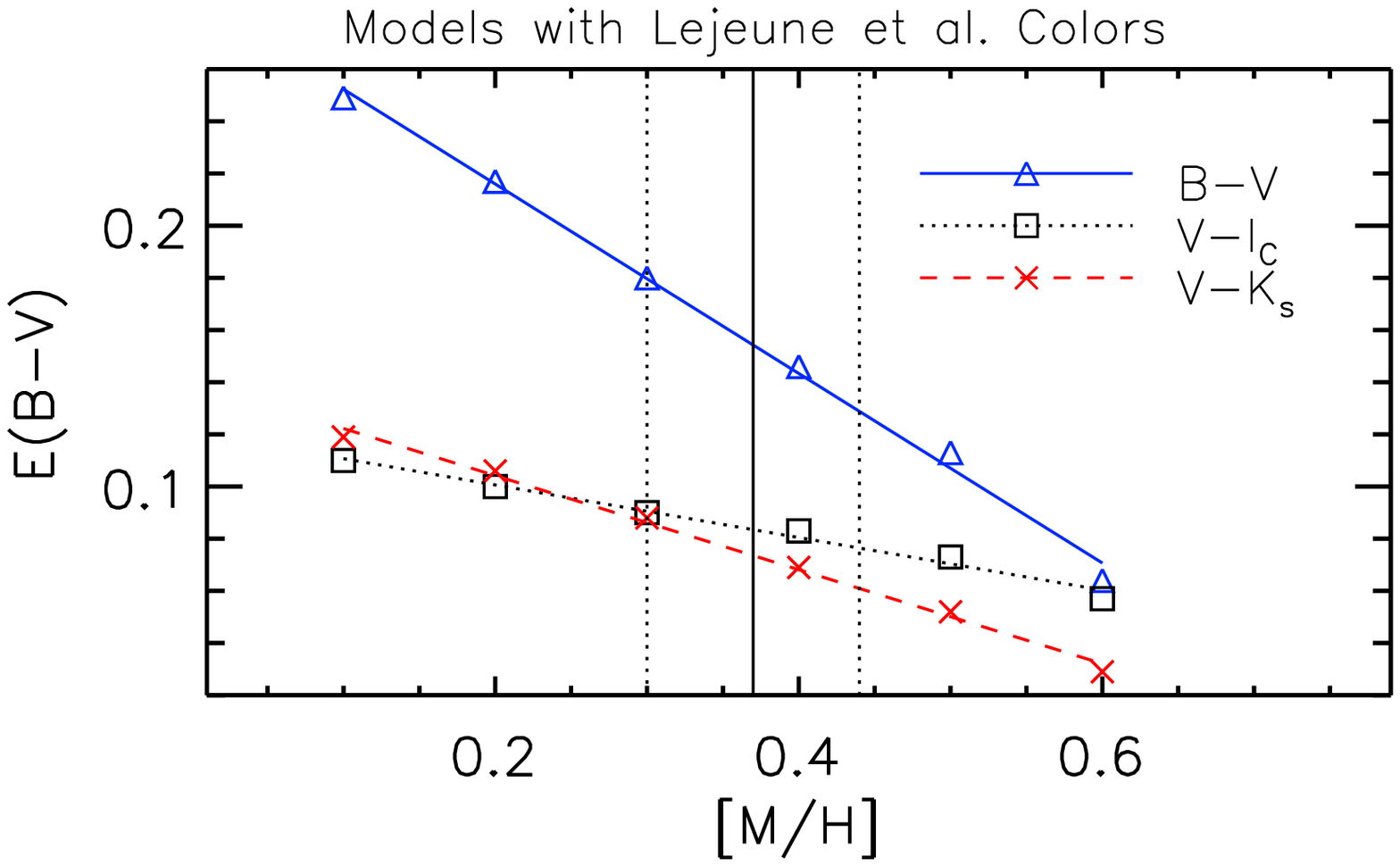}
\caption{{\it Top:} color-excess estimates for NGC~6791 from CMDs with $\bv$ (triangles), $\vi$ (boxes), and $\vk$ (crosses) using $9.5$~Gyr old calibrated isochrones at various metallicity bins. Lines are a linear fit to each set of data points. {\it Bottom:} same as in the top panel, but using models with original color-$\teff$ transformations in \citet{lejeune:97,lejeune:98}. The vertical lines represent the spectroscopic metallicity of the cluster and its $\pm1\sigma$ errors.  \label{fig:fit}} \end{figure}

The top panel in Figure~\ref{fig:fit} shows $\ebv$ estimates as determined from a set of $9.5$~Gyr old calibrated isochrones, spanning a range of input metallicity from $+0.1$ to $+0.6$~dex. Following the procedure described in Paper~IV, we derived a color excess on each of $BV$, $VI_C$, and $VK_s$ CMDs, by shifting a model along a reddening vector at a given distance.  We restricted isochrone fits to $17.8 \leq V \leq 18.5$, making our estimates of color excess insensitive to both age and distance. As described in \S~\ref{sec:model}, we adopted extinction laws with an explicit dependence on colors, which produce slightly redder colors for MS stars in NGC~6791 than those without color terms: $0.008$, $0.005$, $0.008$, and $0.001$~mag in $\bv$, $\vi$, $\vk$, and $\jk$, respectively, when $\ebv=0.1$. The $\ebv$ estimates displayed in Figure~\ref{fig:fit} are those for zero-color stars.

We computed a distance using weighted median statistics, restricting model fits to $1.00 \leq \bv \leq 1.10$ and $1.05 \leq \vi \leq 1.15$ ($4900$~K $\la \teff \la 5200$~K).  This part of the MS has a relatively shallow slope on CMDs, where a distance can be constrained with little dependence on reddening. We used a mean distance modulus obtained from the $BV$ and $VI_C$ CMDs when estimating a color excess on the $VK_s$ CMD.  In this way, we iteratively solved for both color excess and distance of the cluster. The results are shown as triangles, boxed points, and cross marks for $BV$, $VI_C$, and $VK_s$ CMDs, respectively. The solid, dotted, and dashed lines are a linear fit to MS-fitting results on each CMD, as a good approximation of the observed trend in the above metallicity range.

In the top panel of Figure~\ref{fig:fit} the line from the $BV$ CMD has a slope that is steeper than those from the other two color indices. In this preliminary fitting exercise, we used differential sensitivities of colors on metallicity ($\bv$ versus $\vi$ and $\bv$ versus $\vk$) to constrain the cluster's metallicity and foreground reddening (see Papers~III and IV). The line from the $BV$ CMD crosses the $VI_C$ CMD line at [M/H]$=+0.459\pm0.012$, and the reddening at this metallicity corresponds to $\ebv=0.095$. On the other hand, the intersection between the $BV$ and $VK_s$ lines has [M/H]$=+0.369\pm0.009$ and $\ebv=0.128$. A weighted mean photometric metallicity from these two color combinations yields [M/H]$=+0.40\pm0.05$, which is in excellent agreement with the spectroscopic metallicity of the cluster ([Fe/H]$=+0.37\pm0.07$).  The error represents half of the difference between the two photometric estimates, while an error propagated from fitting a model to a CMD is about six times smaller than this value. At the mean photometric metallicity, the reddening of the cluster is $\ebv=0.115\pm0.010$ (scatter from the three CMDs) $\pm0.009$ (propagated from the photometric metallicity error).

Our photometric solution for reddening can also be constrained by adopting the accurate cluster metallicity in the literature.  At the mean spectroscopic abundance ([Fe/H]$=+0.37\pm0.07$), as shown by vertical lines in Figure~\ref{fig:fit},\footnote{Since a number of spectroscopic studies adopted color-$\teff$ relations to calculate temperatures of stars, their abundances are correlated with a foreground reddening of the cluster adopted in each study. In Figure~\ref{fig:fit} we simply use vertical lines to represent a range of spectroscopic metallicity determinations, neglecting a correlation between [Fe/H] and $\ebv$.} the mean reddening from $BV$, $VI_C$, and $VK_s$ CMDs becomes $\ebv=0.121\pm0.011$ (scatter from the three CMDs) $\pm0.014$ (propagated from the metallicity error). The dispersion in $\ebv$ from the three CMDs may reflect small systematic errors unaccounted for in the current isochrone calibration. However, other sources of systematic errors are also likely. For example, \citet{carney:05} made a comparison of their photometry with bright 2MASS stars, and found that their transformed $K_s$-band photometry is $0.014\pm0.005$~mag brighter than 2MASS.  If we allow this change to $\vk$ colors, $\ebv$ from the $VK_s$ CMD would become smaller by $\Delta \ebv = 0.006$, reducing the scatter in the estimates of color excess from the three CMDs at the spectroscopic metallicity of the cluster.

Our $\ebv$ is found within a broad range of previous estimates in the literature (see \S~\ref{sec:intro}). Reassuringly, it is also comparable to or slightly lower than the integrated value [$\ebv=0.133$] from the dust emission map \citep{schlegel:98,schlafly:10}. The cluster is located high above the Galactic plane ($|z| \sim 750$~pc), which is at least five times the scale height of the dust layer in the Galactic disk. Therefore, most of the dust should be found between the Sun and the cluster, and the integrated line-of-sight reddening should serve as a good approximation for the cluster reddening.

We also attempted to estimate $\ebv$ from the $JK_s$ CMD, but our values were always too high compared to those obtained using other color indices: at [M/H]$\sim+0.4$, the $JK_s$ CMD yields $\ebv\sim0.24\pm0.01$. Since $\ejk/\ebv=0.56$, $\Delta \ebv = 0.12$~mag requires a $\sim0.07$~mag change in the zero point of $\jk$ colors.  While a transformation from the UKIRT $JHK$ system to the CIT or 2MASS system is not trivial, \citet{carney:05} found that their $\jk$ photometry is only $0.007\pm0.005$~mag redder than 2MASS. The comparison with IRFM (Figure~\ref{fig:irfmcomp}) shows that our model $\jk$ colors are too blue by $0.02$--$0.03$~mag, but the differences are still too small to explain the large color excess on the $JK_s$ CMD. This may indicate a problem either in the IR color-$\teff$ relations at high metallicity or in the zero point of the photometry.  We leave the discussion on the calibration of IR colors to the next paper of this series.

In contrast, models without empirical color corrections result in a significant inconsistency in the derived reddening as shown in the bottom panel of Figure~\ref{fig:fit}. The $\ebv$ estimates from the $BV$ CMD are systematically higher, while those from the $VI_C$ and $VK_s$ CMDs are lower than those obtained using the calibrated isochrones (top panel). Our empirical corrections do not include metallicity terms, and therefore the slopes of the lines in the bottom panel are similar to those in the top panel. On the other hand, the intercepts of these lines are modified by our color-$\teff$ corrections, leading to an improved internal consistency of $\ebv$ estimates at the spectroscopic metallicity of the cluster (top panel).

\begin{figure}
\centering
\includegraphics[scale=0.5]{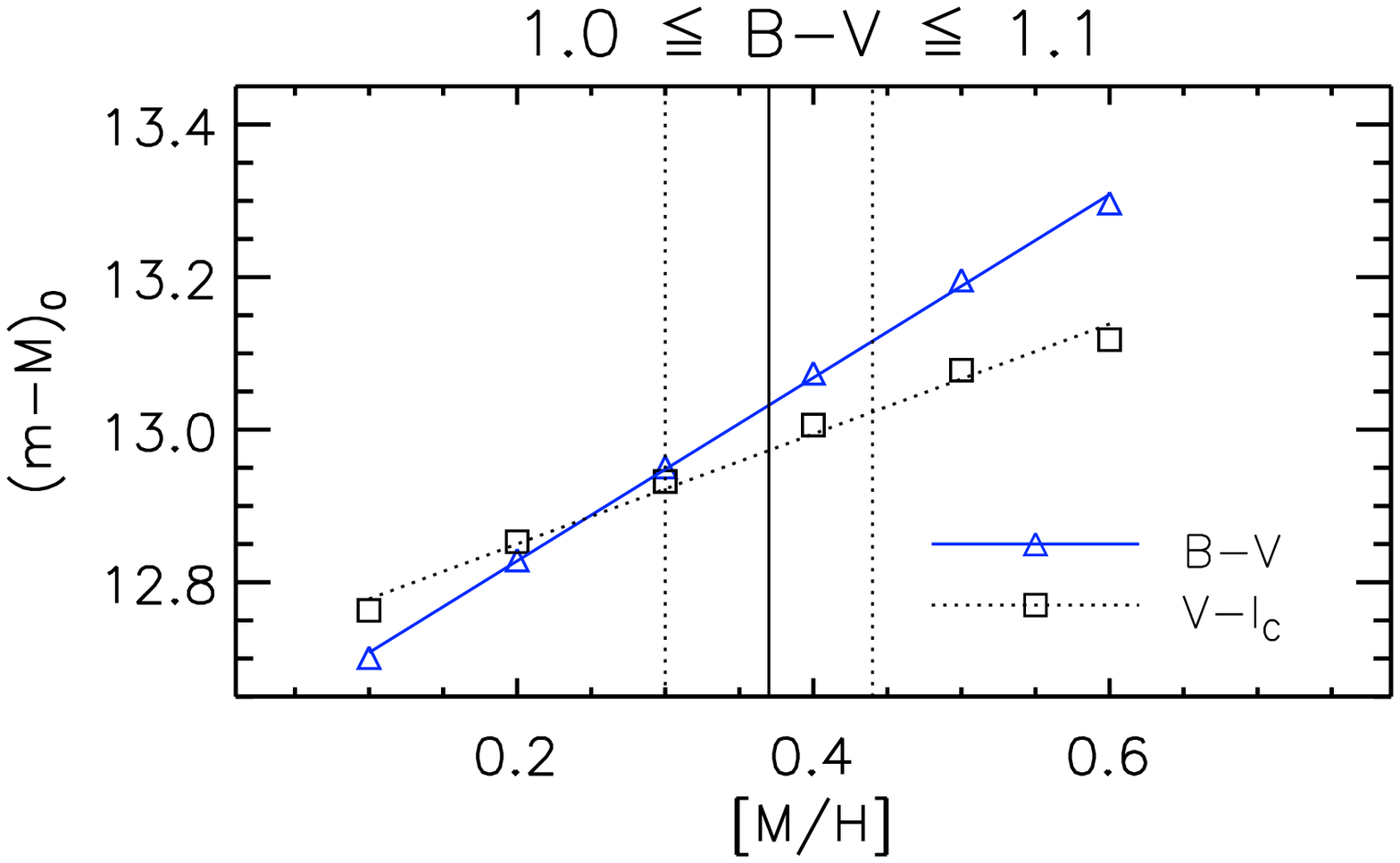}
\includegraphics[scale=0.5]{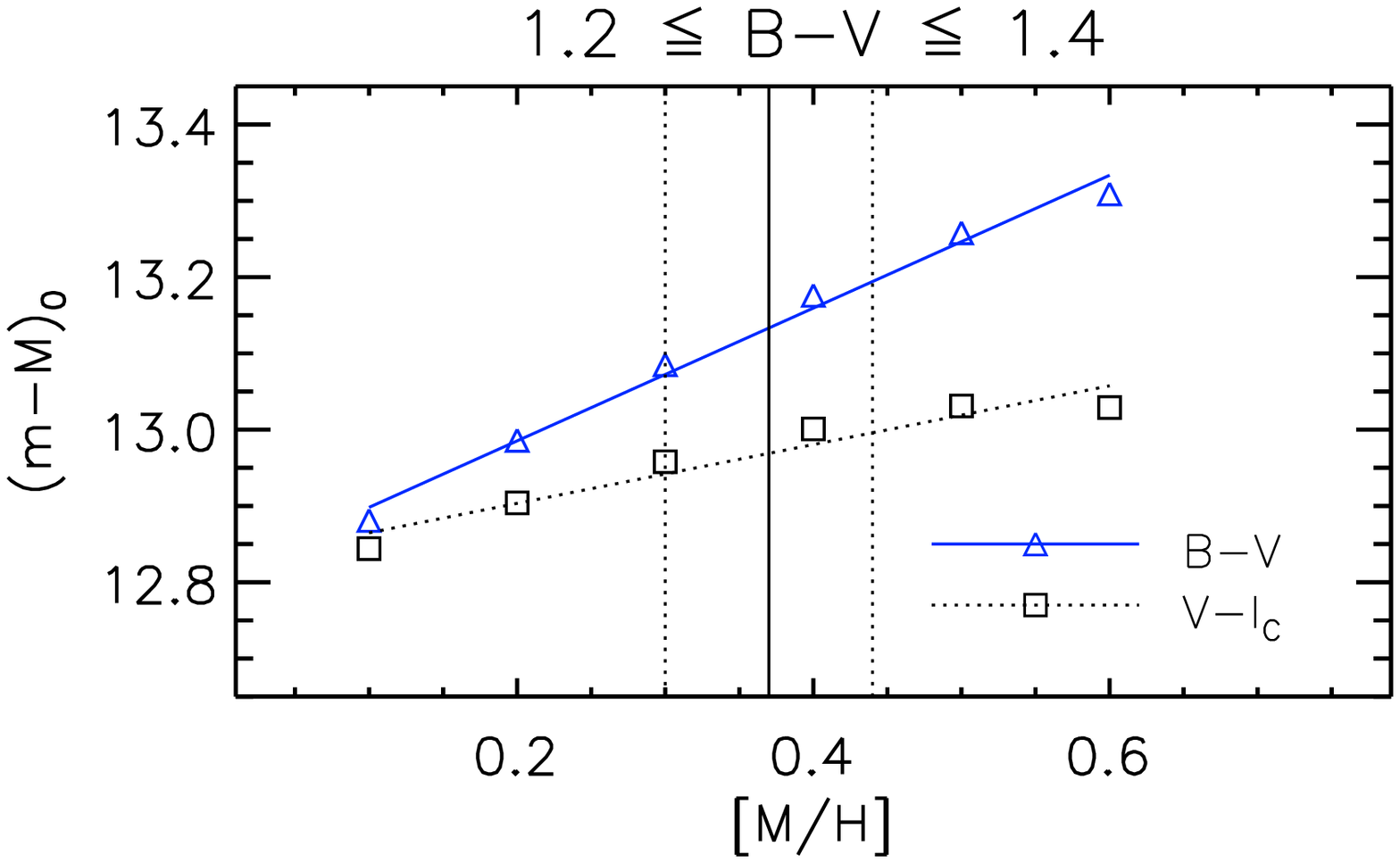}
\caption{Distance estimates for NGC~6791 from CMDs with $\bv$ (triangles) and $\vi$ (boxes) colors in various metallicity bins. The $9.5$~Gyr old isochrones with empirical color corrections were used in two different color bins: $1.0 \leq \bv \leq 1.1$ (top) and $1.2 \leq \bv \leq 1.4$ (bottom), and corresponding color ranges in $\vi$. The foreground reddening was fixed at $\ebv=0.11$.  \label{fig:meh.change}} \end{figure}

Similarly, the top panel in Figure~\ref{fig:meh.change} shows a distance modulus as a function of metal abundance in the model. We fixed the foreground reddening at $\ebv=0.11$ to be consistent with the above result and assumed an age of $9.5$~Gyr. The distance modulus was derived using stars with $1.00 \leq \bv \leq 1.10$ and $1.05 \leq \vi \leq 1.15$ in the $BV$ and $VI_C$ CMDs, respectively.  Triangles show distance moduli from the $BV$ CMD, while boxed points are those from the $VI_C$ CMD. We excluded $\vk$ because its CMD only covers the vertical part of the MS.  Solid lines are a linear fit to each set of the data points.  At higher metallicities, models become brighter, and a greater distance is derived from MS fitting. However, the $BV$ CMD is more sensitive to a metallicity change than the $VI_C$ CMD, and consequently has a steeper slope in Figure~\ref{fig:meh.change}.

At the photometric metallicity of the cluster ([M/H]$=+0.40$; see Figure~\ref{fig:fit}), isochrone fitting in the top panel of Figure~\ref{fig:meh.change} yields $\dmn=13.074\pm0.005$ and $13.006\pm0.005$ from the $BV$ and $VI_C$ CMDs, respectively, where the errors represent a scatter of data points from the best-fitting isochrone. Within the fitting errors alone, distances from these two color indices are inconsistent with each other. As noted above, this difference can originate from small systematic errors in our models, but it can also be easily caused by various systematic errors involved in the MS fitting (\S~\ref{sec:error}). The two lines cross at a lower metallicity than the spectroscopic metallicity of the cluster, indicating that the photometric metallicity is [M/H]$\approx+0.26$.  In \S~\ref{sec:chi} we seek a global solution of the cluster's parameters from all available features on CMDs, where the difference from a model in these color indices is minimized. At the spectroscopic metallicity of the cluster, the average distance modulus from the $\bv$ and $\vi$ color indices is $\dmn=13.002\pm0.042$ (from the difference between the two CMDs) $\pm0.067$ (propagated from the metallicity error). Our distance estimate is within the range found in previous studies in the literature, $12.9 \la \dmn \la 13.1$ \citep[e.g.,][among others]{salaris:04,carney:05,twarog:07,brogaard:11}.

However, the mutual agreement of our models in $\bv$ and $\vi$ breaks down when we go to cooler stars. The bottom panel in Figure~\ref{fig:meh.change} shows the case when fits were made using redder stars in NGC~6791 ($1.20 \leq \bv \leq 1.40$ and $1.25 \leq \vi \leq 1.65$, or $4300$~K $\la \teff \la 4750$~K) than those shown in the top panel. A strong inconsistency between $\bv$ and $\vi$ colors is seen even after the process of empirical color correction. Distance moduli from the $BV$ CMD are systematically larger by $\sim0.1$~mag than those obtained from bluer stars, while distance estimates in $\vi$ remain almost unaffected. This implies that systematic errors in $\bv$ colors, either in the photometry or in the model in the above temperature range, are significantly larger than in $\vi$. This is unfortunate, since fitting along the MS, in the presence of such systematic offsets, could reduce the accuracy in our distance and metallicity estimates considerably.

\begin{figure*}
\centering
\includegraphics[scale=0.5]{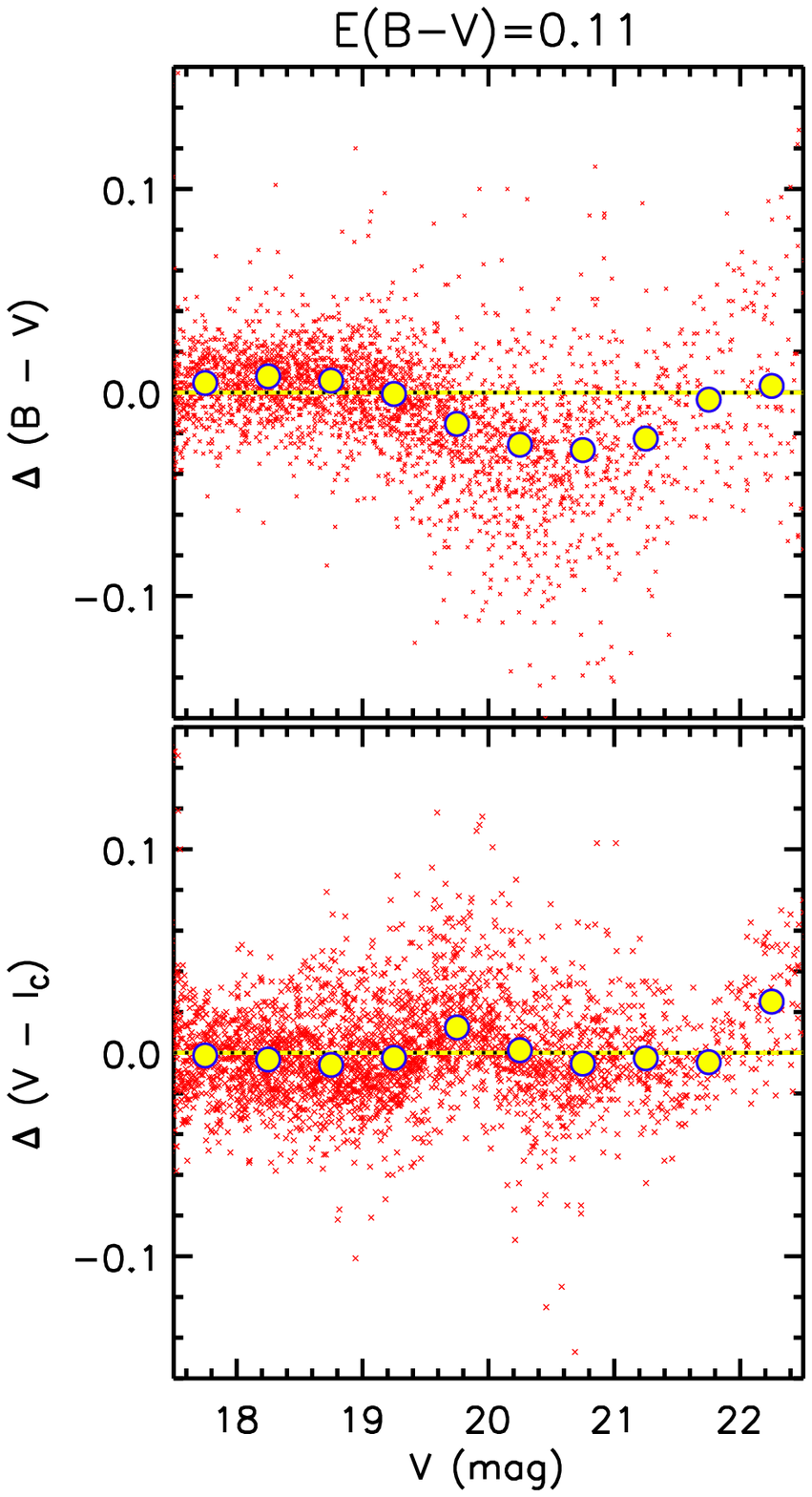}
\includegraphics[scale=0.5]{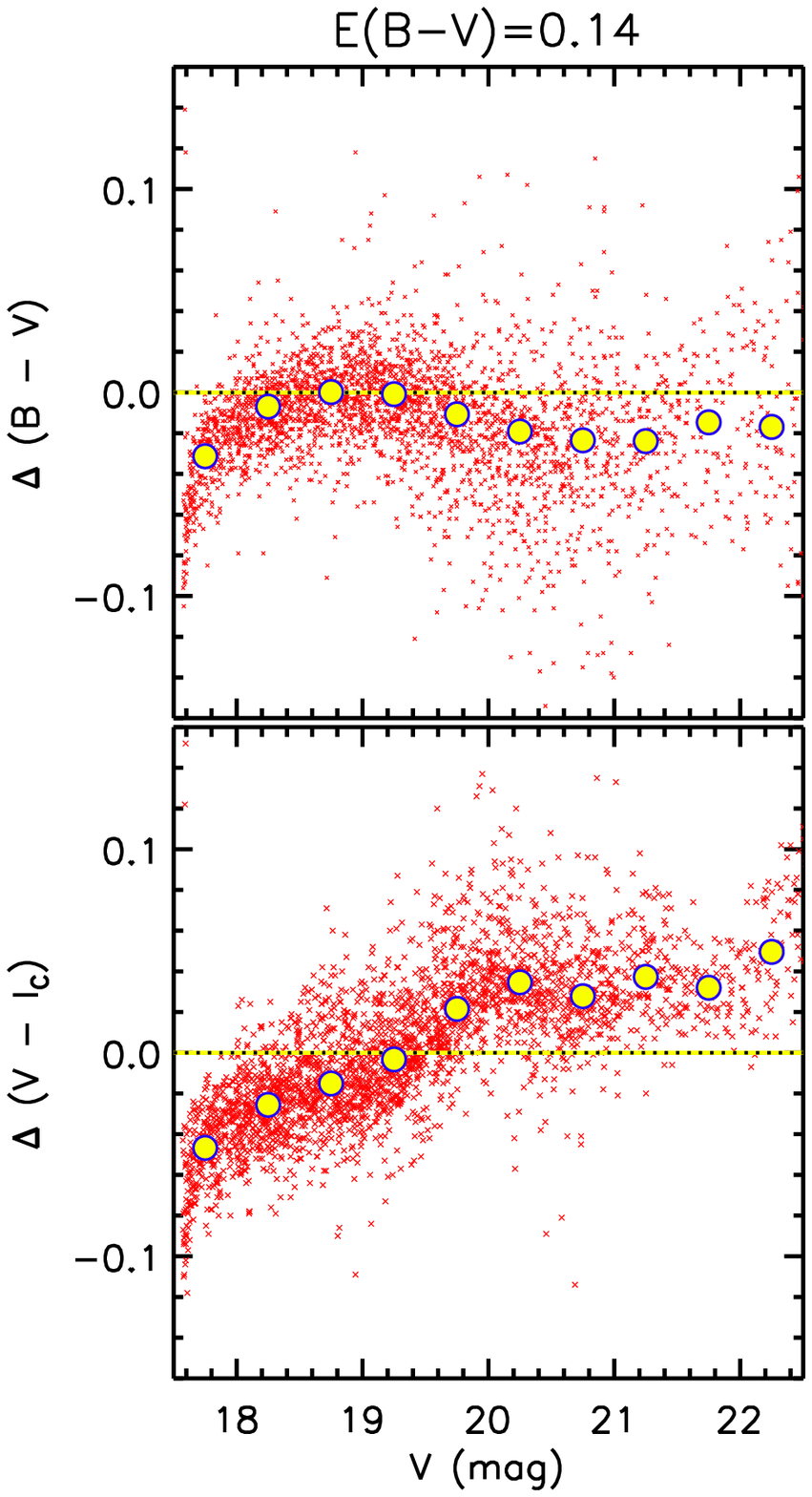}
\caption{The $\bv$ and $\vi$ color residuals of stars in NGC~6791 in the sense of observed colors minus calibrated models at [M/H]$=+0.37$. The left and right panels show comparisons with $\ebv=0.11$ and $0.14$, respectively, at a distance derived on each CMD.  Only those points left from the photometric filtering are shown.  Large circles are average color differences in bins of $\Delta V = 0.5$~mag.  Their error bars are smaller than the size of a symbol.  \label{fig:delta}} \end{figure*}

The left panels in Figure~\ref{fig:delta} show color residuals from the best-fitting isochrones for individual stars in $\bv$ (top panels) and $\vi$ (bottom panels). We used $9.5$~Gyr old models at the spectroscopic metallicity of the cluster ([Fe/H]$=+0.37$).  Colors were corrected for the reddening with $\ebv=0.11$, where we adopted a distance modulus as derived using stars in $1.0 \leq \bv \leq 1.1$ on each CMD: $\dmn=13.037$ in $BV$ and $\dmn=12.983$ in the $VI_C$ CMD. Only those data points that were left from the photometric filtering routine (see Figure~\ref{fig:cmd}) are shown.

\input{tab4.tex}

As seen in the top left panel of Figure~\ref{fig:delta}, there is a significant residual trend in $\bv$ at $19.5\la V \la 21.5$. Large circles represent moving-averaged differences in bins of $\Delta V=0.5$~mag, which are tabulated in Table~\ref{tab:colordiff}. The largest color deviation ($\sim 0.03$~mag) is seen at $V \sim 20.75$~mag or at $\bv \sim 1.3$, in the sense that our best-fitting model is redder than the observed MS. Given that the slope of the MS is about $4.5$ in the $BV$ CMD, an average offset of $\sim 0.02$~mag in $\bv$ is consistent with the $\sim0.1$~mag larger distance moduli obtained using redder stars in Figure~\ref{fig:meh.change}. The observed offsets persisted even if we adopted different metallicities ($+0.3 \la {\rm [M/H]} \la +0.5$) of models.  On the other hand, the $\vi$ color differences in the bottom left panel are found within $\Delta (\vi) \la 0.015$~mag, down to $V\sim22$~mag ($\vi\sim2.0$).  Even at the maximum displacement in $\bv$, the match to the data in $\vi$ is excellent. That the model matches the observed MS in the $VI_C$ CMD relatively well suggests that the systematic residuals in $\bv$ were probably not caused by errors in the stellar interior models.

In order to check that the large residual trend in $\bv$ was not caused by the adopted reddening of the cluster, we show color residuals for an alternative reddening [$\ebv=0.14$] in the right panels of Figure~\ref{fig:delta}. As in the left panels, we derived and adopted distances from individual CMDs using stars in $1.0 \leq \bv \leq 1.1$: $\dmn=13.096$ in $BV$ and $\dmn=13.092$ in the $VI_C$ CMD.  As seen in these panels, the large $\bv$ color residuals do not disappear, but color residuals in $\vi$ become significantly large. The $\vi$ color index reacts more sensitively to $\ebv$, suggesting that $\ebv=0.14$ is too high.

There are a few remaining possibilities that could have made the observed MS in $\bv$ deviate from the model. Firstly, the systematic color offset could originate from inaccurate color-$\teff$ relations in the model. The higher metallicity causes stars to have greater line blanketing at shorter wavelengths.  The effect of line blanketing would be enhanced at lower temperatures, so we can hypothesize that the \citet{lejeune:97} relations failed to reproduce fluxes for cool MS stars in NGC~6791 because of the difficulty in modeling complicated spectral energy distributions in shorter wavelength passbands, such as in $B$. The effect of line blanketing would be less in $V$ and certainly less in $I_C$, so $\vi$ colors would be less affected. Second, it is possible that the problem lies with systematic errors in the cluster photometry.  \citet{stetson:03} claimed that zero-point errors of their photometry are at least $0.0014$~mag in $V$ and $I_C$, and $0.0025$~mag in $B$; the corresponding zero-point error in $\bv$ is at a minimum of $0.003$~mag. In addition, there could exist color-scale errors in the photometry. As mentioned above, the expected amount of the color-scale error is $\sim0.02$~mag in $\bv$.  Below we proceed to look for effects that could produce the discrepancy by using alternative color-$\teff$ transformations and independent cluster photometry.

\subsection{IRFM Color-$\teff$ Relations}\label{sec:irfm}

\begin{figure}
\centering
\includegraphics[scale=0.5]{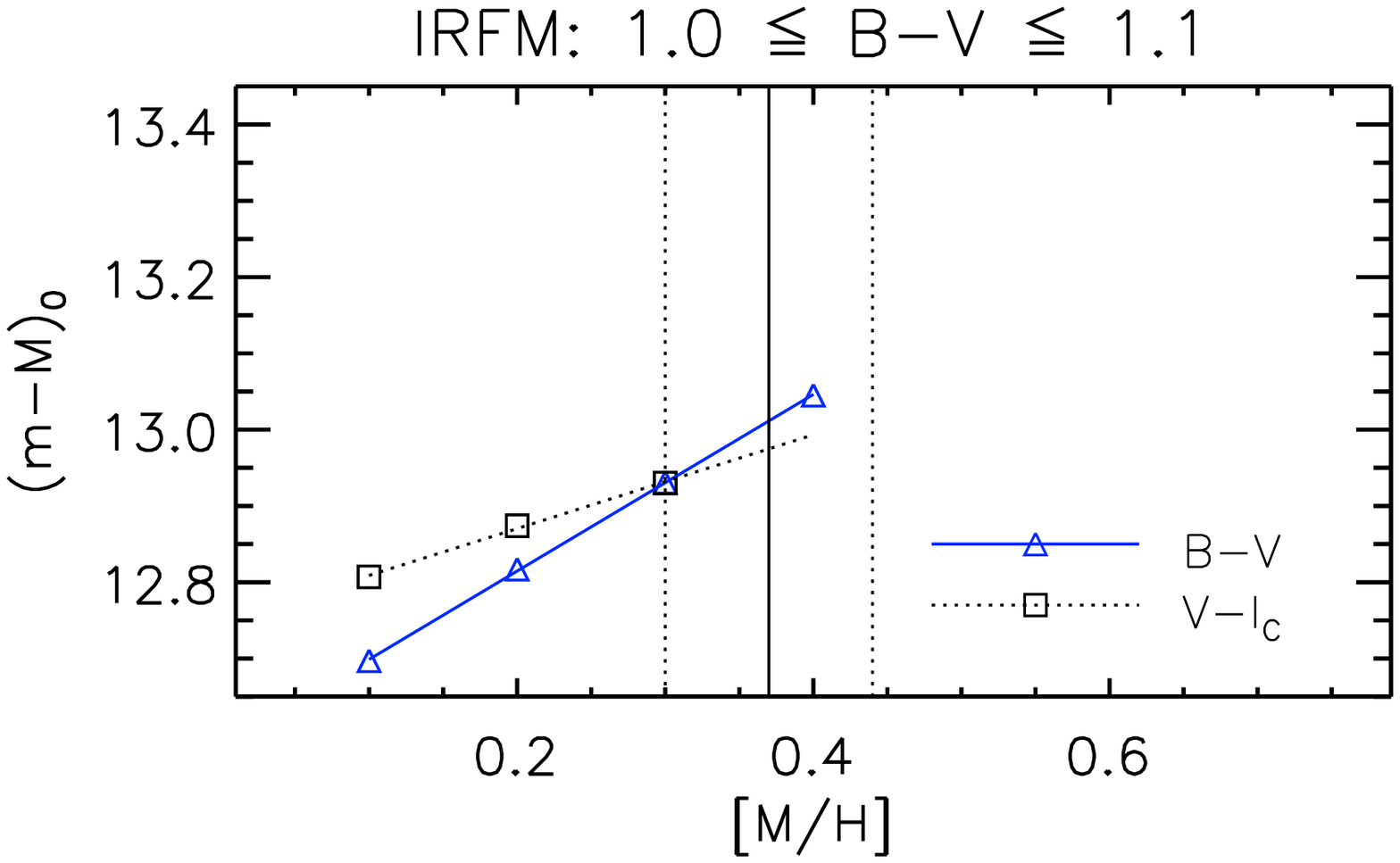}
\includegraphics[scale=0.5]{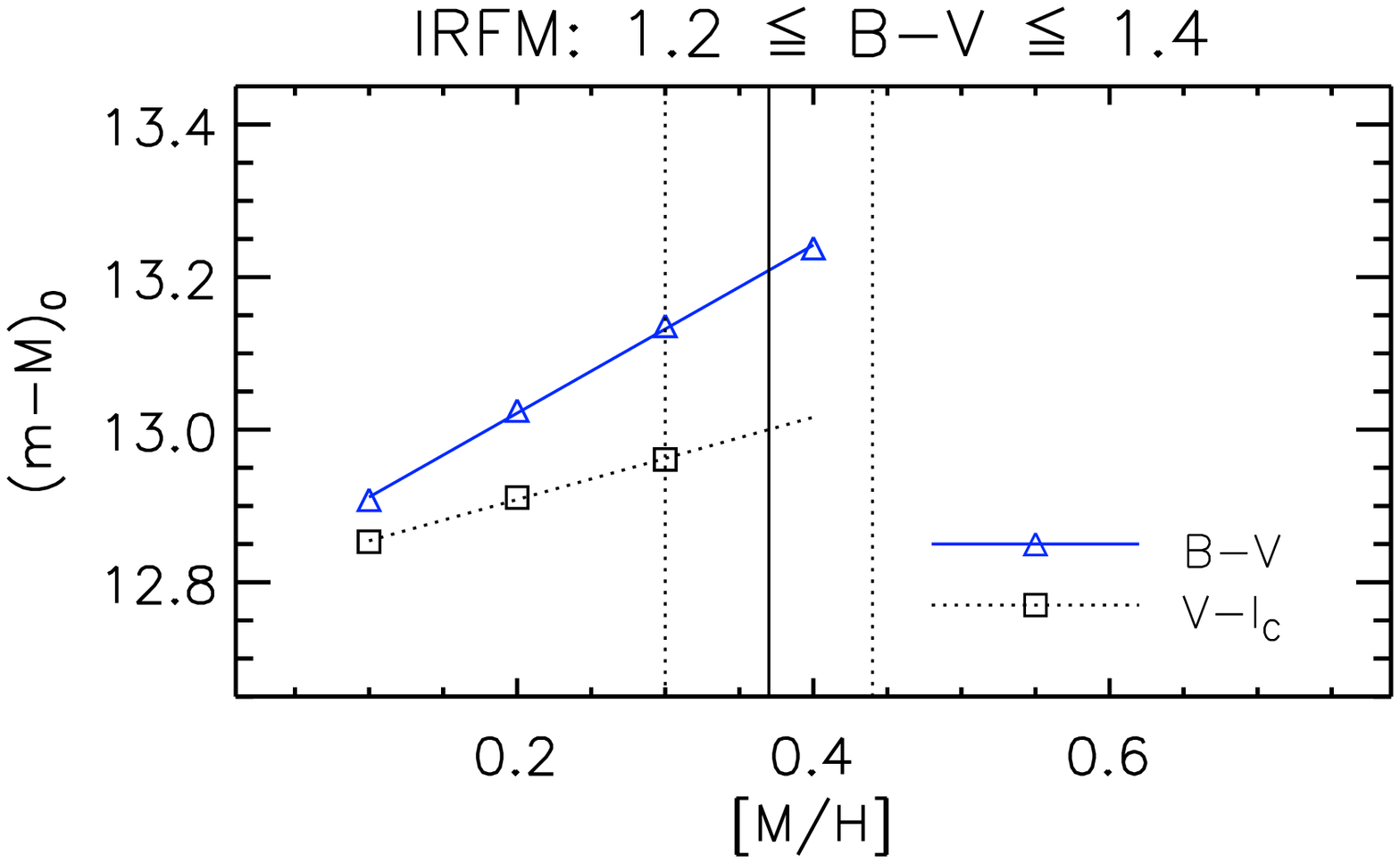}
\caption{Same as in Figure~\ref{fig:meh.change}, but distance moduli estimated using our base models with \citet{casagrande:10} IRFM relations.  \label{fig:meh.change.irfm}} \end{figure}

We utilized color-$\teff$ relations in \citet{casagrande:10} to test whether the $\bv$ color offset is found independently of models employed. Figure~\ref{fig:meh.change.irfm} is the same as Figure~\ref{fig:meh.change}, but shows IRFM-based slopes and intercepts. We used the same base isochrones, but after transforming model $\teff$ into $\bv$ and $\vi$ using the \citet{casagrande:10} relations. Given that their color-$\teff$ relations are mostly empirical with little dependence on models, an extra step of empirical calibration as described in this paper is not needed. The same set of stars as in Figure~\ref{fig:meh.change} was used in the top ($1.00 \leq \bv \leq 1.10$) and the bottom ($1.20 \leq \bv \leq 1.40$) panels.

In the top panel, the IRFM method differs from what our models predict in that the metallicity dependence and the zero points are slightly different. The lines have slopes of $1.16\pm0.02$~mag~dex$^{-1}$ for $\bv$ and $0.62\pm0.04$~mag dex$^{-1}$ for $\vi$, while those in our models (top panel in Figure~\ref{fig:meh.change}) have $1.20\pm0.01$~mag dex$^{-1}$ and $0.72\pm0.01$~mag~dex$^{-1}$, respectively. As a consequence, the IRFM-based photometric solution to metallicity is somewhat different from what we obtained above, and yields [Fe/H]$=+0.30$. Since it is within a $1\sigma$ bound of the spectroscopic metallicity of the cluster, we consider that there are no alarming signs of mismatch with the data.

As shown in the bottom panel of Figure~\ref{fig:meh.change.irfm}, however, a significant discrepancy in distance is seen between $BV$ and $VI_C$ CMDs when fits were limited to cooler stars ($1.20 \leq \bv \leq 1.40$ or $4300$~K $\la\teff\la4750$~K).  The observed behavior is similar to those seen in Figure~\ref{fig:meh.change} with our calibrated isochrones. In other words, the two independent color-$\teff$ relations show that a distance modulus from the $BV$ CMD strongly depends on the color range chosen for MS fitting, but the distance from the $VI_C$ CMD is certainly less affected by the choice.

\begin{figure}
\centering
\includegraphics[scale=0.75]{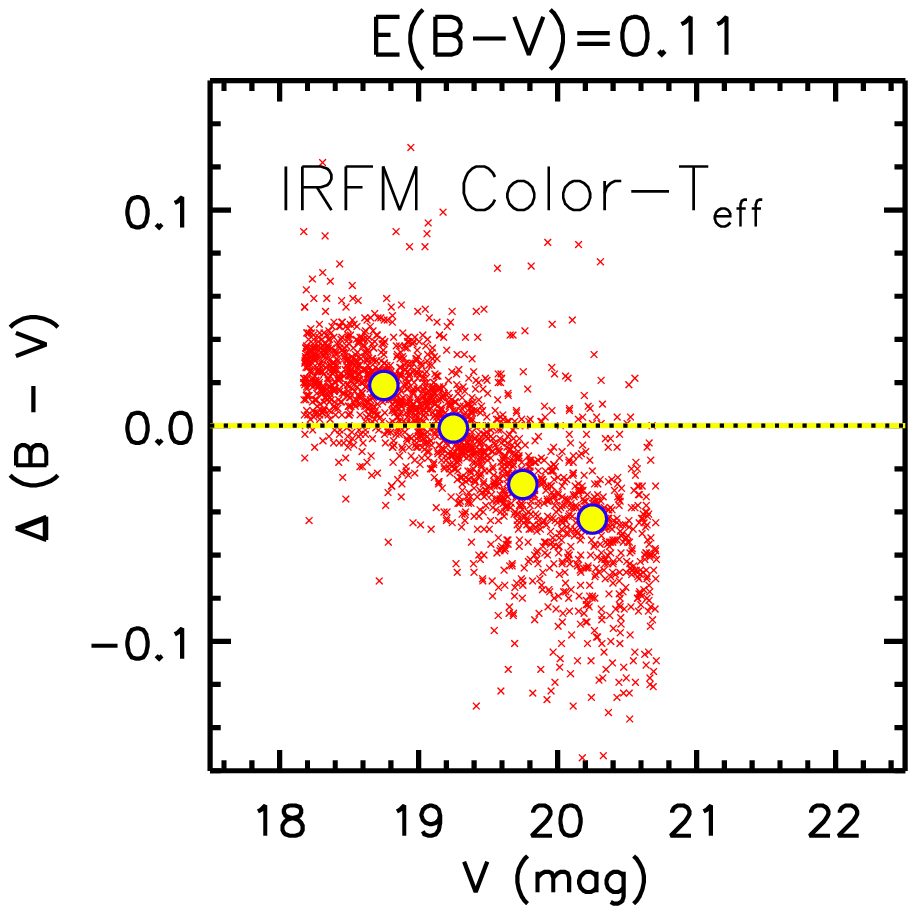}
\includegraphics[scale=0.75]{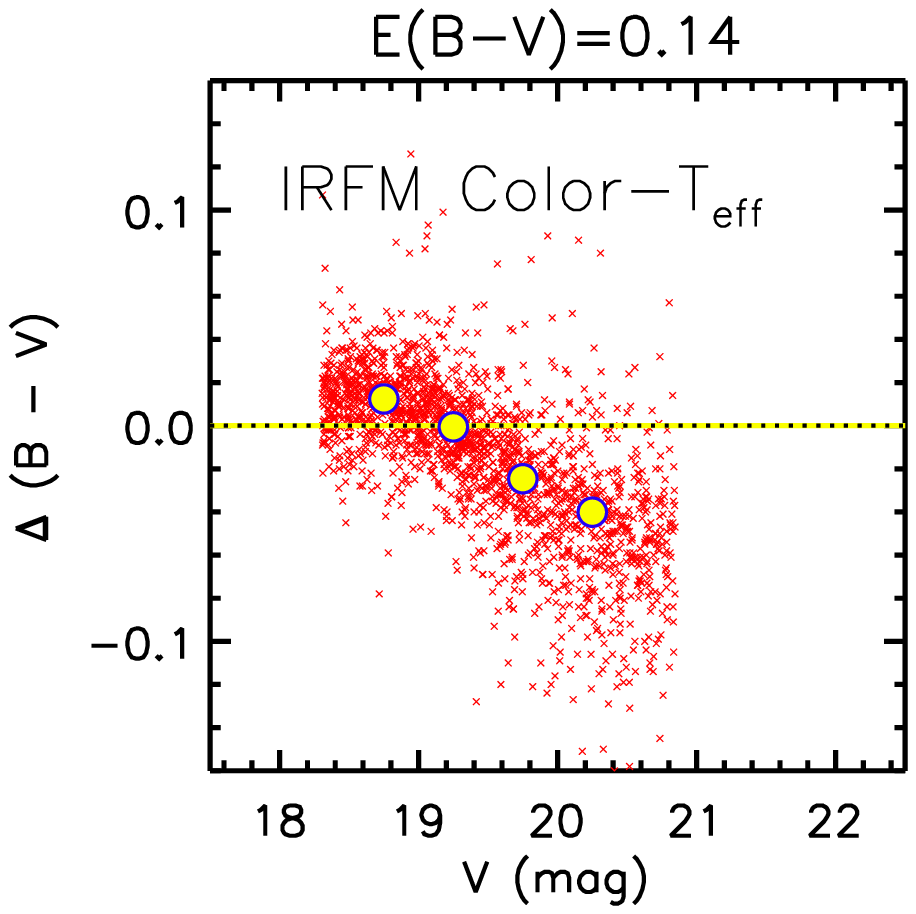}
\caption{Same as in Figure~\ref{fig:delta}, but showing color residuals in $\bv$ from a [M/H]$=+0.37$ model with the \citet{casagrande:10} IRFM relation. The left and the right panels show comparisons at $\ebv=0.11$ and $\ebv=0.14$, respectively. A comparison in $\vi$ is not shown because of the metallicity limit in the IRFM relation.  \label{fig:delta3}} \end{figure}

Figure~\ref{fig:delta3} shows color differences from a $9.5$~Gyr old model at [M/H]$=+0.37$ with $\ebv=0.11$, after transforming model $\teff$ into colors using the IRFM relations in \citet{casagrande:10}.  The comparison is limited to $V\sim20.5$~mag due to the $\teff$ limit set in the IRFM work. We used a distance modulus [$\dmn=13.010$] that best matches the model with the IRFM relation on the $BV$ CMD in $1.00 \leq \bv \leq 1.10$. In the right panel, we also show the case when an alternative value of reddening [$\ebv=0.14$] was adopted at its best-fitting distance [$\dmn=13.048$]. As seen in Figure~\ref{fig:delta3}, the systematic residual trend from the IRFM-based models also shows that the $\bv$ photometry is too blue for fainter stars ($V > 19.5$~mag), independently of the cluster reddening adopted. Interestingly, the systematic trend is even stronger than those seen in Figure~\ref{fig:delta} based on our calibrated models. This is not surprising, as the empirical IRFM relations are not as well defined for extreme metallicity and temperature.

\begin{figure*}
\centering
\includegraphics[scale=0.65]{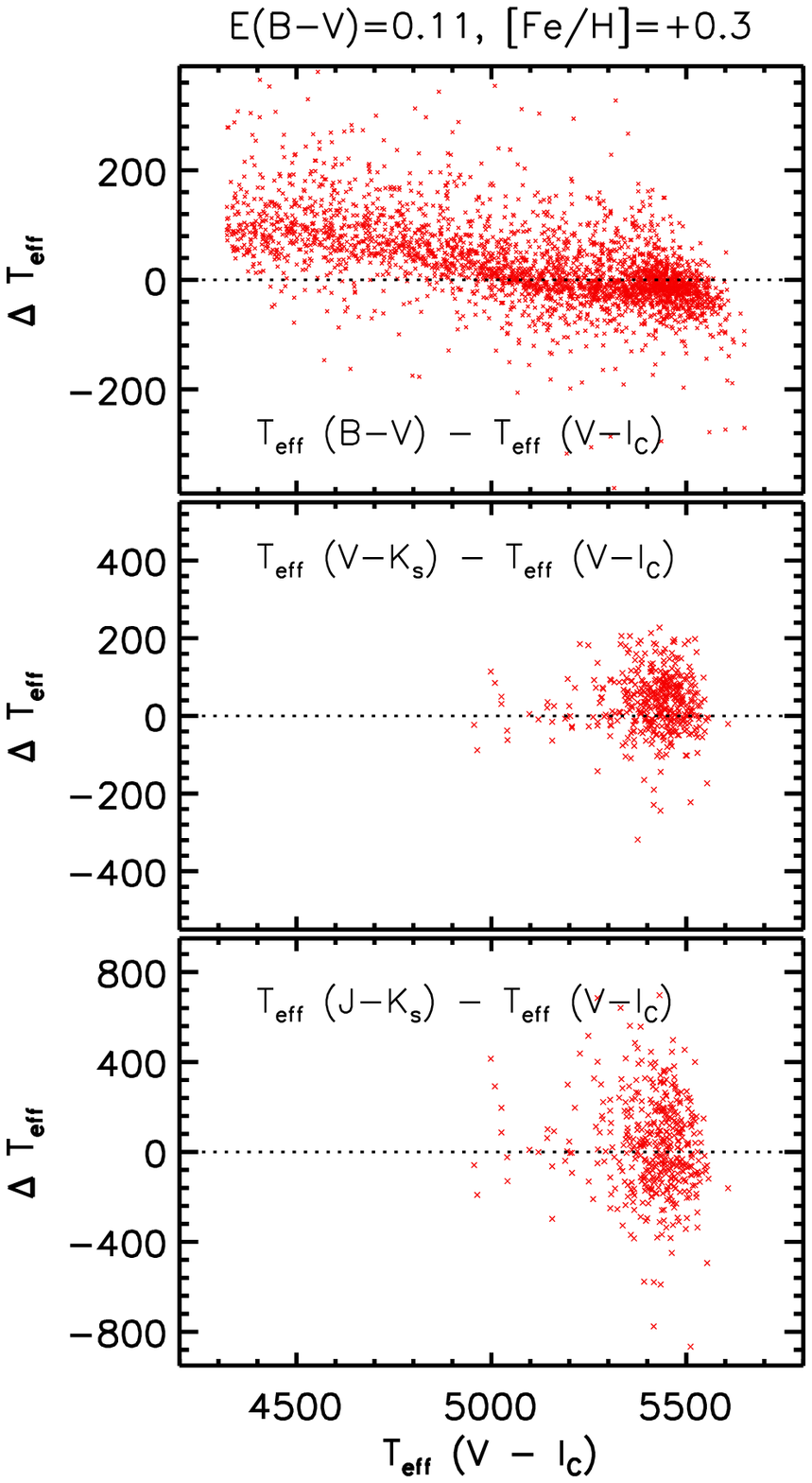}
\includegraphics[scale=0.65]{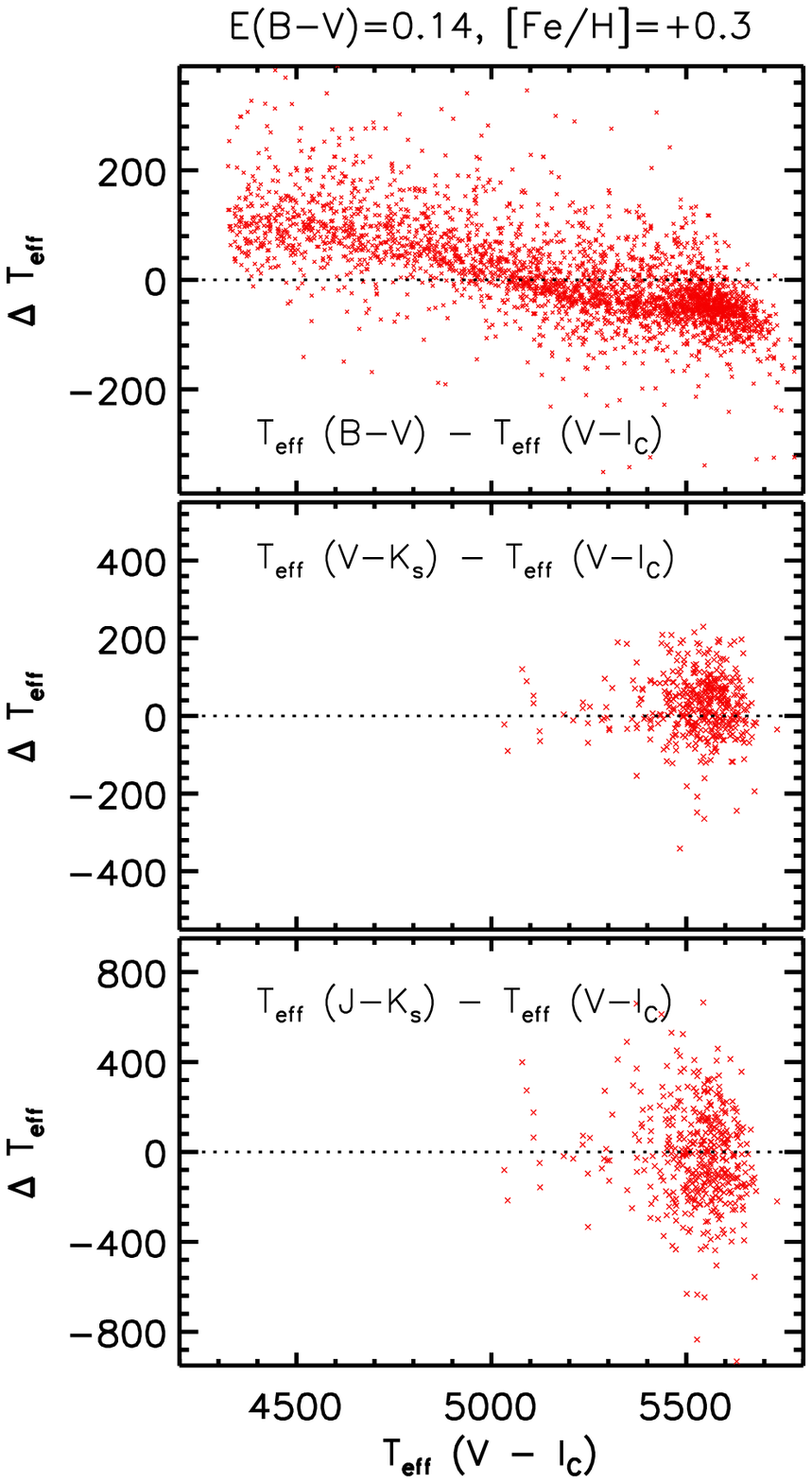}
\caption{Differences in $\teff$ as inferred from the \citet{casagrande:10} IRFM relations in $\bv$, $\vi$, $\vk$, and $\jk$. Only those selected as single cluster members in NGC~6791 are shown. Observed colors were corrected for the cluster foreground reddening with $\ebv=0.11$ and $0.14$ in the left and right panels, respectively. Because the IRFM relation in $\vi$ is defined at [Fe/H]$\leq+0.3$, temperatures were estimated at [M/H]$=+0.3$ in all color indices.\label{fig:irfm}} \end{figure*}

A strong internal inconsistency of IRFM colors for cool dwarfs can also be seen without relying on YREC interior models. Figure~\ref{fig:irfm} compares $\teff$ estimates from $\bv$, $\vk$, and $\jk$ with those from $\vi$ based solely on IRFM relations.  Only those remaining from the photometric filtering (black cross points in Figure~\ref{fig:cmd}) are shown.  Observed colors were corrected for the cluster's foreground reddening, before $\teff$ were estimated using the \citet{casagrande:10} relations.  We assumed $\ebv=0.11$ in the left panels and $\ebv=0.14$ in the right panels. Because the \citet{casagrande:10} relations were defined up to [Fe/H]$=+0.3$ in $\vi$, we assumed [Fe/H]$=+0.3$ in the above comparisons.  However, metallicity effects are relatively small in the differential $\teff$ comparisons.  For $\vk$ and $\jk$, we only included stars with $V < 18.5$~mag to avoid a bias below the detection limit in $K$ (dashed line in Figure~\ref{fig:cmd}).

Figure~\ref{fig:irfm} clearly shows that there exists a strong systematic trend in the temperature difference ($\Delta \teff$) between $\bv$ and $\vi$ colors.  Temperatures inferred from $\bv$ are on average $80$~K higher than those from $\vi$ at $\teff \la 4800$~K, while temperatures from $\bv$ and $\vi$ are found on nearly the same scale ($<10$~K) at $\teff \ga 5000$~K. The internal inconsistency of $\teff$ persists even at the spectroscopic metallicity of the cluster. From the metallicity sensitivity of the IRFM relations, we estimated that a $\sim0.1$~dex increase in [Fe/H] results in a $\sim30$~K increase in temperature from both $\bv$ and $\vi$ colors, but the change in metallicity does not make any appreciable change in the systematic $\teff$ difference. The higher temperatures in $\bv$ than in $\vi$ at $\teff < 4800$~K imply that $\bv$ colors are too blue at a given $\vi$ color, and are consistent with systematically larger distance moduli from $\bv$ in Figure~\ref{fig:meh.change}. Unless both our and IRFM color-$\teff$ relations are in error, the internally inconsistent $\teff$ suggests that the problem lies in the cluster photometry of cool MS stars.

\subsection{Other Cluster Photometry}\label{sec:kr95}

We compared Stetson's cluster photometry with those in \citet{kaluzny:95}. \citeauthor{kaluzny:95} obtained deep $UBV$ photometry of stars in the central part of the cluster using the KPNO $2.1$ m telescope, and took shallower $UBVI_C$ data over the entire cluster field using the KPNO $0.9$ m telescope. Below, the former data set is referred to as Table~$1$, and the latter as Table~$2$, following the file numbering conventions in the VizieR archives,\footnote{{\tt http://cdsarc.u-strasbg.fr/}} from which we downloaded their photometric tables. We used a matched coordinate list in their Table~$2$ and relatively bright stars in \citet{stetson:03} to derive our own astrometric plate solutions and to transform pixel positions into celestial coordinates for their data sets.

\begin{figure}
\epsscale{1.15}
\plotone{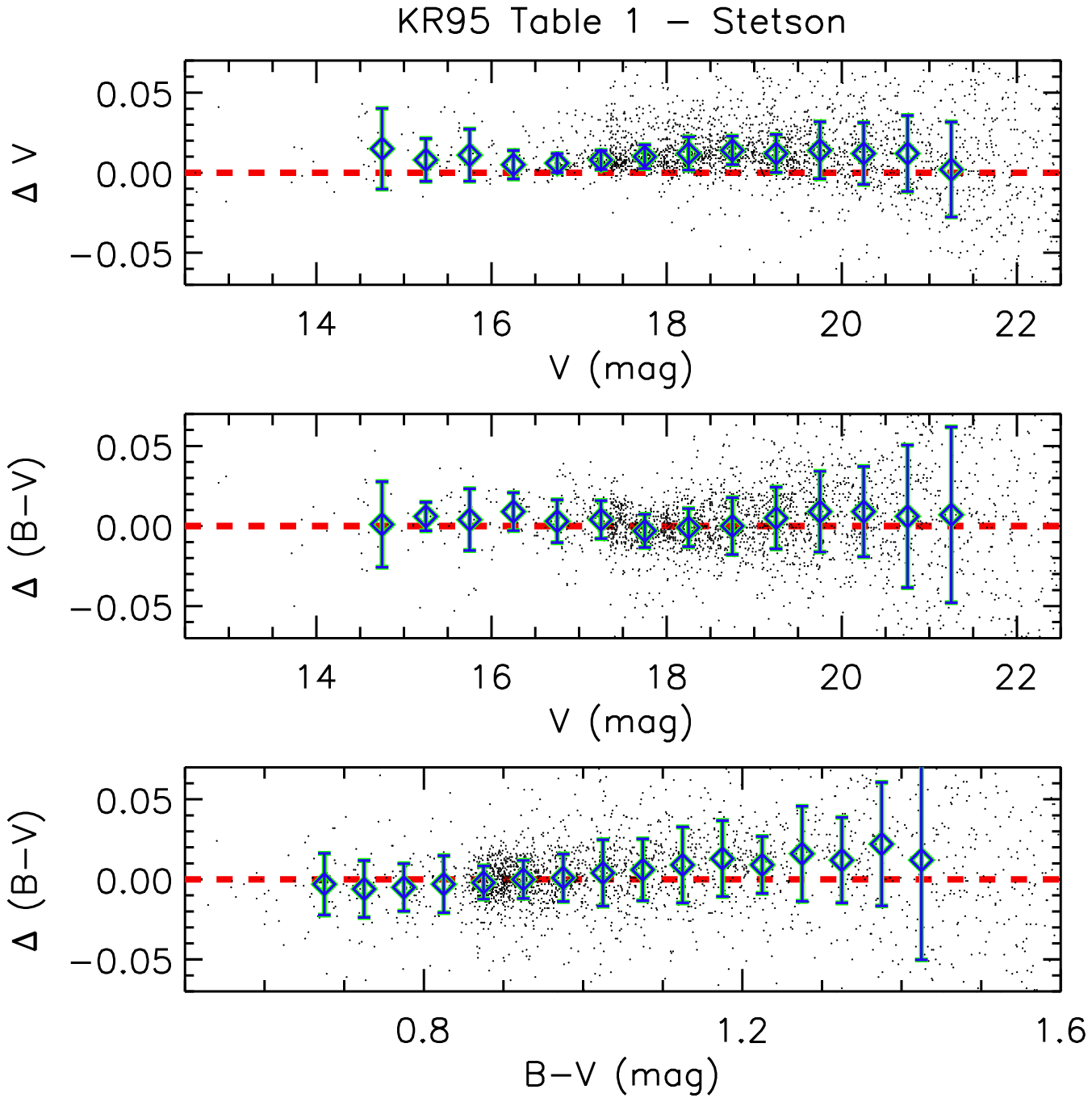}
\caption{Comparisons of the revised \citeauthor{stetson:03} photometry with values in Table~$1$ in \citet{kaluzny:95}. The difference is in the sense \citeauthor{kaluzny:95} values minus those in \citeauthor{stetson:03}\ The blue diamond points are median differences, and the error bars indicate a median absolute deviation of the difference multiplied by $1.483$. A standard error in the mean difference is typically smaller than the size of a diamond symbol.  The red dashed line represents a zero difference. \label{fig:kr95}} \end{figure}

\begin{figure}
\epsscale{1.15}
\plotone{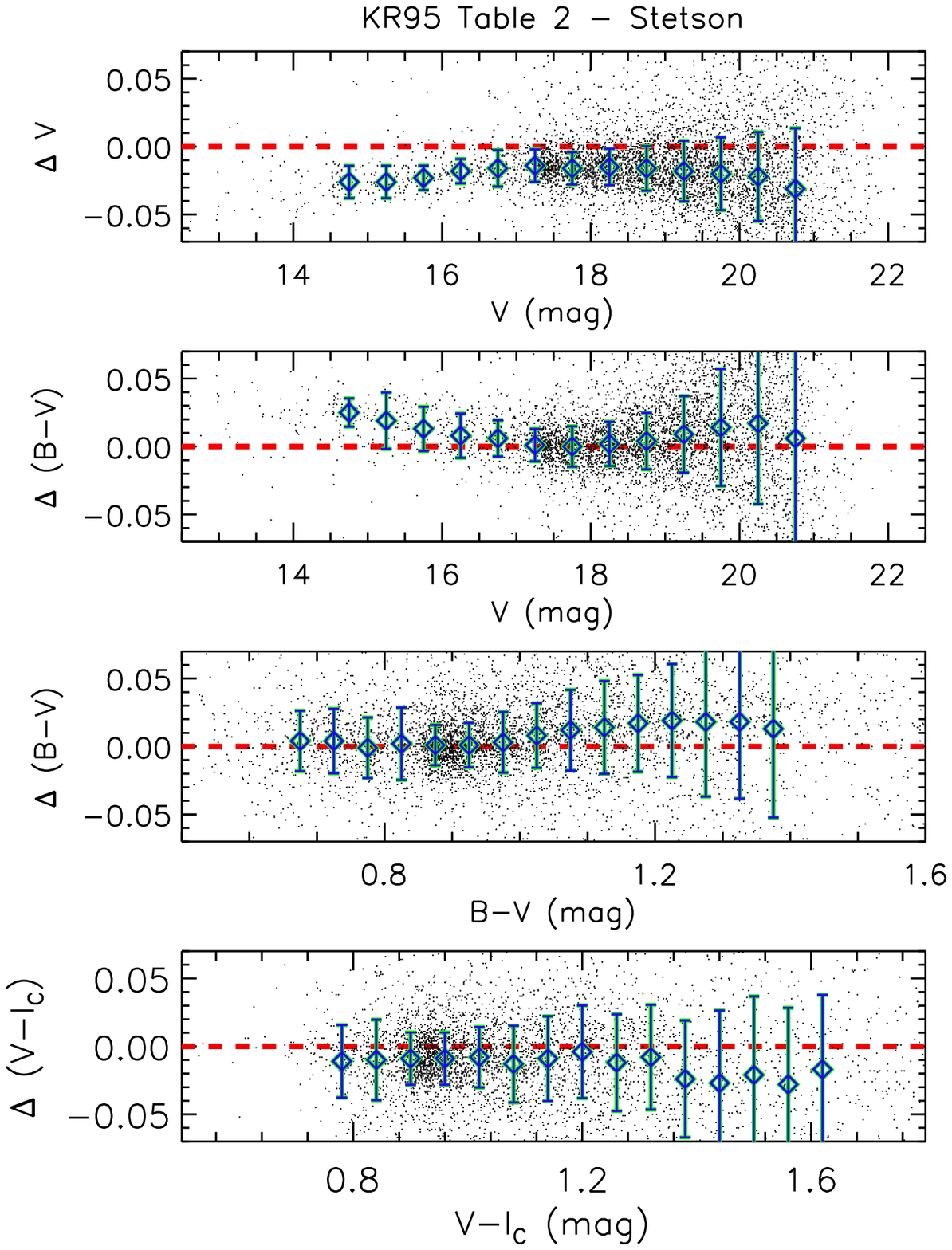}
\caption{Same as in Figure~\ref{fig:kr95}, but showing comparisons with photometry in Table~$2$ in \citet{kaluzny:95}.\label{fig:kr95b}} \end{figure}

Figure~\ref{fig:kr95} shows magnitude and color differences between Table~$1$ in \citet{kaluzny:95} and \citet{stetson:03}, in the sense of the former minus the latter. Similarly, Figure~\ref{fig:kr95b} shows differences from Table~$2$ in \citeauthor{kaluzny:95}. The blue diamond point is a median difference in $0.5$~mag intervals in $V$, $0.05$~mag in $\bv$, and $0.06$~mag in $\vi$. Error bars represent a median of all absolute deviations (MAD) multiplied by $1.483$ to make the estimator consistent with the standard deviation for a normal distribution.  The scatter of data points is large, but a standard error in the mean difference ($\sigma/\sqrt{N}$) is typically smaller than the size of a diamond symbol. In these comparisons, we imposed the same selection criteria as in \citeauthor{stetson:03}, who have also done star-by-star comparisons with photometry in \citeauthor{kaluzny:95}. These were based on $\chi$, {\tt sharp}, {\tt separation} indices, photometric errors, as well as a photometric quality flag in \citeauthor{kaluzny:95}. Instead of selecting stars with good quality flags in both \citeauthor{kaluzny:95} tables, however, we only used quality flags from each table when comparing it with \citeauthor{stetson:03} photometry. This seemed to make more stars available in the photometric comparisons, especially fainter stars at $V > 20$, which were not included in the original comparisons in \citeauthor{stetson:03}.

Figures~\ref{fig:kr95} and \ref{fig:kr95b} clearly show that there are systematic errors present in either \citeauthor{stetson:03} or \citeauthor{kaluzny:95} photometry. For stars near the cluster's turn-off ($V\sim17.5$ or $\bv\sim0,9$), differences in $V$ are of the order of $0.02$~mag. Besides photometric zero-point errors, Figures~\ref{fig:kr95} and \ref{fig:kr95b} demonstrate that differences in photometry depend on the magnitude or color of a star. Near the cluster's turn off, $\bv$ colors agree with each other among the three data sets. However, the difference increases toward fainter magnitudes and redder colors for both \citeauthor{kaluzny:95} tables, which amounts to $0.02$~mag at $\bv\sim1.3$.  Interestingly, this offset can explain the difference we found in Figure~\ref{fig:meh.change} between our calibrated isochrones and the \citeauthor{stetson:03} photometry.  Although it is difficult to make a fair judgment about which photometry is correct based on the photometric comparison alone, one might imagine that a color-scale error in the Stetson's photometry has not been completely removed in the newer data reduction \citep{stetson:05}. If Stetson's $\bv$ colors are systematically bluer at fainter magnitudes, higher IRFM temperatures in $\bv$ than those from $\vi$ (Figure~\ref{fig:delta3}) can be naturally explained. The $\vi$ colors also show large differences between \citeauthor{stetson:03} and \citeauthor{kaluzny:95} (bottom panel in Figure~\ref{fig:kr95b}). However, our models show no signs of large systematic color residuals in $\vi$ (left panel in Figure~\ref{fig:delta}), and distances derived using bluer stars are consistent with those from redder ones (Figure~\ref{fig:meh.change}).

\subsection{Color-$\teff$ Relations in SDSS Colors}\label{sec:sdss}

\begin{figure}
\centering
\includegraphics[scale=0.45]{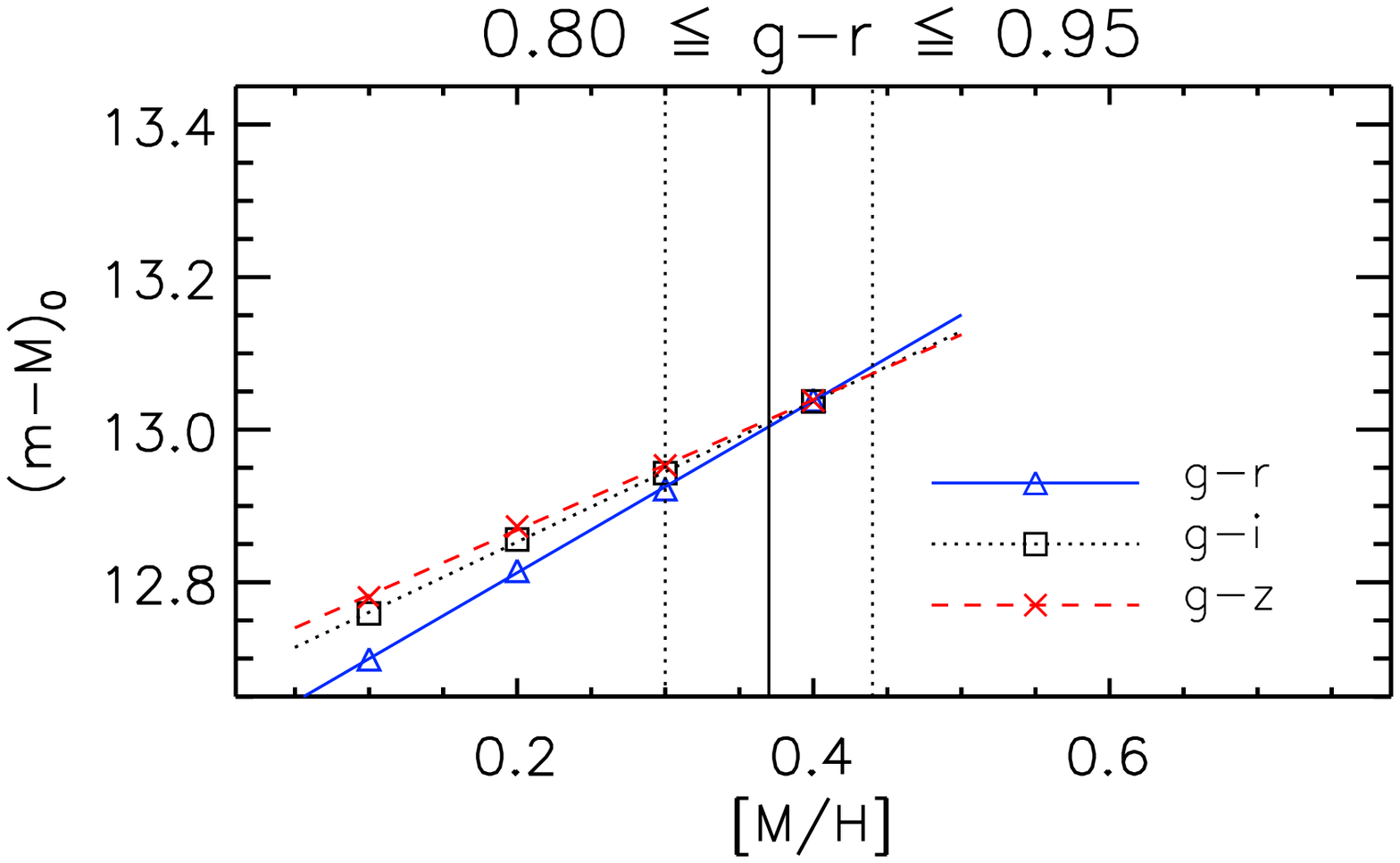}
\includegraphics[scale=0.45]{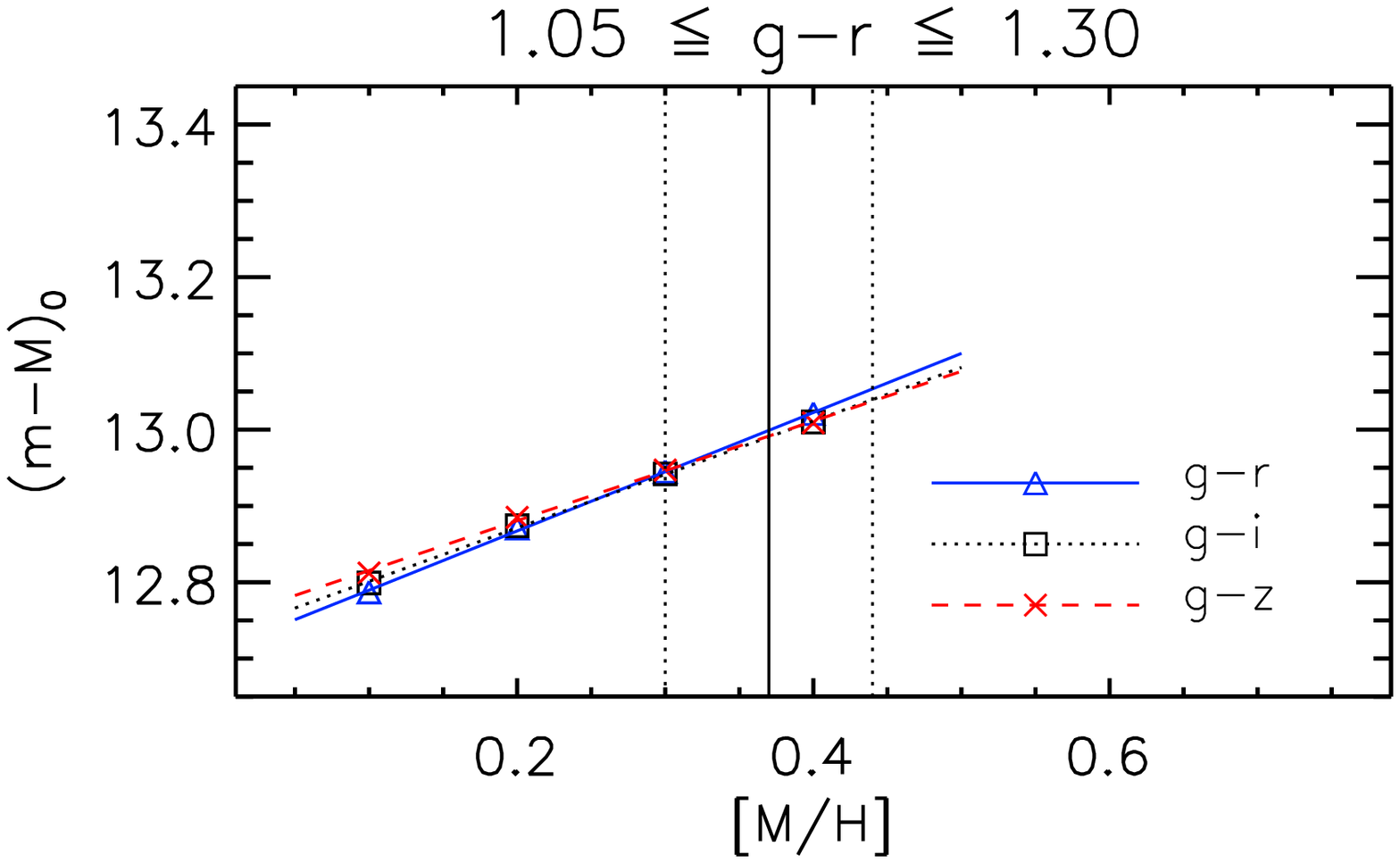}
\caption{Same as in Figure~\ref{fig:meh.change}, but showing MS-fitting results from the application of calibrated isochrones in the SDSS $griz$ colors \citep{an:09b}.  \label{fig:meh.change.sdss}} \end{figure}

We further checked the accuracy of model colors in the SDSS system.  Figure~\ref{fig:meh.change.sdss} is the same as Figure~\ref{fig:meh.change}, but showing metallicity sensitivities in SDSS colors. We took $griz$ photometry of the cluster from \citet{an:08}. As described in \citeauthor{an:08}, the cluster photometry was extracted from the original SDSS imaging frames using the DAOPHOT/ALLFRAME suite of programs \citep{stetson:87,stetson:94}, and was put on the natural $ugriz$ system in SDSS.  After applying photometric filtering on $gr$, $gi$, and $gz$ CMDs, we employed isochrones in \citet[][see Table~\ref{tab:library}]{an:09b} to derive a distance modulus on each CMD using the $r$-band as a luminosity index. The interior models are the same as those used in the current study, but the colors are those corrected based on observations of M67, on top of theoretical colors computed using MARCS \citep{gustafsson:08}. We assumed the same cluster age and reddening as in Figure~\ref{fig:meh.change}.

Figure~\ref{fig:meh.change.sdss} shows results when our fits were limited to the same color ranges as in Figure~\ref{fig:meh.change}. In the top panel, MS fitting included stars in $0.80 \leq \gr \leq 0.95$, which is similar to $1.0 \leq \bv \leq 1.1$, and the corresponding color ranges in $\gi$ and $\gz$. In the bottom panel, results are shown for stars in $1.05 \leq \gr \leq 1.30$ or $1.2 \leq \bv \leq 1.4$. In both panels, we assumed $\ebv=0.11$.  Unlike in the Johnson-Cousins and 2MASS systems, differential sensitivities in $griz$ colors are limited; in other words, the lines from different colors in Figure~\ref{fig:meh.change.sdss} have similar slopes, leading to a large uncertainty in the estimate of photometric metallicity \citep[see][]{an:09a,an:13}.

Nevertheless, Figure~\ref{fig:meh.change.sdss} clearly shows that the distance estimates from the three SDSS color indices are in good agreement with each other at the spectroscopic metallicity of the cluster.  In addition, almost the same distance modulus of the cluster is obtained independently of color ranges chosen for the fits. At the spectroscopic metallicity of the cluster, the mean distance moduli from the three color indices are $\dmn=13.009\pm0.034$ and $12.994\pm0.050$ from the blue (top panel) and the red (bottom panel) color fitting ranges, respectively.  The errors are dominated by the error in the spectroscopic metallicity, while individual fitting errors and the dispersion in distance among the three SDSS color indices are small ($\sigma < 0.01$~mag). Within the errors, both distance estimates are also in excellent agreement with our $BVI_C$-based photometric solution from the blue part of the CMD [$\dmn=13.002$; see top panel in Figure~\ref{fig:meh.change}].  In Figure~\ref{fig:delta}, the \citet{stetson:03} photometry shows a maximum deviation from the theoretical model at $\bv \approx 1.3$ ($\teff \sim 4600$~K), which corresponds to $\gr \approx 1.1$ \citep{an:13}.  However, such an offset is not seen on the $gr$ CMD \citep[see Figure~$18$ in][]{an:09b}.

To summarize, the good agreement between photometric and spectroscopic metallicity (top panels in Figures~\ref{fig:fit} and \ref{fig:meh.change}) suggests that our isochrones in $\bv$ and $\vi$ are mutually consistent after our process of empirical color correction.  The distance and reddening we found are also close to values obtained in the literature. However, the mutual agreement breaks down for cooler stars (lower panel of Figure~\ref{fig:meh.change}). Isochrones using empirical corrections are too red in $\bv$ compared to the photometry of stars in NGC~6791. We also found a similar color offset using IRFM color-$\teff$ transformations. The observed offset could be due to a color-scale error in the cluster photometry, as we found that the $\bv$ photometry in \citet{stetson:03} is systematically bluer for cool MS stars in comparison with alternative cluster photometry. Alternatively, the observed difference could be caused by differences between Stetson's and our compiled cluster photometry of the Hyades and Praesepe. It can also originate from errors in the models, but we obtained internally consistent fitting results in the SDSS colors using models with similar input physics and color calibration. To avoid the large color offset, we excluded $\bv$ photometry of cool stars ($\bv > 1.10$) in the following estimation of the cluster parameters.

\subsection{Joint Determination of Cluster Parameters}\label{sec:chi}

As discussed above, differential sensitivities of broadband colors can be used to constrain the reddening, distance, and metallicity of NGC~6791. In this section, we simultaneously solved for all these parameters and the age of the cluster, by directly comparing calibrated models with the observed cluster sequence (open circles in Figure~\ref{fig:cmd}). The fit covered all parts of the CMDs including specific features that were used to constrain foreground reddening, metallicity, and distance (\S~\ref{sec:prelm}), but excluded $\bv$ photometry of cool stars ($\bv > 1.10$) as discussed above. In the next section, we provide a robust error analysis on our parameter estimates.

To simultaneously determine the cluster's metallicity, reddening, distance, and age, we searched for the minimum $\chi^2$ value as defined below: \begin{equation} \chi^2_{\rm tot} = \sum_{j=1}^3 \sum_{i=1}^{N_j} \chi^2_{ij} = \sum_{j=1}^3 \sum_{i=1}^{N_j} \left[ \frac {(V_{ij} - V_{{\rm model},j})^2}{\sigma_{V,ij}^2} +  \frac {(X_{ij} - X_{{\rm model},j})^2}{\sigma_{X,ij}^2} \right], \label{eq:chifit} \end{equation} where the subscript $i$ indicates individual points of the observed sequence (open circles in Figure~\ref{fig:cmd}) on the $j$th CMD (each with $\bv$, $\vi$, and $\vk$, respectively). $N_j$ is the number of points included in the fit. $V_{ij}$ is a star's $V$ magnitude, and $X_{ij}$ is its color, while $V_{{\rm model},j}$ and $X_{{\rm model},j}$ represent the magnitude and color of a model, respectively.  The $\sigma_{V,ij}$ and $\sigma_{X,ij}$ are photometric errors in $V$ magnitude and color, respectively.

We performed a grid search in {\rm [M/H]}, $\ebv$, and age to find a minimum $\chi^2_{\rm tot}$.  The average distance modulus was determined using the $BV$ and $VI_C$ CMDs.  In the distance estimation, we restricted model fits to $\bv \leq 1.10$, but included redder stars in the $VI_C$ CMD with $\vi \leq 2.0$.  The $\vi=2.0$ limit was set to avoid regions where our empirical corrections are poorly defined. We further restricted fits to $V \geq 18$~mag in the distance estimation to exclude subgiant stars in the cluster.  Adding the $VK_s$ CMD in the fitting process helped to better constrain the cluster's age, metallicity, and reddening with its longer wavelength baseline (see the top panel in Figure~\ref{fig:fit}), although it was not used in the distance estimation.

\begin{figure}
\epsscale{0.95}
\plotone{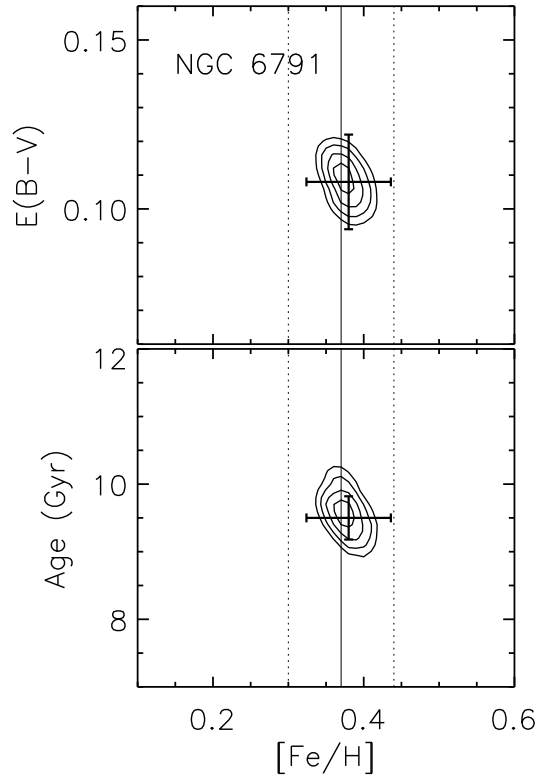}
\caption{The $\Delta \chi^2$ distribution from MS fitting for NGC~6791.  Likelihood contours are shown at $\Delta \chi^2_{\rm tot} = 4.7$, $9.7$, $16.3$, and $23.5$ ($68.3\%$, $95.4\%$, $99.73\%$, and $99.99\%$ confidence levels for four degrees of freedom) relative to a minimum $\chi^2_{\rm tot}$.  Error bars indicate $\pm1\sigma$ total errors in Table~\ref{tab:error} at the minimum $\chi_{\rm tot}^2$.  The solid and dotted vertical lines represent the spectroscopic metallicity of the cluster and its $\pm1\sigma$ errors.  \label{fig:chi}} \end{figure}

The top and bottom panels in Figure~\ref{fig:chi} show likelihood contours in {\rm [Fe/H]} versus $\ebv$ and {\rm [Fe/H]} versus age, respectively.  The $\chi^2$ surfaces were smoothed using a Gaussian kernel.  Contours are shown at $\Delta \chi^2_{\rm tot} = 4.7$, $9.7$, $16.3$, and $23.5$ ($68.3\%$, $95.4\%$, $99.73\%$, and $99.99\%$ confidence levels for four degrees of freedom) relative to the minimum value of $\chi^2_{\rm tot}$.  The minimum $\chi^2_{\rm tot}$ value was large ($\chi^2_{\rm tot}\sim1000$), suggesting that photometric errors were underestimated and/or that there were remaining systematic errors in the model calibration.  For a conservative error estimate, we normalized photometric errors by multiplying by a factor of $3.5$ in all CMDs (see Paper~III) to make $\chi^2_{\rm tot} \approx 89$, which is approximately equal to the total number of degrees in the fitting ($\nu=96-4=92$).

At the global $\chi^2$ minimum, our solution yields [M/H]$=+0.38$, $\ebv=0.108$, $9.5$~Gyr, and a corresponding cluster distance modulus is $\dmn=13.00$.  Our best-fitting cluster parameters are in good agreement with those from our earlier discussions in \S~\ref{sec:prelm}. Importantly, Figure~\ref{fig:chi} shows that our photometric metallicity estimate is in excellent agreement with the spectroscopic metallicity of the cluster. Our best-fitting age based on non-diffusion models ($9.5$~Gyr) could be $\sim1$~Gyr too old \citep[e.g.,][]{proffitt:91,castellani:99,chaboyer:01} compared to the case with microscopic diffusion, although an accurate age scale is not the main concern of this paper.

\input{tab6.tex}

We adopted a conservative $1\sigma$ fitting error from the size of contours at $\Delta \chi^2_{\rm tot} = 9.7$, yielding $\sigma ({\rm [Fe/H]})=0.015$, $\sigma [\ebv] = 0.004$, and $0.2$~Gyr in age. For a $1\sigma$ fitting error in distance modulus, we propagated these errors assuming $\Delta \dmn / \Delta {\rm [Fe/H]} \approx 1$ and $\Delta \dmn / \Delta \ebv \approx 2$ (Paper~III), along with a fitting error on each CMD.  However, final errors are dominated by the sum of various systematic errors as described in the next section. Best-fitting cluster parameters from the global $\chi^2$ minimization and their {\it total} errors are listed in the first row of Table~\ref{tab:par}.

\begin{figure*}
\epsscale{0.95}
\plotone{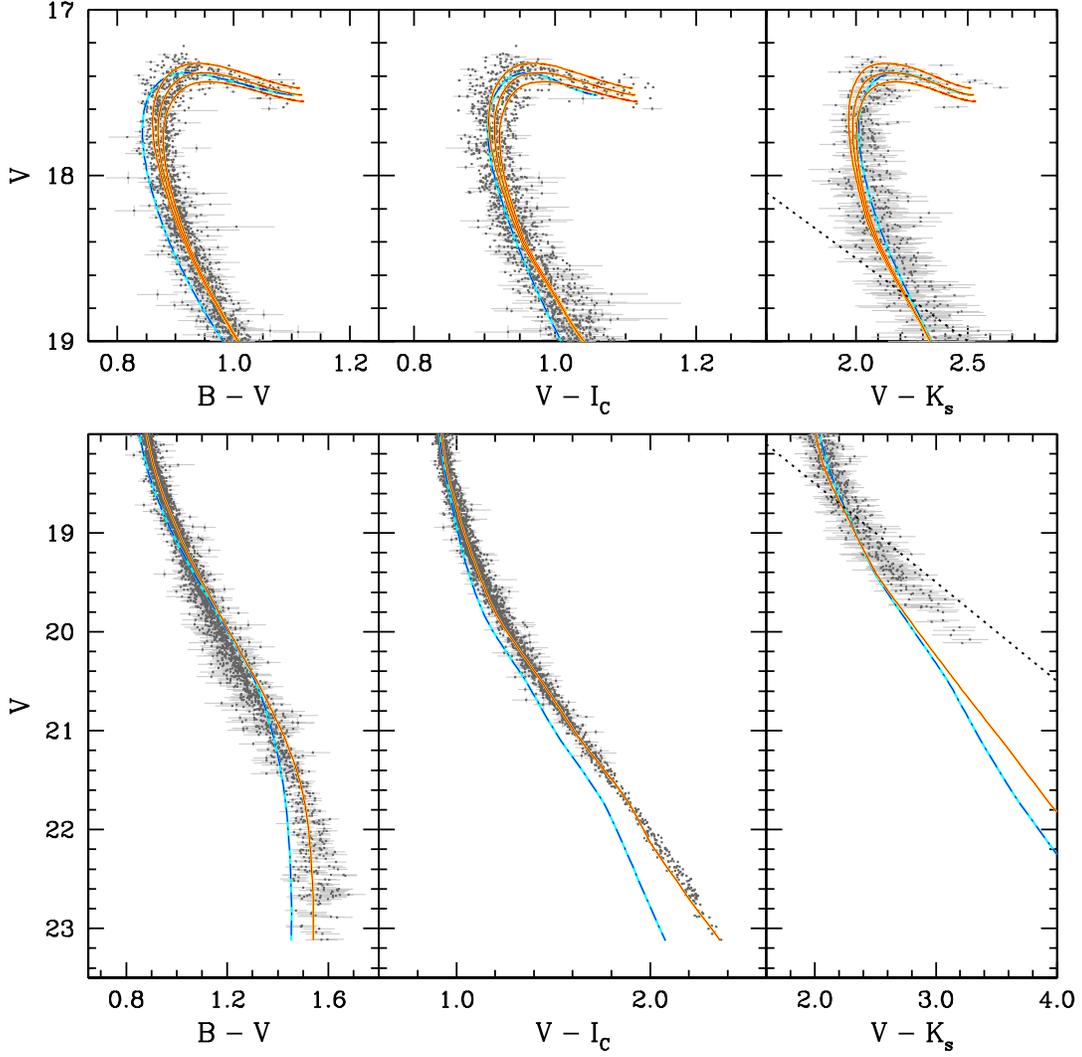}
\caption{Isochrone fits on CMDs of NGC~6791. Only those left from the photometric filtering routine are shown with error bars, representing a single-star cluster sequence. Solid red lines are calibrated isochrones with the best-fitting parameters (without binary corrections): [M/H]$=+0.38$, $\ebv=0.108$, $\dmn=13.00$, and the age of $9.5$~Gyr. The same models with $\pm0.5$~Gyr in age are additionally shown by solid red lines in the top panels. The dotted-dashed blue lines are the same theoretical models, but with \citet{lejeune:97,lejeune:98} colors. The diagonal dotted lines on the right panels represent the completeness limit in $K$ in \citet{carney:05}.  \label{fig:cmd2}} \end{figure*}

Figure~\ref{fig:cmd2} displays CMDs with $\bv$, $\vi$, and $\vk$ indices, where the best-fitting isochrones are shown as solid red lines. Models with $\pm0.5$~Gyr age differences are also displayed to show the quality of fits with varying ages. The dotted-dashed blue lines are the same theoretical model, but with the original \citet{lejeune:97,lejeune:98} colors (i.e., without color calibration). Compared to the calibrated isochrones, original model colors are generally bluer, and the difference becomes a few hundredths of a magnitude at the bottom of the MS (see bottom panels). In the $VK_s$ CMD, the best-fitting models tend to have bluer colors than the observed MS near MS turn-off (see top panels), which could be reconciled by an older age and/or a higher reddening.  This is partly because our global $\chi^2$ solution was heavily weighted by lower MS stars, in particular by stars in the $VI_C$ CMD with its longer color baseline than the other color indices. As noted above, we restricted fits to $\bv \le 1.1$, $\vi \le 2.0$, and $\vk \la 2.2$.

\subsection{Errors in the Cluster Parameters}\label{sec:error}

\input{tab5.tex}

We took a $95.4\%$ confidence interval in Figure~\ref{fig:chi} as a conservative measure of a $1\sigma$ fitting error in the cluster parameter estimates.  These are shown in the first row (``fitting'' error) in Table~\ref{tab:error}. However, a total error budget for the cluster parameters is no longer dominated by fitting errors, by means of our empirical color corrections, but by the sum of various systematic errors such as those listed in the first column of Table~\ref{tab:error}.  They include systematic errors in the helium enrichment parameter $({\Delta Y} / {\Delta Z})$, reddening laws [$R_V$, $R_{VI} \equiv \evi / \ebv$, and $R_{VK} \equiv \evk / \ebv$], zero points in the cluster photometry [$\Delta V$, $\Delta K_s$, $\Delta (\bv)$, and $\Delta (\vi)$], cluster parameters of the calibrating systems, and the bolometric correction of the Sun ($BC_{V,\odot}$).  For alternative values of each of these errors, we repeated the above search for the minimum $\chi_{\rm tot}^2$ value using a downhill simplex method \citep{press:92}, and estimated a size of errors in ${\rm [M/H]}$, age, $\ebv$, and $\dmn$, as listed in the third through sixth columns.  At the bottom of the table, we list total errors as a quadrature sum of these error contributions.  Below we provide detailed information on each of these error sources.

The helium abundance ($Y$) sensitively affects isochrone luminosity by ${\Delta M_V} / {\Delta Y} \sim 3$ at fixed $\teff$ and heavy-element mass fraction ($Z$).  To explore effects of helium abundance on our derived cluster parameters, we compared our MS fitting results with those using models with different helium enrichment parameters, ${\Delta Y} / {\Delta Z} = 2.0$ and $3.5$ (G.\ Newsham \& D.\ M.\ Terndrup 2007, private communication).  These models were constructed from the same YREC suite of programs, and many input parameters are similar to those used in our models, except that ${\Delta Y} / {\Delta Z} = 1.2$ was assumed in our base models. At fixed $\teff$, the above models with higher ${\Delta Y} / {\Delta Z}$ are approximately $0.1$ and $0.3$~mag fainter in bolometric magnitudes than our base models at [M/H]$=+0.4$.  The $\alpha$-element abundances in their models were set based on the abundance patterns observed for giants in the Galactic bulge \citep{fulbright:07}, but [$\alpha$/Fe] ratios at the metallicity of NGC~6791 are nearly solar.

We applied the same empirical color-$\teff$ corrections (Table~\ref{tab:corr}) to the alternative models with ${\Delta Y} / {\Delta Z} = 2.0$ and $3.5$, and derived cluster parameters from the $\chi_{\rm tot}^2$ minimization. We obtained an error dependence by fitting a line to values of each cluster parameter as a function of ${\Delta Y} / {\Delta Z}$. Errors in each quantity were then estimated assuming $\sigma ({\Delta Y} / {\Delta Z}) = 0.17$, as derived from the primordial helium abundance, the initial solar value, and helium abundances of the Hyades and the Pleiades (see Paper~III). A higher ${\Delta Y} / {\Delta Z}$ results in a fainter MS at a fixed color or a shorter distance to the cluster.  However, the effect of helium on stellar isochrones is nearly monochromatic, meaning that photometric metallicity and reddening estimates are weakly dependent on ${\Delta Y} / {\Delta Z}$.

The next three rows in Table~\ref{tab:error} show effects of errors in the reddening laws.  We adopted the same size of errors in $R_V$, $R_{VI}$, and $R_{VK}$ as in Paper~III.  The next four rows list errors resulting from zero-point errors in the photometry. Within the fitting range ($\bv \leq 1.10$), we assumed $0.005$~mag as a measure of systematic errors in the optical photometry. \citet{carney:05} found that a weighted mean difference between their CIT $K$-band photometry and 2MASS values transformed onto the CIT system \citep{carpenter:01} is $0.014\pm0.005$~mag, with no apparent dependence on magnitude. This difference is larger than the calibration uncertainty ($0.007$~mag) that was specified in the explanatory supplement to the 2MASS All Sky Data Release.\footnote{See http://www.ipac.caltech.edu/2mass/releases/allsky/doc/explsup.html} We took an error in $K_s$ as a quadrature sum of these two errors. The sense in Table~\ref{tab:error} is that if $\bv$ photometry were systematically underestimated by $0.005$~mag (bluer) than true values, a true metallicity is $0.028$~dex higher than the photometric metallicity.

Calibration errors in Table~\ref{tab:error} are a quadrature sum of parameter errors from calibrating cluster systems employed in this study. Since models matched to NGC~6791 CMDs were primarily calibrated using both the Hyades and Praesepe, we took errors from these two clusters: for the Hyades, we adopted errors of $\sigma({\rm [Fe/H]})=0.01$, $\sigma[\ebv]=0.002$, $\sigma[\dmn]=0.01$. For Praesepe, we adopted $\sigma({\rm [Fe/H]})=0.02$, $\sigma[\ebv]=0.002$, $\sigma[\dmn]=0.04$.  We simply took $100$~Myr for an error in age from the consideration that both clusters are about $550$--$650$~Myr old.

There can be a zero-point shift in the absolute $V$ magnitude scale, depending on the adopted absolute visual magnitude of the Sun. The $M_V$ of the Sun in our solar-metallicity model is $M_{V,\odot}=4.86$ at $T_{{\rm eff},\odot}=5777$~K assuming a solar age of $4.57$~Gyr. This value is $0.05$~mag larger than $M_{V,\odot}=4.81\pm0.03$ in \citet{torres:10}. We adopted the difference as a measure of a systematic error in the bolometric correction ($BC_{V,\odot}$), and included it in Table~\ref{tab:error}.

In addition to the above systematic errors, unresolved cluster binaries and/or photometric blends can modify MS-fitting parameter estimates, as they are brighter than single MS stars at a given color, forming a skewed distribution above a single MS.  Although our photometric filtering routine reduced a number of blended sources on CMDs, low mass-ratio binaries in particular could remain and affect our parameter estimation. \citet{bedin:08} estimated a binary fraction\footnote{The binary fraction is defined as the number of binaries divided by the total number of systems.} of NGC~6791 along the MS, and found $32\pm3\%$ in the core of the cluster. Because the fraction is rather uncertain, we adopted a $40\%$ binary fraction like those used for nearby systems in Paper~III. At a $40\%$ binary fraction, results from artificial cluster tests (Paper~III) showed that photometric metallicities are underestimated by $\sim0.036$~dex, while $\ebv$ are overestimated by $\sim0.003$~mag even after the photometric filtering. The corresponding change in distance modulus is $\Delta \dmn \approx +0.04$~mag.  The effect on age was not included in the simulation, but is likely to be small.

Together with the calibration and the bolometric correction errors in Table~\ref{tab:error}, the binary correction is one of the major systematic uncertainties in the determination of the cluster's distance. However, we caution that binary corrections are uncertain owing to the poorly constrained nature of binaries in the cluster, such as a binary fraction, a mass function for secondaries, and a distribution of binary mass ratios.  Nevertheless, the above biases are systematic in nature, and we list adjusted values of the cluster's metallicity, reddening, and distance with binary corrections in the second row of Table~\ref{tab:par}. We added in quadrature the size of the binary corrections to the total error budget.  The cluster CMDs are relatively free of foreground/background stars unlike the visual impression of CMDs.  We contend that effects of these non-cluster members are small compared to the other systematic errors.

\section{Summary and Discussions}\label{sec:summary}

We extended our effort to calibrate stellar isochrones in the Johnson-Cousins and 2MASS filter systems, using a set of photometric data for well-studied open cluster systems. Our base models were constructed using the YREC suite of programs, where model $\teff$ were initially converted into observable colors using color-$\teff$ transformations in \citet{lejeune:97,lejeune:98}.  In our earlier series of papers, we defined color corrections as a function of $\teff$ to match photometry of the Hyades and the Pleiades. In this work, we added Praesepe, extending earlier color calibration to cooler MS stars. We focused our analysis on the test of the models, and thus our empirical color corrections, using cool and metal-rich stars in NGC~6791.

There exists no geometric distance measurement for NGC~6791. As an alternative, we utilized an accurate metallicity of the cluster from a number of recent high-resolution spectroscopic studies to validate our calibrated models. Since optical and near-IR color indices have different sensitivities to metallicity, we examined whether the cluster's distance and reddening estimates from different CMDs converge on each other at the spectroscopic metallicity of the cluster. When the original \citet{lejeune:97,lejeune:98} color-$\teff$ relations were used to transform theoretical $\teff$ into colors, we found internally inconsistent results in distance and reddening. However, we demonstrated that our empirical color corrections to the \citeauthor{lejeune:97} relations bring about the consistency, improving the accuracy of MS-fitting distance and reddening estimates. We also found that YREC models are in satisfactory agreement with the observed mass-radius and mass-luminosity relations from the two binary systems in NGC~6791, when reasonable assumptions on $\teff$ scale and helium abundance were made. The above tests support our earlier conclusion that discrepancies found between models and observed CMDs mainly originate from systematic errors in the color-$\teff$ transformations. Our calibrated isochrones are valid at least within a metallicity range covered by the calibrating cluster systems and NGC~6791 ($0 \la {\rm [Fe/H]} \la +0.4$).

The accuracy of MS fitting can be improved by our empirical color-$\teff$ corrections. In particular, differential sensitivities of broadband colors enabled us to constrain reddening, distance, and metallicity of NGC~6791. Our photometric solution yields $\ebv=0.105\pm0.014$, [M/H]$=+0.42\pm0.07$, $\dmn = 13.04\pm0.09$, and the age of $9.5\pm0.3$~Gyr, after excluding $\bv$ colors of the cool MS stars in the cluster. Errors in these quantities include all potential systematic errors involved in the MS fitting. Our best-fitting age may be $\sim1$~Gyr too old since our models do not incorporate microscopic diffusion.  Our $\ebv$ estimate is found at the lower boundary of earlier $\ebv$ estimates in the literature, but the lower reddening also leads to a better agreement of models with the eclipsing binary data, when temperatures are estimated from the IRFM relations in \citet{casagrande:10}.

However, we found that the best-fitting models from the upper MS do not match well observed $\bv$ colors in the lower MS of NGC~6791. We could not fully resolve this unfortunate circumstance.  Nevertheless, an excellent agreement in $\vi$ for the same set of stars, as well as tests using IRFM color-$\teff$ relations and independent cluster photometry, suggests that the problem could originate from a color-scale error in the \citeauthor{stetson:03} photometry. While the photometric difference of $\sim0.02$~mag could still be considered to be acceptable in many applications, our empirical corrections are probably better defined. For example, the $\bv$ correction in the top panel of Figure~\ref{fig:corr} shows that the scatter of individual points around the mean line is $\sim0.016$~mag at $\teff\sim4500$~K. If our adopted cluster photometry is correct, the dispersion translates into an error in the mean of $0.007$~mag in a $4$--$8$ point moving window.

If stars with $\sim0.02$~mag photometric offsets in $\bv$ were included in MS fitting, our photometric solution of the cluster parameters would yield a smaller metallicity ($\Delta {\rm [M/H]}\sim0.1$~dex), a larger foreground reddening ($\Delta \ebv \sim 0.01$), and a shorter distance modulus ($\Delta \dmn \sim 0.1$), according to our error matrix in Table~\ref{tab:error}. Since the presence of such photometric offsets could reduce the accuracy of MS fitting, we excluded $\bv$ colors of the red MS stars in the above parameter estimates. Until more evidence is collected, our calibrated isochrones can be used with limited accuracy for the cool and metal-rich MS stars.

Our best-fitting distance modulus for NGC~6791 is in good agreement with \citet{brogaard:11}, who found $\dmn=13.01\pm0.08$ from eclipsing binary systems. They inferred luminosities of individual eclipsing binary components from measured stellar radii and spectroscopic temperatures, and adopted bolometric corrections in $V$ from MARCS models to derive absolute $V$ magnitudes of stars. In their derivation of the cluster's distance, they assumed $\ebv=0.160\pm0.025$ from a comparison between spectroscopic and photometric temperatures of the binary components.  Based on the distance estimate in \citet{brogaard:11}, \citet{brogaard:12} extended their analysis of NGC~6791 by combining the cluster's CMDs. They employed Victoria-Regina and DSEP isochrones \citep{dotter:08}, and found that the cluster is about $8.3$~Gyr old, and has a helium mass fraction of $Y=0.30\pm0.01$ when the metallicity is [Fe/H]$=+0.35$ with \citet{grevesse:98} abundance mixtures. From the isochrone fitting on the $VI_C$ CMD, they derived $\ebv=0.14\pm0.02$ for the global mean of the color excess.

More recent work \citep{casagrande:14,vandenberg:14} claimed that equally acceptable isochrone fits can be obtained with slightly different parameter inputs from those in \citet{brogaard:12}, although they did not fully explore all available parameter space. For example, \citet{casagrande:14} found that an $8.5$~Gyr old Victoria-Regina model at $Y=0.28$ and [Fe/H]$=+0.30$ with \citet{asplund:09} abundance mixtures shows nearly indistinguishable fits when $\ebv=0.16$ is assumed on $BVIJ$ and $ugriz$ CMDs. Similarly, \citet{vandenberg:14} used the same model as in \citeauthor{casagrande:14} to find a good fit with $\ebv=0.14$ from the $VJ$ CMD, and attributed the difference in $\ebv$ from \citeauthor{casagrande:14} to a potential zero-point error in $J$-band photometry.

These recent $\ebv$ estimates, although not statistically significant given the size of our estimated {\it total} errors, are systematically higher than our best-fitting solution [$\ebv=0.105\pm0.014$] and other previous measurements \citep{janes:84,montgomery:94,chaboyer:99,stetson:03,carraro:06,montalto:07}.  The $\sim0.03$--$0.05$~mag difference in $\ebv$ is partly due to different model isochrones employed in each of the studies. Although Victoria-Regina \citep{vandenberg:14} and YREC models agree with each other on the bolometric magnitude versus temperature plane at the metallicity of NGC~6791 (assuming the same helium abundance), the former models with color-$\teff$ conversions from MARCS produce slightly bluer ($\sim0.02$~mag) and brighter magnitudes ($\sim0.1$~mag) near the cluster's MS turn-off in the $BVI_C$ CMDs.

However, it should be emphasized that our empirical color corrections were defined to bring the isochrones to match the observed MS of nearby clusters, instead of relying solely on theoretical computations \citep[e.g.,][]{casagrande:14}, and therefore should provide more reliable predictions on colors and magnitudes of stars. In addition, we note that isochrone fitting in the above studies \citep{casagrande:14,vandenberg:14} focused on relatively bright stars in the cluster ($V \la 20$ or $\teff \ga 4700$~K in $VI_C$ CMD), while their best-fitting isochrones show systematic departures for cooler stars.  On the other hand, our work in this paper concerns model matches down to $\vi \sim 2.0$ or $V\sim22$~mag.

Distances estimated using our calibrated models are ultimately tied to the {\it Hipparcos} distance to the Hyades and the mean geometric distance to the Pleiades (see Paper~III). The distance to Praesepe adopted in the current work was derived using the Hyades-based calibration. Our luminosity calibration helps not only to determine distances to clusters and individual stars, but also to make an accurate estimate of their foreground reddening. Both quantities are necessary to have a proper luminosity calibration of astrophysical objects. Nevertheless, there is an important limitation in our methodology that the current empirical corrections rely on a set of calibrating clusters near solar metallicity ([Fe/H]$\approx+0.04$ or $+0.14$). Although we demonstrated the accuracy of our models at [Fe/H]$\sim+0.4$ using different color indices, our calibrations should ultimately be tested and confirmed using metal-rich stars with accurately known distance and reddening. Currently, there are only a few star clusters that can be used for accurate calibrations of stellar colors and magnitudes with precise cluster parameters.  However, there will be more cluster systems available  in the era of Gaia \citep{perryman:01}, over a wider range of stellar mass, metallicity, and age.

\acknowledgements

D.\ A.\ thanks Franck Delahaye for checking his models employed in this work. D.\ A.\ and J.-W.\ L.\ acknowledge support provided by the National Research Foundation of Korea to the Center for Galaxy Evolution Research (No.\ 2010-0027910). D.\ A.\ was partially supported by Basic Science Research Program through the National Research Foundation of Korea (NRF) funded by the Ministry of Education (2010-0025122). M.\ P.\ and D.\ T.\ acknowledge support from NASA grant NNX15AF13G and NSF grant AST-1411685. This publication makes use of data products from the 2MASS, which is a joint project of the University of Massachusetts and the Infrared Processing and Analysis Center/California Institute of Technology, funded by the National Aeronautics and Space Administration and the National Science Foundation.

{}

\end{document}

%% file: tab1.tex
\begin{deluxetable*}{llccll}
\tablewidth{0pt}
\tabletypesize{\scriptsize}
\tablecaption{A Library of Empirical Calibrated Isochrones\label{tab:library}}
\tablehead{
  \colhead{} &
  \colhead{Interior} &
  \colhead{Base Color-$\teff$} &
  \colhead{Filter} &
  \colhead{Calibrating} &
  \colhead{Valid Range} \nl
  \colhead{Reference} &
  \colhead{Models} &
  \colhead{Relations} &
  \colhead{Passbands} &
  \colhead{Clusters} &
  \colhead{in $\teff$}
}
\startdata
Papers~II and III & YREC (OPAL) & \citeauthor{lejeune:97} & $BVI_CK_s$ & Hyades                      & $4,000$--$8,200$~K \nl
Paper~IV          & YREC (OPAL) & \citeauthor{lejeune:97} & $BVI_CJHK_s$ & Hyades, Pleiades            & $4,000$--$12,000$~K \nl
This work         & YREC (OP)   & \citeauthor{lejeune:97} & $BVI_CJHK_s$ & Hyades, Pleiades, Praesepe  & $3,600$--$12,000$~K \nl
\hline
\citet{an:09b} & YREC (OP) & MARCS & $ugriz$ & M67 & $4,000$--$6,000$~K \nl
\citet{an:13} & YREC (OP) & MARCS & $ugriz$ & Primarily M67 and M92 & $4,000$--$7,000$~K \nl
\enddata
\end{deluxetable*}

%% file: tab2.tex
\begin{deluxetable}{lccc}
\tablewidth{0pt}
\tablecaption{Metallicity of NGC~6791 from High-Resolution Spectroscopy
\label{tab:feh}}
\tablehead{
  \colhead{Reference} &
  \colhead{{\rm [Fe/H]}} &
  \colhead{$\sigma$ (internal)} &
  \colhead{$\sigma$ (systematic)}
}
\startdata
\citet{gratton:06}   & $+0.47$ & $0.04$ & $0.08$   \nl
\citet{carraro:06}   & $+0.39$ & $0.01$ & \nodata  \nl
\citet{origlia:06}   & $+0.35$ & $0.02$ & $\la0.1$ \nl
\citet{brogaard:11}  & $+0.29$ & $0.03$ & $0.07$   \nl
\citet{geisler:12}   & $+0.42$ & $0.01$ & \nodata  \nl
\citet{cunha:15}     & $+0.34$ & $0.06$ & \nodata  \nl
\citet{boesgaard:15} & $+0.30$ & $0.02$ & $\la0.1$ \nl
\hline
Unweighted Mean      & $+0.37$ & $0.07$\tablenotemark{a} & \nl
\enddata
\tablenotetext{a}{Unweighted standard deviation.}
\end{deluxetable}

%% file: tab3.tex
\begin{deluxetable}{cccccc}
\tablewidth{0pt}
\tabletypesize{\scriptsize}
\tablecaption{Empirical Color Corrections\label{tab:corr}}
\tablehead{
  \colhead{$T_{\rm eff}$} &
  \colhead{$\Delta (B\, -\, V)$} &
  \colhead{$\Delta (V\, -\, I_C)$} &
  \colhead{$\Delta (V\, -\, K_s)$} &
  \colhead{$\Delta (J\, -\, K_s)$} &
  \colhead{$\Delta (H\, -\, K_s)$} \\
  \colhead{(K)} &
  \colhead{(mag)} &
  \colhead{(mag)} &
  \colhead{(mag)} &
  \colhead{(mag)} &
  \colhead{(mag)}
}
\startdata
 $3600$ & $ 0.0877$ & $ 0.2857$ & $ 0.3947$ & $ 0.0810$ & $ 0.0200$ \nl
 $3650$ & $ 0.0877$ & $ 0.2834$ & $ 0.3947$ & $ 0.0810$ & $ 0.0200$ \nl
 $3700$ & $ 0.0877$ & $ 0.2546$ & $ 0.3817$ & $ 0.0845$ & $ 0.0241$ \nl
 $3750$ & $ 0.0868$ & $ 0.2231$ & $ 0.3548$ & $ 0.0905$ & $ 0.0291$ \nl
 $3800$ & $ 0.0856$ & $ 0.1948$ & $ 0.3256$ & $ 0.0955$ & $ 0.0311$ \nl
\enddata
\tablecomments{Only a portion of this table is shown here to demonstrate its form and content. A machine-       readable version of the full table is available.}
\end{deluxetable}

%% file: tab4.tex
\begin{deluxetable*}{crrrr}
\scriptsize
\tablewidth{0pt}
\tablecaption{Average Color Differences from Calibrated Isochrones in NGC~6791\label{tab:colordiff}}
\tablehead{
  \colhead{$V$} &
  \multicolumn{2}{c}{$\ebv=0.11$} &
  \multicolumn{2}{c}{$\ebv=0.14$} \\
  \cline{2-3}
  \cline{4-5}
  \colhead{(mag)} &
  \colhead{$\Delta (\bv)$} &
  \colhead{$\Delta (\vi)$} &
  \colhead{$\Delta (\bv)$} &
  \colhead{$\Delta (\vi)$}
}
\startdata
$17.75$ & $ 0.0047\pm0.0007$ & $-0.0013\pm0.0009$ & $-0.0313\pm0.0011$ & $-0.0467\pm0.0011$ \\
$18.25$ & $ 0.0080\pm0.0005$ & $-0.0033\pm0.0007$ & $-0.0069\pm0.0006$ & $-0.0256\pm0.0007$ \\
$18.75$ & $ 0.0059\pm0.0005$ & $-0.0060\pm0.0008$ & $ 0.0002\pm0.0005$ & $-0.0151\pm0.0008$ \\
$19.25$ & $-0.0007\pm0.0006$ & $-0.0025\pm0.0008$ & $-0.0008\pm0.0006$ & $-0.0033\pm0.0008$ \\
$19.75$ & $-0.0154\pm0.0007$ & $ 0.0124\pm0.0009$ & $-0.0107\pm0.0007$ & $ 0.0216\pm0.0010$ \\
$20.25$ & $-0.0256\pm0.0009$ & $ 0.0012\pm0.0010$ & $-0.0192\pm0.0009$ & $ 0.0346\pm0.0009$ \\
$20.75$ & $-0.0283\pm0.0012$ & $-0.0053\pm0.0011$ & $-0.0235\pm0.0012$ & $ 0.0278\pm0.0011$ \\
$21.25$ & $-0.0226\pm0.0019$ & $-0.0027\pm0.0014$ & $-0.0239\pm0.0019$ & $ 0.0373\pm0.0014$ \\
$21.75$ & $-0.0035\pm0.0031$ & $-0.0046\pm0.0019$ & $-0.0147\pm0.0031$ & $ 0.0320\pm0.0016$ \\
$22.25$ & $ 0.0032\pm0.0048$ & $ 0.0251\pm0.0027$ & $-0.0169\pm0.0048$ & $ 0.0496\pm0.0029$
\enddata
\tablecomments{The sense is observed minus model colors. Errors represent a standard error in the mean color difference.}
\end{deluxetable*}

%% file: tab6.tex
\begin{deluxetable*}{lcccc}
\tablewidth{0pt}
\tablecaption{Summary of Cluster Parameters for NGC~6791\label{tab:par}}
\tablehead{
  \colhead{} &
  \colhead{[M/H]} &
  \colhead{Age} &
  \colhead{$E(B - V)$} &
  \colhead{$(m - M)_0$} \\
  \colhead{Solution} &
  \colhead{(dex)} &
  \colhead{(Gyr)} &
  \colhead{(mag)} &
  \colhead{(mag)}
}
\startdata
Global $\chi^2$ Minimum        & $0.38\pm0.06$ & $9.5\pm0.3$ & $0.108\pm0.014$ & $13.00\pm0.07$ \nl
Bias-corrected for Binaries    & $0.42\pm0.07$ & $9.5\pm0.3$ & $0.105\pm0.014$ & $13.04\pm0.08$ \nl
\enddata
\end{deluxetable*}

%% file: tab5.tex
\begin{deluxetable*}{llcccc}
\tablewidth{0pt}
\tablecaption{Error Budget for the Cluster Parameters of NGC~6791\label{tab:error}}
\tablehead{
  \colhead{Source of} &
  \colhead{Adopted} &
  \colhead{$\Delta {\rm [M/H]}$} &
  \colhead{$\Delta$Age} &
  \colhead{$\Delta E(B - V)$} &
  \colhead{$\Delta (m - M)_0$} \\
  \colhead{Error} &
  \colhead{Size of Error} &
  \colhead{(dex)} &
  \colhead{(Gyr)} &
  \colhead{(mag)} &
  \colhead{(mag)}
}
\startdata
Fitting\dotfill             & \nodata    & $\pm0.015$ & $\pm0.20$ & $\pm0.004$ & $\pm0.023$ \nl
$\Delta Y/\Delta Z$\dotfill & $\pm0.17$  & $\pm0.001$ & $\pm0.02$ & $\pm0.002$ & $\mp0.016$ \nl
$R_V$\dotfill               & $\pm0.30$  & $\pm0.022$ & $\mp0.07$ & $\mp0.007$ & $\mp0.006$ \nl
$R_{VI}$\dotfill            & $\pm0.07$  & $\pm0.027$ & $\mp0.16$ & $\mp0.007$ & $\pm0.025$ \nl
$R_{VK}$\dotfill            & $\pm0.12$  & $\pm0.002$ & $\pm0.07$ & $\mp0.002$ & $\mp0.002$ \nl
$\Delta V$\dotfill          & $\pm0.005$ & $\mp0.002$ & $\mp0.02$ & $\pm0.001$ & $\pm0.005$ \nl
$\Delta (B-V)$\dotfill      & $\pm0.005$ & $\pm0.028$ & $\mp0.04$ & $\mp0.004$ & $\pm0.004$ \nl
$\Delta (V-I)_C$\dotfill    & $\pm0.005$ & $\mp0.021$ & $\pm0.10$ & $\pm0.006$ & $\mp0.018$ \nl
$\Delta K_s$\dotfill        & $\pm0.016$ & $\pm0.005$ & $\pm0.06$ & $\mp0.002$ & $\pm0.001$ \nl
Calibration\dotfill         & \nodata    & $\pm0.022$ & $\pm0.10$ & $\pm0.003$ & $\pm0.041$ \nl
$BC_{V,\odot}$\dotfill      & $\pm0.04$  & \nodata    & \nodata   & \nodata    & $\pm0.040$ \nl
\hline
Total\dotfill               & \nodata    & $\pm0.056$ & $\pm0.32$ & $\pm0.014$ & $\pm0.071$ \nl
\enddata
\end{deluxetable*}